\documentclass[10pt]{article}
\usepackage{amsmath}
\usepackage{graphicx,psfrag,epsf}
\usepackage{enumerate}
\usepackage{natbib}
\usepackage{url} 
\usepackage{subcaption}
\newcommand{\blind}{1}

\newtheorem{theorem}{Theorem}
\newtheorem{defn}{Definition}

\newtheorem{assump}{Assumption}
\newtheorem{result}{Result}
\newtheorem{cond}{Condition}
\newtheorem{lemma}{Lemma}
\newtheorem{remark}{Remark}

\usepackage{hyperref}
\hypersetup{
    colorlinks=true,
    linkcolor=blue,
   filecolor=magenta,
    urlcolor=blue,
}
\usepackage[utf8]{inputenc}
\usepackage{float}
\usepackage{amssymb}
\begin {document}

\def\spacingset#1{\renewcommand{\baselinestretch}%
	{#1}\small\normalsize} \spacingset{1}


\if1\blind
{
	\title{ Adapting BH to One- and Two-Way Classified Structures of Hypotheses}
	\author{Shinjini Nandi \thanks{
			Shinjini Nandi is a PhD student at Department of Statistical Science, Temple University (email: shinjini.nandi@temple.edu);}\\
		and \\
		Sanat K. Sarkar\thanks{
		Sanat K. Sarkar is Professor of Department of Statistical Science, Temple University, Philadelphia, PA 19122 (email: sanat@temple.edu). This work is based on Nandi's doctoral research. Sarkar’s research was supported by NSF Grants DMS-1208735 and DMS-1309273.}\hspace{.2cm}\\
	Department of Statistical Science, Temple University}
	\date{}
	\maketitle
} \fi

\if0\blind
{
	\bigskip
	\bigskip
	\bigskip
	\begin{center}
		{\LARGE\bf Adapting BH to One- and Two-Way Classified Structures of Hypotheses}
	\end{center}
	\medskip
} \fi

\bigskip
\begin{abstract}
Multiple testing literature contains ample research on controlling false discoveries for hypotheses classified according to one   criterion, which we refer to as one-way classified hypotheses. Although simultaneous classification of hypotheses according to two different criteria, resulting in two-way classified hypotheses, do often occur in scientific studies, no such research has taken place yet, as far as we know, under this structure. This article produces procedures, both in their oracle and data-adaptive forms, for controlling the overall false discovery rate (FDR) across all hypotheses effectively capturing the underlying one- or two-way classification structure. They have been obtained by using results associated with weighted Benjamini-Hochberg (BH) procedure in their more general forms providing guidance on how to adapt the original BH procedure to the underlying one- or two-way classification structure through an appropriate choice of the weights. The FDR is maintained non-asymptotically by our proposed procedures in their oracle forms under positive regression dependence on subset of null $p$-values (PRDS) and in their data-adaptive forms under independence of the $p$-values. Possible control of FDR for our data-adaptive procedures in certain scenarios involving dependent $p$-values have been investigated through simulations. The fact that our suggested procedures can be superior to contemporary practices has been demonstrated through their applications in simulated scenarios and to real-life data sets. While the procedures proposed here for two-way classified hypotheses are new, the data-adaptive procedure obtained for one-way classified hypotheses is alternative to and often more powerful than those proposed in \cite{Huetal2010}.

\end{abstract}

\noindent%
\vfill

\newpage
\spacingset{1.2} 
\section{Introduction}
Large-scale multiple testing problems often involve classifying a set of hypotheses into several groups. In some cases, the families/groups might be formed naturally due to characteristics of the underlying scientific experiment. In other situations, a certain feature attributable to each hypothesis might serve as the basis of partition. Grouping of hypotheses due to any well defined argument, whether natural or artificial, benefits analyses. In all such instances, since the classification is due to a single criterion, we refer to the setup as `one-way classification' of hypotheses. Multiple testing procedures adapted to such arrangement of hypotheses incorporate the information of similar characteristics within each group. Consequently, they can address problems specific to such structures, they usually have more power and better control over false discoveries than their counterparts that ignore group structures. One-way classified hypotheses have been widely investigated in the literature. Existing multiple testing procedures have been revamped (in  \cite{Pacifico2004}, \cite{BenjHeller2007}, etc.) to accommodate such layout. \cite{Huetal2010} introduced a weighted BH procedure for one-way  grouped hypotheses that assigns weights to each group proportional to the number of null hypotheses in it, before applying the BH procedure to the weighted hypotheses pooled together across all groups.
\cite{Ignatiadis2016} suggested a data driven weighted procedure to test similarly classified hypotheses. Any set of weights that depend on external covariates and satisfy some simple constraints can be considered for a testing procedure similar to the method suggested in \cite{Huetal2010}. The optimum set of weights are chosen subject to maximization of power using data-based optimization techniques.

In many situations, a set of hypotheses might be classified according to more than one norm of classification. Just like one-way classified hypotheses, multiple interesting features or nature of the experiment may determine the norms. For example, brain imaging studies involving fMRI data (\cite{2015arXiv151203397F}), geographical studies involving data collected through satellite remote sensing (\cite{clements2014}), studies in genetics involving microarray time course experiments (\cite{SunWei2011}), etc, comprise of spatio-temporal data. The multitude of hypotheses arising out of such data can be clustered into groups formed through aggregation of neighboring spatial units, and/or related time points.
Other examples can be found in bioinformatics, studies involving association between genes and proteins, and genomewide association studies that involve analysis of association of SNPs with different regions of the brain (\cite{STEIN20101160}). Examples where more than two types of classification are imposed simultaneously on a set of hypotheses are very rare. If a set of hypotheses is classified in exactly two different ways, we call it a set of `two-way classified hypotheses'.

In such cases, researchers are most interested in the hypotheses that emerge as significant when effects due to both classifications are factored in. The scope of existing multiple testing procedures is limited to one-way classified data and such methods are incapable to gauge the simultaneous effect of two-way classification. Some efforts made to study such structures in \cite{STEIN20101160}, \cite{SunWei2011}, etc. involve repetitive application of one-way classification multiple testing procedures. Broadly speaking, in the first step, one of the two classifications is prioritized over the other. Considering the hypotheses as classified only due to this factor, significant groups and/or individual hypotheses are determined. In the second step, these significant elements are further tested for significance due to the second grouping criterion and finally the set of significant hypotheses is determined. \cite{2015arXiv151203397F}, \cite{2017arXiv170306222R} discuss multi-way classification and suggest an algorithm that recursively applies BH procedure to all partitions created and selects the set of hypotheses as significant which are rejected in all partitions.

The goal of this article is to suggest a new data-adaptive multiple testing procedure for one-way classified hypotheses and broaden the scope of multiple testing procedures to two-way classified hypotheses.
In Section 2, we describe existing methodologies that serve as the background for developing our new methods. In section 3, this is followed by description of the one-way  classification model, the weighted multiple testing procedure and our proposed data-adaptive version of it. In Section 4, we introduce the two-way classification model and modify the multiple testing procedures for one-way classified hypotheses to suit to the new layout. We also discuss the corresponding data-adaptive procedures that can be applied to multiple hypotheses subjected to such classification. We establish that our proposed data-adaptive methods (both one-way and two-way) are adequate for finite sets of hypotheses, at least under independence. Section 5 demonstrates through simulation studies that the performances of our suggested methods are superior to existing practices in most practical scenarios. Though our suggested data-adaptive procedures are proven to control false discoveries for independent hypotheses, simulations show that for suitable choices of parameters, they are also applicable to positively dependent hypotheses, in certain scenarios involving high density of signals. To illustrate its utility, our proposed method is applied to a dataset on prevalence of microbial communities involving two-way classified hypotheses in section 6, and its performance is compared with that of an existing method. We end our paper with some concluding remarks in Section 7.

\section{Preliminaries and Basic Methodologies}  In this section, we recall some existing results on weighted and data-adaptive weighted $p$-value based FDR controlling procedures for testing a set of hypotheses with no specific group structure and present them in their general forms to set the stage for developing similar procedures in the larger domain of one- and two-way classified hypotheses. The discussions surrounding these results will provide ideas on the basic methodological steps that we will take to develop our proposed newer procedures in the next section.

Consider simultaneous testing of a set of $N$ hypotheses $H_1, \ldots, H_N$ based on their respective p-values $P_1, \ldots, P_N$ subject to a control over $$\text{FDR} = E \left [\frac{V_N}{\max\{R_N,1\}} \right ], $$ with $R_N$ and $V_N$ being the total numbers of rejected and falsely rejected null hypotheses, respectively, under the following assumption:

\begin{assump}\label{assump1} $P_i \sim U(0,1)$ for each $i \in I_0$, with $I_0$ being the set of indexes of null hypotheses. \end{assump}

Regarding dependence among the p-values, we assume that they are positively regression dependent on subset (PRDS) of null p-values, as defined below generally for any set of random variable $X_1, \ldots, X_k$:

\begin{cond}\label{cond1} A set of random variable $X_1, \ldots, X_k$ is said to be positively regression dependent on a particular subset $S$ of these random variables if $E\left[\phi(X_1, \ldots, X_k)| X_i = x\right]$ is non-decreasing in $x$, for each $X_i \in S$ and for any (coordinatewise) non-decreasing function $\phi$ of $(X_1, \ldots, X_k)$. \end{cond}

Clearly, independent p-values are PRDS. For examples of non-independent p-values satisfying Condition \ref{cond1}, the readers are referred to \cite{BY2001} and \cite{Sarkar2002}, \cite{Sarkar2008}. A weaker form of positive dependence condition, with $E\left[\phi(X_1, \ldots, X_k)| X_i = x\right]$ replaced by $E\left[\phi(X_1, \ldots, X_k)| X_i \le x\right]$, is often assumed in the literature in the context of BH type FDR controlling procedures [\cite{finner2009}, \cite{Sarkar2008}]. This condition could have been used instead of Condition \ref{cond1} in this paper without affecting our results relying on such a condition.

Let us now recall the definition of the BH procedure for a single group of hypotheses in its more general form in terms of weighted p-values.
\begin{defn}\label{defn1}
For a set of $N$ hypotheses, suppose that the $i$th p-value $P_i$ is assigned a non-stochastic weight $w_i \ge 0$, for $i=1, \ldots, N$. The weighted BH procedure at level $\alpha$ corresponding to these weights is a stepup procedure with the critical constants $i \alpha/N$, $i=1, \ldots, N$, i.e., it orders the weighted p-values $P_i^w = w_iP_i, i=1, \ldots, N$, in increasing order as $P_{(i)}^w$, $i=1, \ldots, N$, and rejects the hypotheses $H_{(1)}, \ldots, H_{(R)}$ corresponding to $P_{(1)}^w, \ldots, P_{(R)}^w$ where $$R = \max \left \{1\leq j \leq N: P_{(j)}^w \leq \frac{j\alpha}{N} \right \},$$ provided the maximum exists; otherwise, it rejects none.
\end{defn}

\begin{result}\label{result1}
The FDR of the weighted BH procedure based on p-values satisfying the PRDS condition is bounded above by $\frac{\alpha}{N}\sum\limits_{i \in I_0}\frac{1}{w_i}$.
\end{result}

A proof of this result using techniques from \cite{Sarkar2002} is provided in Appendix.

Result \ref{result1} serves as our foundation. It leads to systematic development of our proposed procedures in their oracle forms through appropriate choice of weights suited to either one- or two-way classification structures levied on the set of hypotheses, before we construct their appropriate data-adaptive versions. More specifically, one can determine weights that satisfy
\begin {eqnarray}\label{e1} \sum\limits_{i \in I_0}w_i^{-1} = N, \end{eqnarray}
and appropriately capture the underlying classification structure to develop a weighted BH procedure in its oracle form that controls the FDR at $\alpha$ across all hypotheses, conservatively under PRDS and exactly under independence, before constructing an appropriate  data-adaptive version of it. For instance, the choice of weights, $w_i=\pi_0 = |I_0|/N$, for all $i = 1, \ldots, N$, satisfying this condition yields the single-group BH procedure in its oracle form.
A data-adaptive version of it is the one that uses the existing data to estimate $\pi_0$. There are several such data-adaptive single-group BH procedures that have been put forward in the literature, for example, \cite{BH2000}, \cite{Storeyetal2004}, \cite{Sarkar2008}, and \cite{Blanchard:2009:AFD:1577069.1755880}.

The same condition is also satisfied by the weights chosen by \cite{Huetal2010} in  their construction of a weighted BH procedure in its oracle form, referred to as the one-way grouped BH procedure, in the context of testing one-way classified hypotheses. Its control over the FDR under PRDS is also proven and can now be seen to follow from Result \ref{result1}, which is more general. It will be revisited in the next section where we introduce a data-adaptive version of it that is not only different from the data-adaptive procedures originally proposed in \cite{Huetal2010}, but also more preferred in a non-asymptotic setting, where its FDR is theoretically shown to be controlled, at least under independence. 

The next section will also contain newer procedures in their oracle as well as data-adaptive forms that we introduce in this article for testing two-way classified hypotheses. The non-asymptotic FDR control of all these new data-adaptive procedures under independence is established using the following result.

\begin{result}\label{result2} The FDR of a data-adaptive weighted BH procedure with (co-ordinate wise) non-decreasing estimated weight functions $\hat{w}_i(\mathbf{P}) > 0$, $i=1,\ldots, N$, is bounded above by $\alpha$ under independence if \begin{align}\label{e2}
	E\left[\sum\limits_{i \in I_0} \frac{1}{\hat{w}_i(\mathbf{P}^{(-i)},0)}\right] \leq N
	\end{align} where $\hat{w}_i(\mathbf{P}^{(-i)},0)$ represents $\hat{w}_i$ as a function of $\mathbf{P}^{(-i)} = \{P_1, \ldots, P_N\}\setminus P_i$ with $P_i = 0$.
\end{result}
A proof of this result can be seen in \cite{Sarkar2008}.

We introduce below a newer class of estimates, expressing some of the existing ones in a more general form, which offers a wider scope of data-dependent adaptation of the BH procedure to both one- and two-way group structures of hypotheses with proven non-asymptotic FDR control under independence.
The following lemma will be useful in checking the inequality in (2) for this larger class of estimates:

\begin{lemma}\label{lemma1}
Let $R_N(\lambda) = \sum_{i=1}^N I(P_i \le \lambda)$, $V_N(\lambda) = \sum_{i \in I_0} I(P_i \le \lambda)$, and $R_{N-1}^{(-i)}(\lambda)= \sum_{j (\neq i)=1}^N I(P_j \le \lambda)$, for a fixed $\lambda \in (0,1)$. Then, for any non-negative real valued function $f$ of $R_N(\lambda)$, we have the following result:
\begin {eqnarray} \label{e3}
& &  E \left \{ \sum_{i \in I_0} \frac {(1-\lambda) f((R^{(-i)}_{N-1}(\lambda))}{N -R^{(-i)}_{N-1}(\lambda)} \right \} \le   E \left \{ f(R_N(\lambda)) \right \}. \end {eqnarray}
\end{lemma}
\textbf{Proof:} The inequality in (\ref{e3}) follows from the fact that $N_0-V_N(\lambda) \le N- R_N(\lambda)$, and so the right-hand side of that inequality is greater than or equal to
\begin{eqnarray} & & E\left \{ \frac{[N_{0} - V_{N}(\lambda)] f(R_{N}(\lambda))}{[N- R_{N}(\lambda)] \vee 1} \right \} \nonumber \\ & = &
E \left \{ \sum_{i \in I_0} \frac{I(P_i > \lambda) f(R_{N-1}^{(-i)}(\lambda)+ I(P_i \le \lambda))}{[N- R_{N-1}^{(-i)}(\lambda)- I(P_i \le \lambda)]\vee 1} \right \}, \nonumber \end{eqnarray} which reduces to the left-hand side of (\ref{e3}) since the $P_i$'s are independent.

\begin {remark} \rm It is important to note, as we proceed to use Lemma \ref{lemma1} to develop procedures under more complex structures of hypotheses in the next section, that $R$ will be subscripted differently under different structural settings since its definition should correctly reflect the number of hypotheses involved. \end{remark}

\section{One-Way Grouped BH Procedure: Adapting the BH Procedure to One-Way Classified Hypotheses}
Using the same notations as used in \cite{Huetal2010}, let us suppose that the $N$ hypotheses to be simultaneously tested are split into $m$ non-overlapping groups according to some criterion, with $n_{g \centerdot}$ pairs of hypothesis and the corresponding p-value, $(H_{gi},P_{gi})$, $i = 1, \ldots, n_{g \centerdot}$, falling in group $g$, and $N = \sum\limits_{g=1}^m n_{g \centerdot}$. Let $n_{g0}$ be the number of null hypotheses and $I_{g0} \subseteq \{1, \ldots, n_{g \centerdot}\}$ be the corresponding set of sub-indexes associated with $i$ in group $g$. The set of indexes of all null hypotheses among all hypotheses can then be expressed as $I_0 = \bigcup\limits_{g=1}^m I_{g0}$. Let $\pi_{g0} = n_{g0}/n_{g \centerdot}$ be the proportion of true nulls in group $g$, so that $\pi_0$, the proportion of true nulls in the entire set of $N$ hypotheses, can be expressed as $\pi_0 = \sum\limits_{g=1}^m n_{g \centerdot}\pi_{g0}/N$.

One-way grouped BH, shortly one-way GBH, is an oracle procedure. It is defined by \cite{Huetal2010} as a weighted BH procedure with the weights being formulated in terms $\pi_{g0}$, for $g=1, \ldots, m$, assuming they are known, in a way that allows the BH procedure to effectively adapt to the present structural setting of the hypotheses. We revisit it in the following sub-section, before developing our newly proposed data-adaptive version of it later in this section.

\subsection{Oracle One-Way GBH Procedure} It is a weighted BH procedure with \begin{align}\label{e4}
	w_g = \frac{\pi_{g0}(1- \pi_{0})}{1 - \pi_{g0}}
\end{align}
being assigned as weight to $P_{gi}$, for each $i = 1, \ldots, n_{g \centerdot}$, and $g = 1, \ldots, m$, assuming these proportions are all known. \cite{Huetal2010} referred to it as simply grouped BH, shortly GBH procedure, but as said above, we will refer to it here as one-way GBH procedure. Since
\begin{eqnarray}\label{e5}
\sum_{g=1}^m \sum_{i \in I_{g0}} w_{g}^{-1} = \frac{1}{1-\pi_0}\sum_{g=1}^m \frac{n_{g \centerdot}\pi_{g0}(1-\pi_{g0})}{\pi_{g0}} = \frac{1}{1-\pi_0}\sum_{g=1}^m n_{g \centerdot}(1-\pi_{g0}) = N, \end{eqnarray}
the equality in (\ref{e1}) is satisfied by these weights, and so we have the following theorem, which of course was proved in Hu et al. (2010) using different arguments.

\begin {theorem} \label{th1}
One-way GBH procedure controls the overall FDR under PRDS and Assumption \ref {assump1}. \end {theorem}

There is a Bayesian justification behind the choice of these weights, as articulated by Hu et al. (2010). However, a look at these weights from a different point of view seems to provide further insight into the effectiveness of these weights under the current setting.

For a group with small proportion of true nulls, $\pi_{g0}$ would be small. At the same time, it would have higher odds of being significant relative to other groups, as measured by $(1-\pi_{g0})/(1-\pi_0)$. Consequently, the weight associated with that group gets deflated, facilitating easier rejection of its members when the weighted BH procedure is applied to all the hypotheses.

This sort of interpretation for the weights guides us in understanding how to estimate them, differently from Hu et al. (2010)  and in constructing a data-adaptive version of one-way GBH procedure that we will describe below.

\subsection{Data-Adaptive One-Way GBH Procedure} We propose this procedure by considering the one-way GBH and replacing the weight $w_g$ in it by the following:
\begin{align}\label{e6}
\hat{w}_g = \frac{n_{g \centerdot} - R_{n_{g \centerdot}} + 1 }{N(1 - \lambda)} \cdot \frac{R_N + m -1}{R_{n_{g \centerdot}}}, g=1, \ldots, m,
\end{align}
where, for some fixed $\lambda \in (0,1)$, $R_{n_{g \centerdot}} \equiv R_{n_{g \centerdot}}(\lambda) = \sum\limits_{i=1}^{n_{g \centerdot}} I(P_{gi} \le \lambda)$, and $R_N = \sum\limits_{g= 1}^mR_{n_{g \centerdot}}$. We refer to this procedure as a data-adaptive one-way GBH procedure.

The idea of using this type of estimate for each $w_g$ came from its alternative interpretation noted above. We estimate $\pi_{g0}$ by $\hat{\pi}_{g0} = (n_{g \centerdot} - R_{n_{g \centerdot}} + 1)/n_{g \centerdot}(1 - \lambda)$,
which is a slight adjustment (considered by \cite{Storeyetal2004}), from $n_{g \centerdot} - R_{n_{g \centerdot}}$ to $n_{g \centerdot} - R_{n_{g \centerdot}} + 1$, made in the estimate originally used by \cite{Storey2002} for the proportion of true nulls in the context of single-group multiple testing. To estimate $(1-\pi_{0})/(1-\pi_{g0})$, we propose estimating $(1-\pi_{g0})/(1-\pi_0)$, which is the proportion of false nulls in group $g$ among all false nulls, by $N(R_{n_{g \centerdot}} + m-1)/n_{g}R_{N}$, having made a slight adjustment to its natural estimate $NR_{n_{g \centerdot}}/n_{g \centerdot}R_{N}$, and then inverting this estimate. When $m=1$, the $\hat{w}_g$ in expression (\ref{e6}) reduces to that of \cite{Storeyetal2004}.

\begin {theorem} \label{th2}
The above data adaptive one-way GBH procedure controls the overall FDR under independence among all p-values and Assumption \ref {assump1}. \end {theorem}

Proof. The theorem will follow from Result \ref{result1} if we can show that the estimated weights, treated as functions of all the p-values, used in adaptive one-way GBH satisfy the two conditions in Result \ref{result2} - (i) $\hat{w}_g$ is non-decreasing in $P_{gi}$ for each $g$, and (ii) the inequality in (\ref{e2}) holds.

The first condition follows by noting that both $n_{g \centerdot} - R_{n_{g \centerdot}} + 1$ and
$(R_N + m -1)/R_{n_{g \centerdot}} = 1+ \sum\limits_{g'\neq g}(R_{n_{g \centerdot '}}+ 1)/R_{n_{g \centerdot}}$ are non-increasing in each $R_{n_{g \centerdot}}$, which itself is non-increasing in each $P_{gi}$.

To show that the second condition is also satisfied, let us first express $\hat{w}_g$ as a function of $\mathbf{P}$, the set of all p-values, i.e., as $\hat{w}_g(\mathbf{P})$, for each $g$. Then, note that if we set $P_{gi}$ at $0$ for a particular pair $(g, i \in I_{g0})$, we get $$\hat{w}_g(\mathbf{P}^{-(g,i)}, 0) = \frac{n_{g \centerdot} - R_{n_{g \centerdot}-1}^{(-i)} }{N(1 - \lambda)}\; \frac{R_{n_{g \centerdot}-1}^{(-i)} + \sum_{g^{\prime} \neq g}R_{n_{g_{\centerdot}^{\prime}} } + m}{R_{n_{g \centerdot}-1}^{(-i)}+1}, \; g=1, \ldots, m,$$ where $R_{n_{g \centerdot}-1}^{(-i)} = \sum_{i^{\prime} \neq i} I(P_{gi^{\prime}} \le \lambda)$. Thus, we have
\begin{align}\label{e7}
&\text{E}\left \{  \sum\limits_{g=1}^m\sum\limits_{i \in I_{g0}}\frac{1}{\hat{w}_g(\mathbf{P}^{-(g,i)}, 0)} \right\}
= N  \text{E} \left \{ \sum\limits_{g=1}^m  \sum\limits_{i \in I_{g0}}\frac{(1 - \lambda) f(R_{n_{g \centerdot}-1}^{(-i)}, \sum_{g^{\prime} \neq g}R_{n_{g_{\centerdot}^{\prime}} })}{n_{g \centerdot} - R_{n_{g \centerdot}-1}^{(-i)}} \right\} \nonumber \\ \end{align}
where $$ f \left (x, \sum_{g^{\prime} \neq g}R_{n_{g_{\centerdot}^{\prime}} }\right ) = \frac{x+1}{x+\sum_{g^{\prime} \neq g}R_{n_{g_{\centerdot}^{\prime}} } + m }.$$  Applying Lemma \ref{lemma1} to the expectation in the right-hand side of equation (\ref{e7}) with respect to the p-values in the $g$th group, and completing the expectation with respect to all p-values, we see that the left-hand side of equation (\ref{e7}) is less than or equal to
\begin{align*}
N \text{E} \left \{ \sum\limits_{g=1}^m \frac{R_{n_{g \centerdot}}+ 1}{R_{n_{g \centerdot}} + \sum_{g^{\prime} \neq g}R_{n_{g_{\centerdot}^{\prime}} }+m} \right \} = N \text{E} \left \{ \sum\limits_{g=1}^m \frac{R_{n_{g \centerdot}}+ 1}{R_{N} + m}\right\} = N \text{E} \left \{ \frac{R_{N}+ m}{R_{N} + m} \right \}  = N, \end{align*} i.e., the second condition is also satisfied. Thus, the theorem is proved.

\section{Two-Way Grouped BH Procedure: Adapting the BH method to Two-Way Classified Hypotheses} Suppose, the $N$ hypotheses can be classified simultaneously according to two criteria and can be laid out in an $m \times n$ matrix with $n_{gh} \ge 1$ hypotheses at the intersection of the $g$th row and the $h$th column, i.e., in the $(g,h)$th cell of the matrix. We will consider the two different scenarios, one involving only one hypothesis per cell (i.e., $n_{gh}=1$) and the other involving multiple hypotheses per each cell (i.e., $n_{gh} > 1$), separately in the following two sub-sections. This would give a clearer picture of how our proposed procedures extend from one- to  multiple-hypotheses-per-cell. Also, in each of these scenarios, there is more than one choice of weights to define our proposed procedure in its oracle form, before constructing its data-adaptive version, that captures the underlying two-way structure. However, we will focus on one of them and formally present the corresponding oracle and data-adaptive procedures as our proposed ones to use for further evaluation, and simply point out the scope of deriving similar procedures using other weights.

\subsection{One Hypothesis Per Cell: $n_{gh}=1$} Let $n_{g0}$ be the number of true nulls in the $g$th row, for $g = 1, \ldots m, $, and $m_{0h}$ be the number of true nulls in the $h$th column, for $h = 1, \ldots, n$. The subsets of indexes of true nulls associated with $h$ in the $g$th row and $g$ in the $h$th column are, respectively, $I_{g0}\subseteq \{1, \ldots, n\}$ and $I_{0h} \subseteq \{1, \ldots, m\}$. Consequently, the set of indexes of true nulls among the entire set of hypotheses can be expressed as $I_0 = \bigcup\limits_{g=1}^mI_{g0} = \bigcup\limits_{h=1}^nI_{0h}$.

The proportion of true nulls in the $g$th row is defined as $\pi_{g0} = n_{g0}/n$, and that in the $h$th column as $\pi_{0h} = m_{0h}/m$. The proportion of true nulls in the entire set of $N=mn$ hypotheses is
$\pi_0 = \sum\limits_{g=1}^m \pi_{g0}/m = \sum\limits_{h=1}^n \pi_{0h}/n$.

\subsubsection {Oracle Two-Way GBH Procedure With One Hypothesis Per Cell} The hypothesis $H_{gh}$ at the intersection of the $g$th row and $h$th column is affected upon by its both parent row and column, which motivates us to consider assigning the following weight to $P_{gh}$ corresponding to $H_{gh}$, assuming all these proportions are known:
\begin{align}\label{e8}
	w_{gh} = \left[\frac{1}{2}\left\{\frac{1}{\pi_{g0}}\frac{1-\pi_{g0}}{1 - \pi_0} + \frac{1}{\pi_{0h}}\frac{1-\pi_{0h}}{1 - \pi_0}\right\}\right]^{-1},\; g = 1, \ldots, m,\; h = 1, \ldots, n,
\end{align} to simultaneously account for both row and column effects. The weighted BH procedure applied to the $N$ hypotheses based on the weighted p-values $P_{gh}^w = w_{gh}P_{gh}$, for $g= 1, \ldots, m; h=1, \ldots, n$, is one of our proposed procedures in its oracle form, which we refer to as an Oracle Two-Way GBH$_1$ procedure.

This weight is a simple extension of that from one- to two-way classification setting. If the parent row has a low proportion of true nulls $\pi_{g0}$, it subsequently has a large odds of being significant relative to other rows, as indicated by $(1- \pi_{g0})/(1 -\pi_0)$. This reduces the weight $w_{gh}$, making $H_{gh}$ more likely to be rejected. The weight is similarly affected by the parent column. We assume that both classifications have equal impacts on the individual hypothesis, and so the weight is a function of the simple mean of contributions from each of parent groups.

Noting that $w_{gh}$ can be expressed as $w_{gh}^{-1} = \frac{1}{2}w_{g}^{-1} + \frac{1}{2} w_{h}^{-1}$, with $w_{g}$ along the rows being defined in expression (\ref{e4}) and $w_{h}$ being defined similarly along the columns as $w_{h} = \frac{\pi_{0h}(1- \pi_{0})}{1 - \pi_{0h}},\; h=1, \ldots, n,$ one sees from (\ref{e5}) that
\begin {eqnarray} \label{e9}
\sum_{g=1}^m \sum_{h \in I_{g0}} w_{gh}^{-1} = \frac{1}{2}\sum_{g=1}^m \sum_{h \in I_{g0}} w_{g}^{-1} + \frac{1}{2}\sum_{h=1}^n \sum_{g \in I_{0h}} w_{h}^{-1} = \frac{1}{2} N + \frac{1}{2} N = N, \end {eqnarray} since $\sum_{g=1}^m \sum_{h \in I_{g0}} = \sum_{h=1}^n \sum_{g \in I_{0h}}$. Thus, equality in (\ref{e1}) is satisfied for the weights in expression (\ref{e8}), and so we can state the following theorem from Result \ref{result1} without offering a proof.

\begin{theorem} \label{th3}
	Oracle Two-Way GBH$_1$ procedure based on the weights in (\ref{e8}) controls the overall FDR under PRDS and Assumption \ref {assump1}. \end {theorem}

\begin{remark} \rm The weight in (\ref{e8}) can be customized, still satisfying (\ref{e9}), to suit variable influence of the row and column classifications. As an example, the weight can be adapted to reflect the imbalance between the number of rows and columns as follows: :
	\begin{align} \label{e10}
	w_{gh} = \left[\frac{1}{(m+n)}\left\{\frac{m}{\pi_{g0}}\cdot\frac{1-\pi_{g0}}{1 - \pi_0} + \frac{n}{\pi_{0h}}\cdot\frac{1-\pi_{0h}}{1 - \pi_0}\right\}\right]^{-1},\; g = 1, \ldots, m,\; h = 1, \ldots, n.
	\end{align}
These weights additionally account for the proportion of groups along the rows and also along the columns out of the total number of groups. Clearly, if $m = n$, they reduce to those in (\ref{e8}).
Such choice of weights in adapting BH procedure to two-way classification structure is not unique, and there remains a scope for other choices depending on variable factors or external information. \end{remark}

\subsubsection{Data-Adaptive Two-Way GBH Procedure With One Hypothesis Per Cell} We consider the Oracle Two-Way GBH$_1$ procedure  in Theorem \ref{th3} and estimate the weights in it by the following:
\begin{align}\label{e11}
\hat{w}_{gh} = &\left[\frac{N(1 - \lambda)}{2}\left\{\frac{1}{\{n -R_{n_{g}} + 1\}}\cdot \frac{R_{n_{g}}}{\{R_N + m-1\}} + \frac{1}{\{m -R_{m_{h}} + 1\}}\cdot \frac{R_{m_{h}}}{\{R_N + n -1\}}\right\}\right]^{-1} , \nonumber\\
& g= 1, \ldots, m,\; h = 1, \ldots, n,
\end{align}
where $R_{n_{g}} \equiv R_{n_{g }}(\lambda)= \sum_{h=1}^n I(P_{gh} \le \lambda)$, $R_{m_{h}} \equiv R_{m_{h}}(\lambda)= \sum_{g=1}^m I(P_{gh} \le \lambda)$, and $R_N \equiv R_N(\lambda)= \sum\limits_{g=1}^mR_{n_{g}}(\lambda) = \sum\limits_{h=1}^nR_{m_{h}}(\lambda)$, for some fixed $\lambda \in (0,1)$. The estimated weight assigned to each p-value is similar to that for one-way classified hypotheses; however, it accounts for both parent row and column effects. We refer to this procedure as a Data-Adaptive Two-Way GBH$_1$.

We have the following theorem as an extension of Theorem \ref{th2} from one- to two-way classification setting.

\begin {theorem} \label{th4}
The above Data-Adaptive Two-Way GBH$_1$ procedure controls the overall FDR under independence among all p-values and Assumption \ref {assump1}. \end {theorem}

Proof. This theorem can be proved based on the same arguments that were used to prove Theorem \ref{th2} using Result \ref{result2}. First, note that
$\hat{w}_{gh}^{-1} = \frac{1}{2} \tilde{w}_{g}^{-1} + \frac{1}{2}\tilde{w}_{h}^{-1}$, where

$$ \tilde{w}_{g} = \frac{n-R_{n_{g }}+1}{N(1 - \lambda)} \; \frac{R_N + m -1}{R_{n_{g }}}, $$
and $$\tilde{w}_{h} = \frac{m-R_{m_{ h}}+1}{N(1 - \lambda)} \; \frac{R_N + n -1}{R_{m_{ h}}}.$$  From this we see that $\hat{w}_{gh}$ is non-decreasing in $P_{gh}$, since both $\tilde{w}_{g}$  and $\tilde{w}_{h}$ are non-decreasing in $P_{gh}$, which can be proved exactly the way it was proved for the $\hat{w}_{g}$ in Theorem \ref{th2}.

Moreover, as proved in Theorem \ref{th2} for $\hat {w}_{g}$ using Lemma \ref{lemma1}, we have the following inequalities for $\tilde{w}_{g}$  and $\tilde{w}_{h}$ under the assumption of independence among all p-values:
\begin{align*}
&\text{E} \left \{ \sum\limits_{g=1}^m\sum\limits_{g \in I_{g0}}\frac{1}{\tilde{w}_g(\mathbf{P}^{-(g,h)}, 0)} \right \} \le N , \end{align*}
\begin{align*}
&\text{E} \left \{ \sum\limits_{h=1}^n\sum\limits_{g \in I_{0h}}\frac{1}{\tilde{w}_h(\mathbf{P}^{-(g,h)}, 0)} \right \} \le N , \end{align*}
from which we see that
\begin{align*}
&\text{E} \left \{ \sum \limits_{g=1}^m\sum\limits_{h \in I_{g0}}\frac{1}{\hat{w}_{gh}(\mathbf{P}^{-(g,h)}, 0)} \right \} \nonumber \\
= & \frac{1}{2} E \left \{ \sum\limits_{g=1}^m\sum\limits_{h \in I_{g0}}\frac{1}{\tilde{w}_g(\mathbf{P}^{-(g,h)}, 0)} \right \}  + \frac{1}{2} E \left \{ \sum\limits_{h=1}^n\sum\limits_{g \in I_{0h}}\frac{1}{\tilde{w}_h(\mathbf{P}^{-(g,h)}, 0)} \right \} \le N, \end{align*} i.e., inequality (\ref{e2}) in Result \ref{result2} holds. Thus, the theorem is proved.

\begin {remark} \rm The estimate of weight considered above is stated in its simplest form. Like its oracle counterpart, it can also be modified as
\begin{align}\label{e12}
\hat{w}_{gh} = & \left [ \frac{ N(1 - \lambda)}{m+n} \left \{ \frac{1}{n -R_{n_{g}} + 1 }\cdot \frac{mR_{n_{g }}}{R_N + m-1} + \frac{1}{m -R_{m_{ h}} + 1}\cdot \frac{nR_{m_{ h}}}{R_N + n -1} \right \} \right ]^{-1} , \nonumber \\
& g= 1, \ldots, m,\; h = 1, \ldots, n 
\end{align}
In addition to accounting for the row and column effects, this expression also accounts for the difference in numbers of rows and columns and accordingly emphasizes the corresponding effects on the individual hypothesis.
Theorem \ref{th4} can also be stated in terms of adaptive two-way GBH with one hypothesis per cell in terms of these alternative weights. \end {remark}

\subsection{Multiple Hypothesis Per Cell: $n_{gh} > 1$}
Let $n_{g \centerdot } = \sum\limits_{h=1}^n n_{gh }$ and $n_{\centerdot h } = \sum\limits_{g=1}^m n_{gh }$ be the total numbers of hypotheses, respectively, in the $g$th row and $h$th column, so that $$N = \sum\limits_{g=1}^m n_{g \centerdot } = \sum\limits_{h=1}^n n_{\centerdot h } = \sum\limits_{g=1}^m\sum\limits_{h=1}^n n_{gh }.$$ Let $n_{gh0}$ be the number of true nulls in the $(g,h)$th cell and $I_{gh0} \subseteq \{1, \ldots, n_{gh}\}$ be the corresponding subset of indexes of true nulls.
The overall set of indexes of the true nulls is $$I_0 = \bigcup\limits_{g=1}^m\bigcup\limits_{h=1}^n I_{gh0}.$$

The proportion of true nulls in the $(g,h)$th cell of the $m \times n$ matrix is $\pi_{gh0} = n_{gh0}/n_{gh}$. This helps to define $\pi_{g00} = \sum\limits_{h=1}^nn_{gh}\pi_{gh0}/\sum\limits_{h=1}^n n_{gh }$, $\pi_{0h0} = \sum\limits_{g=1}^mn_{gh }\pi_{gh0}/\sum\limits_{g=1}^m n_{gh }$, and
$$ N \pi_{0} = \sum\limits_{g=1}^m n_{g \centerdot } \pi_{g00} = \sum\limits_{h=1}^n n_{\centerdot h }\pi_{0h0}  =  \sum\limits_{g=1}^m\sum\limits_{h=1}^nn_{gh}\pi_{gh0}$$.

\subsubsection{Oracle Two-Way GBH Procedure With Multiple Hypotheses Per Cell}
 Suppose $P_{ghk}$ is the $k$th p-value in the $(g,h)$th cell, and $H_{ghk}$ is the corresponding hypothesis. Assuming that all the proportions mentioned above are known, we consider assigning the following weights to $P_{ghk}$, for each $k=1, \ldots, n_{gh}$, to capture the underlying two-way classification structure of the hypotheses, and refer to the resulting weighted BH procedure as an Oracle Two-WayGBH$_{>1}$.

\begin{align}\label{e13}
w_{gh} = \left[\frac{1}{4}\left\{\frac{1}{\pi_{gh0}}\left(\frac{1-\pi_{gh0}}{1-\pi_{g00}} + \frac{1-\pi_{gh0}}{1-\pi_{0h0}}\right) + \left(\frac{1}{\pi_{g00}}\cdot\frac{1-\pi_{g00}}{1-\pi_0} + \frac{1}{\pi_{0h0}}\cdot\frac{1-\pi_{0h0}}{1-\pi_0}\right)\right\}\right]^{-1}
\end{align}

Expressing $w_{gh}$ as ${w}_{gh}^{-1} = \frac{1}{4}{w}_{1, gh}^{-1} + \frac{1}{4}{w}_{2, gh}^{-1} + \frac{1}{4}{w}_{g}^{-1} + \frac{1}{4}{w}_{h}^{-1},$ where
\begin{eqnarray}
& & {w}_{1, gh} = \frac{\pi_{gh0}(1-\pi_{g00})} {1-\pi_{gh0}}, \; {w}_{2, gh} =  \frac{\pi_{gh0}(1-\pi_{0h0})} {1-\pi_{gh0}}, \; {w}_{g}  =  \frac{\pi_{g00}(1-\pi_{0})} {1-\pi_{g00}}, \nonumber \\ & & \mbox{and} \quad {w}_{h} =  \frac{\pi_{0h0}(1-\pi_{0})} {1-\pi_{0h0}}, \nonumber \end {eqnarray} one can see that $$ \sum_{g=1}^m \sum_{h=1}^n \sum_{k \in I_{gh0}} \left [ w_{1, gh}^{-1} + w_{2, gh}^{-1} + w_{g}^{-1} +  w_{h}^{-1} \right ] = 4 N, $$ that is, the equality in (\ref{e1}) is satisfied by the weights in (\ref{e13}). Therefore, we can state the following from Result \ref{result1} without a proof

\begin {theorem}\label{th5}  Two-way GBH with multiple hypotheses per cell based on the weights in (\ref{e13}) controls the overall FDR conservatively under PRDS and Assumption \ref {assump1}. \end {theorem}

\begin {remark} \rm Of course, one can consider defining two-way GBH procedure with multiple hypotheses per cell based on other types of weight subject to the equality in (\ref{e1}). For instance, following the preceding case of one hypothesis per cell in the two-way classification setup, a natural choice of weight assigned to hypotheses in the $(g,h)$th cell would be

\begin{align} \label{e14} w_{gh} = \left[\frac{1}{2}\left\{\frac{1}{\pi_{g00}}\cdot\frac{1-\pi_{g00}}{1- \pi_0} + \frac{1}{\pi_{0h0}}\cdot\frac{1-\pi_{0h0}}{ 1- \pi_0}\right\}\right]^{-1}
\end{align}
The choice of weight for the two-way GBH procedure in Theorem \ref{th5} consists of an additional term that depends on the ratio of the proportion of signals in each cell to the same proportions in the parent row and column.
Owing to unequal number of members at the intersections in the two-way layout, further modifications of the weights would be complicated. However, if there are an equal number, say $p>0$, hypotheses at each cell, we can further edit these weights in (\ref{e14}) as
\begin{subequations}
	\begin{align}\label{e15a}
	w_{gh} = \left[\frac{1}{p(m+n)}\left\{\frac{mp}{\pi_{g00}}\cdot\frac{1-\pi_{g00}}{1- \pi_0} + \frac{np}{\pi_{0h0}}\cdot\frac{1-\pi_{0h0}}{ 1- \pi_0}\right\}\right]^{-1},
	\end{align}
and those in the procedure in Theorem \ref{th5} as
	\begin{align}\label{e15b}
		\begin{aligned}
	w_{gh} = \bigg[\frac{1}{p(m+n)}
	\bigg\{&\frac{p}{\pi_{gh0}}\left(\frac{1-\pi_{gh0}}{1-\pi_{g00}} + \frac{1-\pi_{gh0}}{1-\pi_{0h0}}\right)\\
	& + \left(\frac{p(m-1)}{\pi_{g00}}\cdot\frac{1-\pi_{g00}}{1-\pi_0} + \frac{p(n-1)}{\pi_{0h0}}\cdot\frac{1-\pi_{0h0}}{1-\pi_0}\right)\bigg\}\bigg]^{-1}.
	\end{aligned}
	\end{align}
\end{subequations}
\end {remark}

\subsubsection{Data-Adaptive Two-Way GBH Procedure With Multiple Hypotheses Per Cell} Consider the Two-WayGBH$_{>1}$ in Theorem \ref{th5} to replace its weight $w_{gh}$ by
\begin{align}\label{e3.8}
\hat{w}_{gh} =
& \bigg[ \frac{1}{4}\bigg\{\frac{1-\lambda}{n_{gh}-R_{n_{gh}}+1}\left(\frac{n_{g \centerdot}R_{n_{gh}}}{R_{n_{g \centerdot}}+n-1}+ \frac{n_{\centerdot h}R_{n_{gh}}}{R_{n_{\centerdot h}}+m-1}\right) \nonumber \\
 &+ N(1-\lambda)\left(\frac{R_{n_{g \centerdot}}}{\{n_{g \centerdot}-R_{n_{g \centerdot}}+1\}\{R_N+m-1\}}+ \frac{R_{n_{\centerdot h}}}{\{n_{\centerdot h}-R_{n_{\centerdot h}}+1\} \{R_N+n-1\}}\right)\bigg\}\bigg]^{-1}\nonumber \\
 & g = 1\ldots, m,\; h = 1, \ldots, n,
\end{align} where, for some fixed $\lambda$, $R_{n_{gh}} \equiv R_{n_{gh }} (\lambda) =  \sum_{k=1}^{n_{gh }} I(P_{ghk} \le \lambda)$, $R_{n_{g \centerdot }} \equiv R_{n_{g \centerdot}}(\lambda) = \sum_{h=1}^n R_{n_{gh }} (\lambda)$, $R_{n_{\centerdot h }} \equiv R_{n_{\centerdot h }}(\lambda) = \sum_{g=1}^m R_{n_{gh }} (\lambda)$.
\begin{align*}
	R_N = \sum\limits_{g=1}^m R_{n_{g\centerdot}} = \sum\limits_{h=1}^n R_{n_{\centerdot h}} = \sum\limits_{g=1}^m\sum\limits_{h=1}^n R_{n_{gh.}}
\end{align*} It is referred to as a Data-Adaptive Two-WayGBH$_{>1}$.

\begin {theorem} \label{th6} The above Data-Adaptive Two-WayGBH$_{>1}$ controls the overall FDR under independence among all p-values and Assumption \ref {assump1}. \end {theorem}

Proof. Again, this theorem will be proved based on the same arguments that were used to prove Theorem \ref{th4} by verifying that the conditions in Result \ref{result2} are satisfied by the weight functions $\hat{w}_{gh}$; i.e., it is increasing in each $P_{ghk}$ and that the following inequality holds with $P_{ghk}$ being set to $0$ in it: $\sum_{g=1}^m\sum_{h=1}^n \sum_{k \in I_{gh0}} \hat{w}_{gh}^{-1}(\mathbf{P}^{-(g,h,k)},0) \le N$.

As in proving Theorem \ref{th4}, let us consider $\hat{w}_{gh}$ in terms of the following representation: $\hat{w}_{gh}^{-1} = \frac{1}{4}\tilde{w}_{1, gh}^{-1} + \frac{1}{4}\tilde{w}_{2, gh}^{-1} + \frac{1}{4}\tilde{w}_{g}^{-1} + \frac{1}{4}\tilde{w}_{h}^{-1},$ where
\begin{eqnarray}\label{e17}
\tilde{w}_{1, gh} & = &  \frac{n_{gh}-R_{n_{gh }}+1}{n_{g \centerdot}(1-\lambda)} \cdot \frac{R_{n_{g \centerdot}+n-1}}{R_{n_{gh }}}, \; \tilde{w}_{2, gh} = \frac{n_{gh}-R_{n_{gh}+1}}{n_{\centerdot h}(1-\lambda)} \cdot \frac{R_{n_{\centerdot h }+n-1}}{R_{n_{gh }}}, \nonumber \\
\tilde{w}_{g} & = &  \frac{n_{g \centerdot }-R_{n_{g \centerdot}}+1}{N(1-\lambda)} \cdot \frac{R_{N}+m-1}{R_{n_{g  \centerdot}}}, \; \tilde{w}_{h} =   \frac{n_{\centerdot h} -R_{n_{\centerdot h }}+1}{N(1-\lambda)} \cdot \frac{R_{N}+n-1}{R_{n_{\centerdot h }}}, \end {eqnarray}
As argued before in proving Theorems \ref{th2} and \ref{th4}, each of the four weights in (\ref{e17}) can be shown to satisfy the same two properties that we intend to show for $\hat{w}_{gh}$. In other words, $\hat{w}_{gh}$ satisfies the desired two conditions in Result \ref{result2}, and hence the theorem is proved.

\begin {remark} \rm Other choices of weights can be suggested as
\begin{align}\label{e18}
	\hat{w}_{gh} = \left[\frac{N(1-\lambda)}{2}\left\{\frac{1}{n_{g \centerdot}-R_{g.}+1}\frac{R_{g.}}{R_N +m -1} + \frac{1}{n_{\centerdot h}-R_{.h}+1}\frac{R_{.h}}{R_N +n-1}\right\}\right]^{-1}
\end{align}
As in the oracle case, these weights can be further modified to be more informative if there are an equal number of hypotheses $(p >0)$ at each cell. The modified choice corresponding to expression (\ref{e3.8}) would be
\begin{align}\label{e3.11}
	\hat{w}_{gh} =
	& \bigg[ \frac{1}{(m+n)p}\bigg\{\frac{p(1-\lambda)}{n_{gh.}-R_{n_{gh}}+1}\left(\frac{n_{g \centerdot}R_{n_{gh}}}{R_{n_{g \centerdot}}+n-1}+ \frac{n_{\centerdot h}R_{n_{gh}}}{R_{n_{\centerdot h}}+m-1}\right) \nonumber \\
	&+ N(1-\lambda)\left(\frac{p(m-1)R_{n_{g \centerdot}}}{\{n_{g \centerdot}-R_{n_{g \centerdot}}+1\}\{R_N+m-1\}}+ \frac{p(m-1)R_{n_{\centerdot h}}}{\{n_{\centerdot h}-R_{n_{\centerdot h}}+1\} \{R_N+n-1\}}\right)\bigg\}\bigg]^{-1}\nonumber \\
	&\forall g = 1\ldots, m,\; h = 1, \ldots, n
\end{align}
and the modified choice corresponding to expression (\ref{e18}) is
\begin{align}
	\hat{w}_{gh} = \left[\frac{N(1-\lambda)}{(m+n)p}\left\{\frac{mp}{n_{g \centerdot}-R_{g.}+1}\frac{R_{g.}}{R_N +m -1} + \frac{np}{n_{\centerdot h}-R_{.h}+1}\frac{R_{.h}}{R_N +n-1}\right\}\right]^{-1}
\end{align}
\end{remark}

\section{Simulations Studies}

We carried out extensive simulation studies to investigate the performances of our proposed procedures in Theorems 2-6 in terms FDR control and power (expected proportion of correctly rejected false nulls among all false nulls) against their relevant competitors. This section discusses these results.

\subsection{One-Way Classified Hypotheses}
Here, our study was designed
 to compare the performance of the Data Adaptive Two-Way GBH procedure in Theorem \ref{th2} with its following three relevant competitors, the first two of which were considered in \cite{Huetal2010} as extensions to one-way classification setting of the single-group data-adaptive BH procedures proposed, respectively, in \cite{BH2000} and \cite{Benjaminietal2006}.
\vskip 6pt
\noindent {\it LSL (Least-Slope) Grouped BH:} One-Way GBH procedure with $\pi_{g0}$ in equation (\ref{e4}) being estimated by the following, for each $g$:
\begin{align*}
\hat{\pi}_{g0}^{{\rm LSL}} = \min \left(\frac{\lfloor l_{g,i}\rfloor + 1}{n},1\right),\; l_{g,i} = \frac{n -i+1}{1 - P_{g,(i)}},
\end{align*}
such that $l_{g,i} > l_{g, i-1}$, with $P_{g,(i)}$ being the $i$th minimum ordered p-value in the $g$th group.
\vskip 6pt
\noindent {\it TST (Two-Stage) Grouped BH:} One-Way GBH procedure with $\pi_{g0}$ in equation (\ref{e4}) being estimated by the following, for each $g$:
\begin{align*}
\hat{\pi}_{g0}^{{\rm TST}} = \frac{n - r_{g}}{n},
\end{align*}
with $r_g$ being the number of rejections obtained by applying the non-adaptive BH procedure to the p-values in the $g$th group at level $\alpha/(1+\alpha)$.
\vskip 6pt
\noindent {\it Naive Adaptive BH:} The usual data-adaptive BH with the following estimate of $\pi_0$:
\begin{align}\label{e21}
\hat{\pi}_0 = \frac{N - R_N +1}{N(1-\lambda)},\; \mbox{with} \; R_N = \sum\limits_{g=1}^m \sum_{i=1}^n I(P_{g,i} \leq \lambda),
\end{align} for some fixed $\lambda \in (0,1)$, applied to all hypotheses.

\subsubsection{Simulation Setting}
The following steps were taken to simulate values of FDR and power for the aforementioned procedures.

\begin{itemize}

\item [1.] Generate $\theta_{g \centerdot}$, for $g=1, \ldots, m$, as a random sample from Ber$(1-\pi_{\centerdot})$; \

\item [2.] For each $g$ such that $\theta_{g \centerdot}=1$, generate $\boldsymbol{\theta}_{\centerdot |g} = (\theta_{1|g}, \ldots, \theta_{n|g})$ as a random vector of $n$ i.i.d. Ber$(1-\pi)$;\

\item [3.] Given $(\theta_{g \centerdot}, \boldsymbol{\theta}_{\centerdot | g})$, $g=1, \ldots, m$, generate $m$ independent $n$-dimensional random vectors $\mathbf{X}_g = (X_{g,1} \ldots, X_{g,n})$, $g=1, \ldots, m$, as follows:
\begin {eqnarray*}\label{e4.1}
\mathbf{X}_g = \mu \theta_{g \centerdot}\boldsymbol{\theta}_{\centerdot | g} + \sqrt{(1-\rho_g)}\mathbf{Z}_g + \sqrt{\rho_g}Z_{g0}, \end{eqnarray*} for some $0 \leq \rho_g < 1$,
having generated $\{Z_{g0}, \mathbf{Z}_g=(Z_{g1}, \ldots, Z_{gn})^T\}$ as a random vector of $n+1$ i.i.d. $N(0,1)$ samples, for $g=1, \ldots, m$.
\item [4.] Apply each procedure at FDR level $\alpha=0.05$ for testing $H_{g,i}: \text{E}(X_{g,i}) = 0$ against $K_{g,i}: \text{E}(X_{g,i}) > 0$, simultaneously for all $g=1, \ldots, m, i=1, \ldots, n$, in terms of the corresponding p-values, and note the proportions of false rejections among all rejections and correct rejections among all false nulls.
\item [5.] Repeat Steps 1-4 200 times to simulate the values of FDR and power for each procedure by averaging out the corresponding proportions obtained in Step 4.

\end {itemize}

\begin {remark} \rm Our modeling of $\text{E}(\mathbf{X}_g)$ in term of $(\theta_{g \centerdot}, \boldsymbol{\theta}_{\centerdot | g})$ allowed us to split the state of each hypothesis at two levels, group and individual, enabling us to regulate the density of signals in the entire set of hypotheses using the following representation of true nulls among all hypotheses:
\begin {eqnarray} \label{e22} \pi_0 = 1- (1-\pi_{\centerdot})(1-\pi)\;. \end {eqnarray} \end {remark}

\subsubsection{Simulation Findings} We fixed $m = 50$, $n = 100$, $\mu = 0$ for true null hypotheses, and $=3$ for true signals.

We have two main objectives in our simulation study regarding the performance of our proposed Data-Adaptive One-Way GBH procedure in Theorem \ref{th2} - (i) to investigate how well it performs among all four procedures under independence when all of them are theoretically known to control FDR, and (ii) to investigate if it can possibly control FDR under PRDS in view of the fact that such  control is yet to be theoretically proved.

\begin{figure}[H]
\centering
$\begin{array}{rl}
\includegraphics[width=0.365\textwidth]{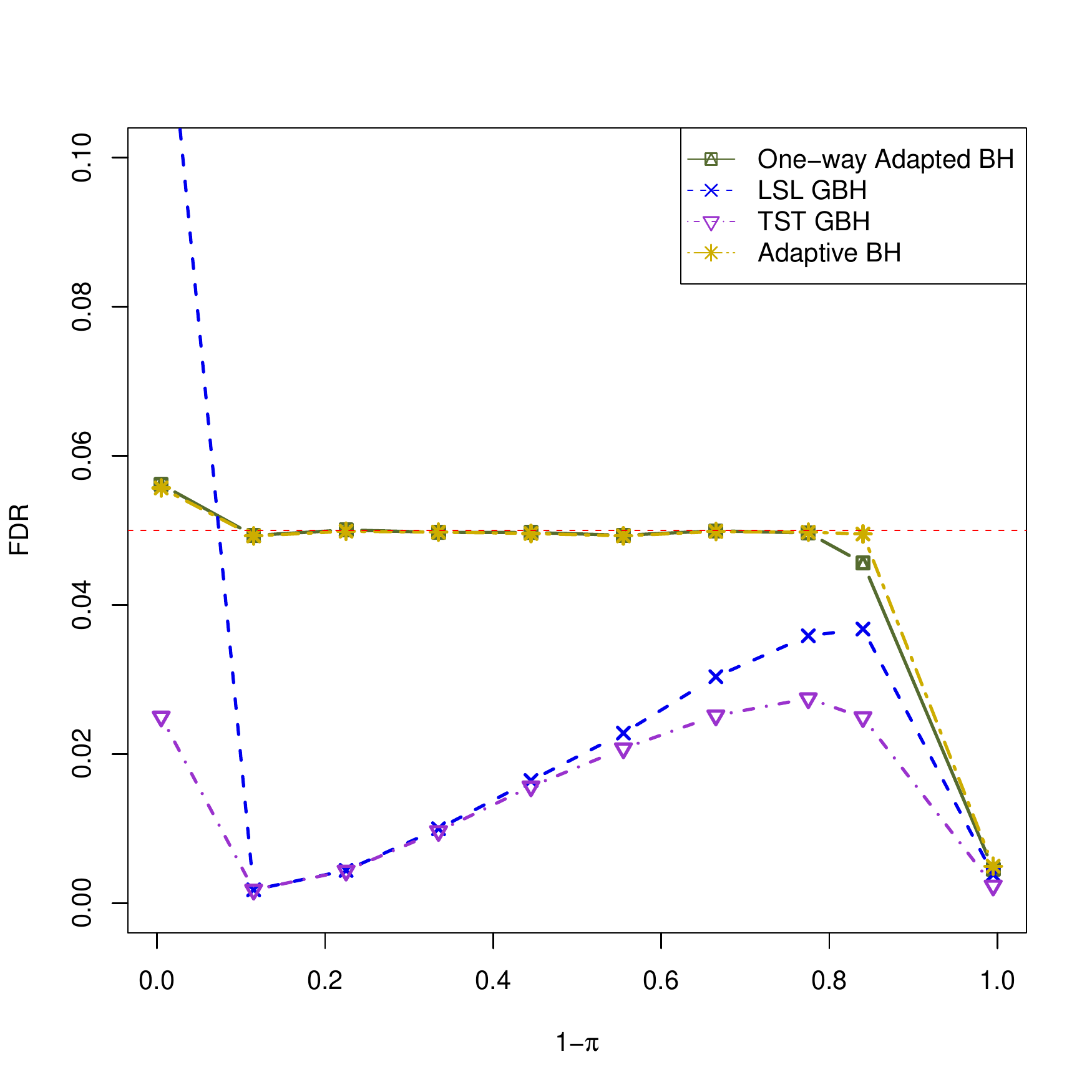} &
\includegraphics[width=0.365\textwidth]{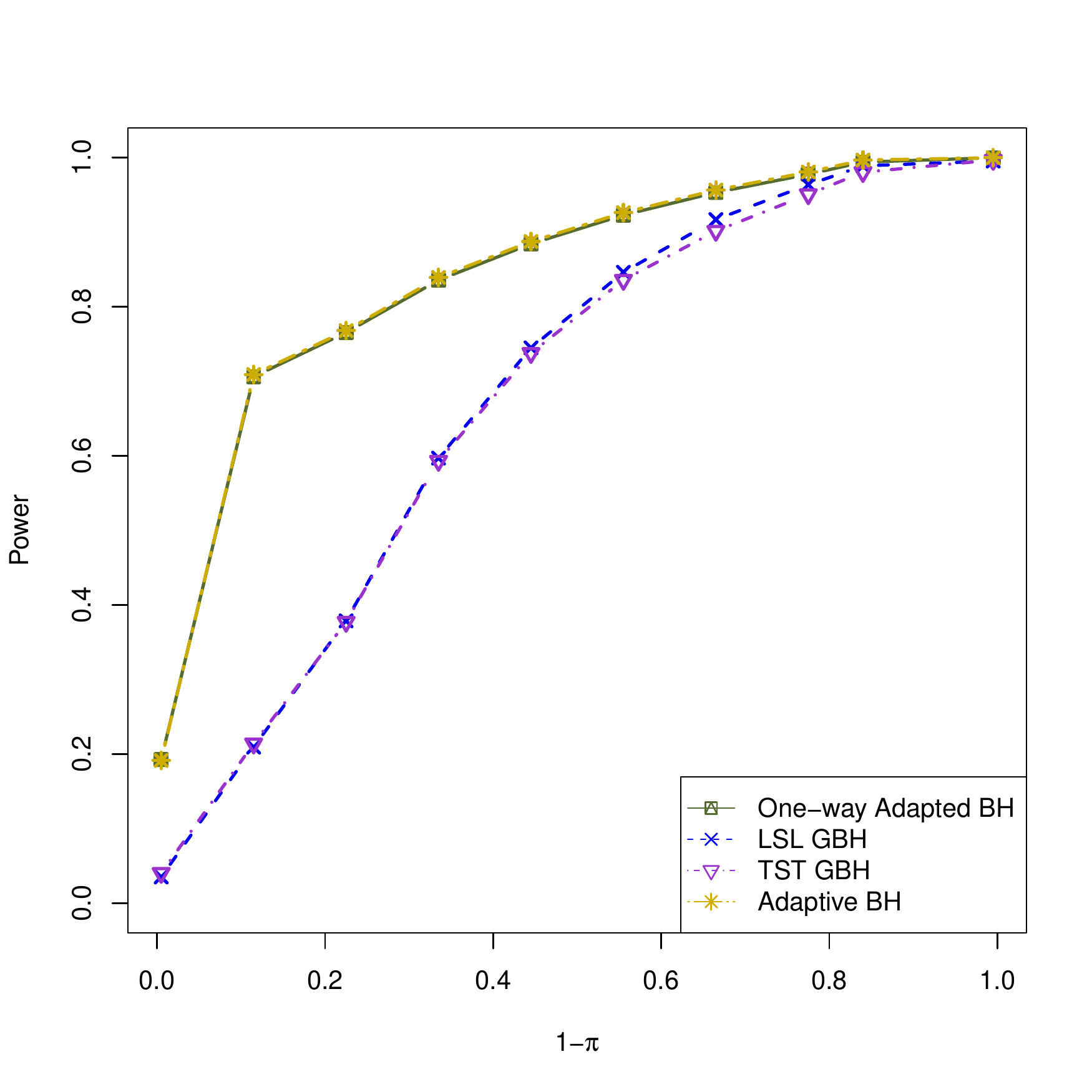}\\
\end{array}$
\caption{FDR and Power comparisons of the Data-Adaptive One-Way GBH proposed with other methods, under independence, ($m=50,n = 100, \rho=0, \pi_{\centerdot} = 0, \pi$)}.
\label{fig1}\end{figure}

\begin{figure}[H]
\centering
$\begin{array}{rl}
\includegraphics[width=0.365\textwidth]{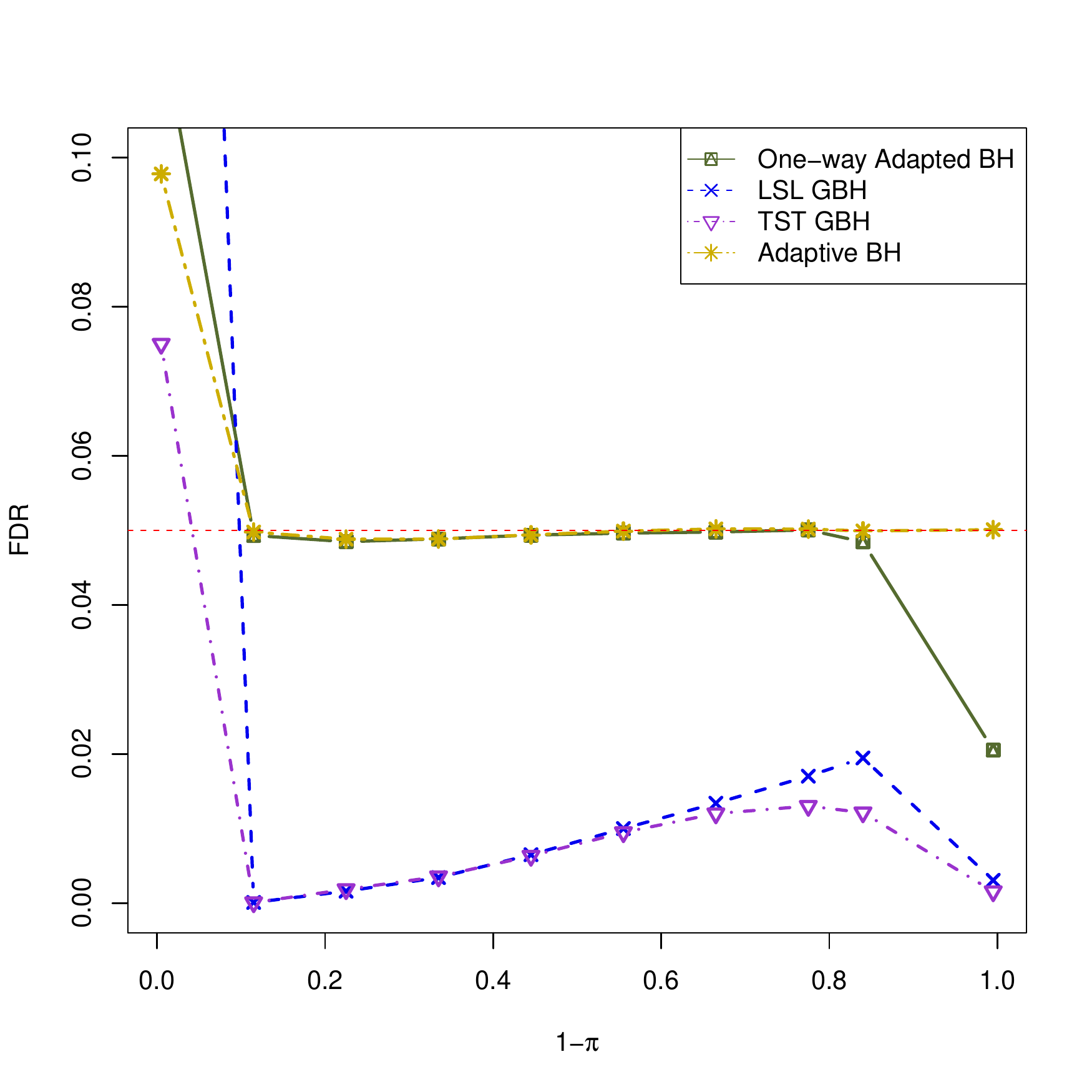} &
\includegraphics[width=0.365\textwidth]{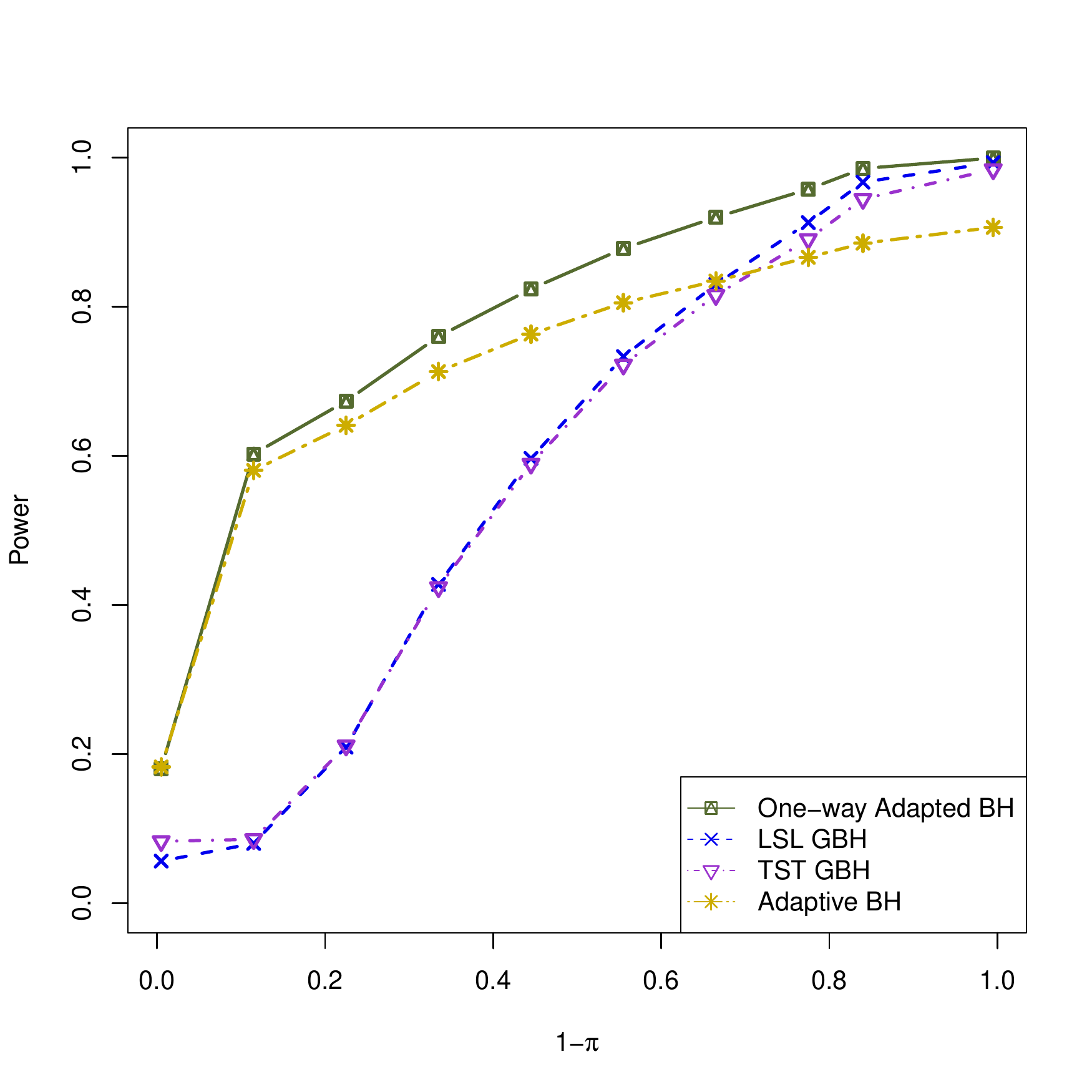}\\
\end{array}$
\caption{FDR and Power comparisons of the Data-Adaptive One-Way GBH with other methods applied to independent one-way classified hypotheses when true signals are unevenly distributed ($m=50,n = 100, \rho=0, \pi_{\centerdot} = 0.5, \pi $)}.
\label{fig2}\end{figure}

Figures 1-2 display the findings of the first type of investigation, with $\lambda =0.5$. Figure \ref{fig1} considers situations where signals are distributed evenly across all groups, with $\pi_{\centerdot} = 0$ and $\pi$ (i.e., $\pi_0$ by equation (\ref{e22})) being allowed to vary. As seen from this figure, our proposed procedure performs better than the LSL and TST GBH procedures. It controls FDR less conservatively and is more powerful than its counterparts at all levels of density of true signals. However, its performance is quite similar to the adaptive BH procedure owing to the signals being uniformly distributed across all groups. Since signals may potentially be non-uniformly distributed across the groups, we considered a scenario where only half the groups may contain significant members; see Figure \ref{fig2}. Our proposed procedure is remarkably more powerful in this case than the other methods.

\begin{figure}[H]
\centering

%

\begin{subfigure}[b]{\textwidth}
\centering
\includegraphics[width=0.22\textwidth]{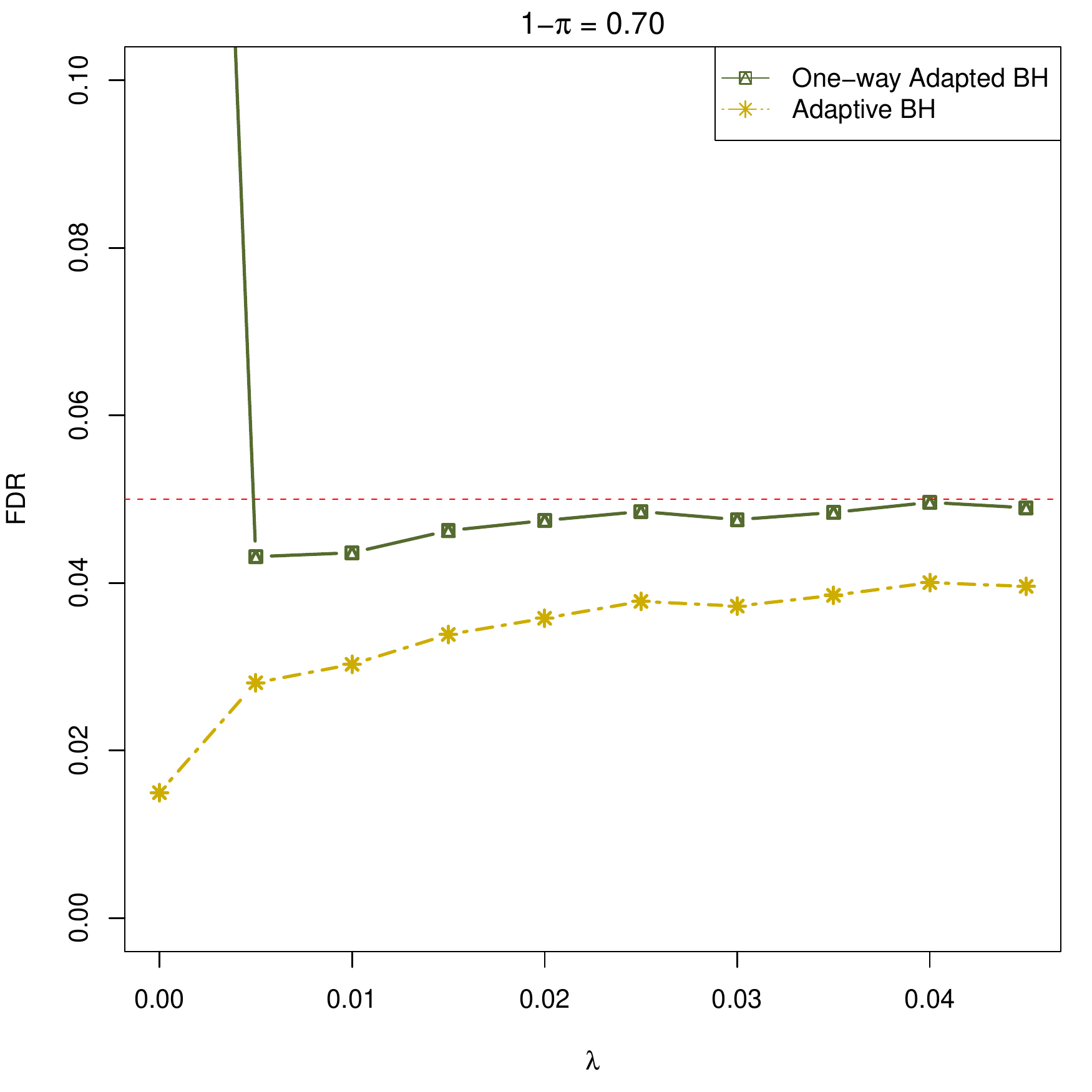}	
\includegraphics[width=0.22\textwidth]{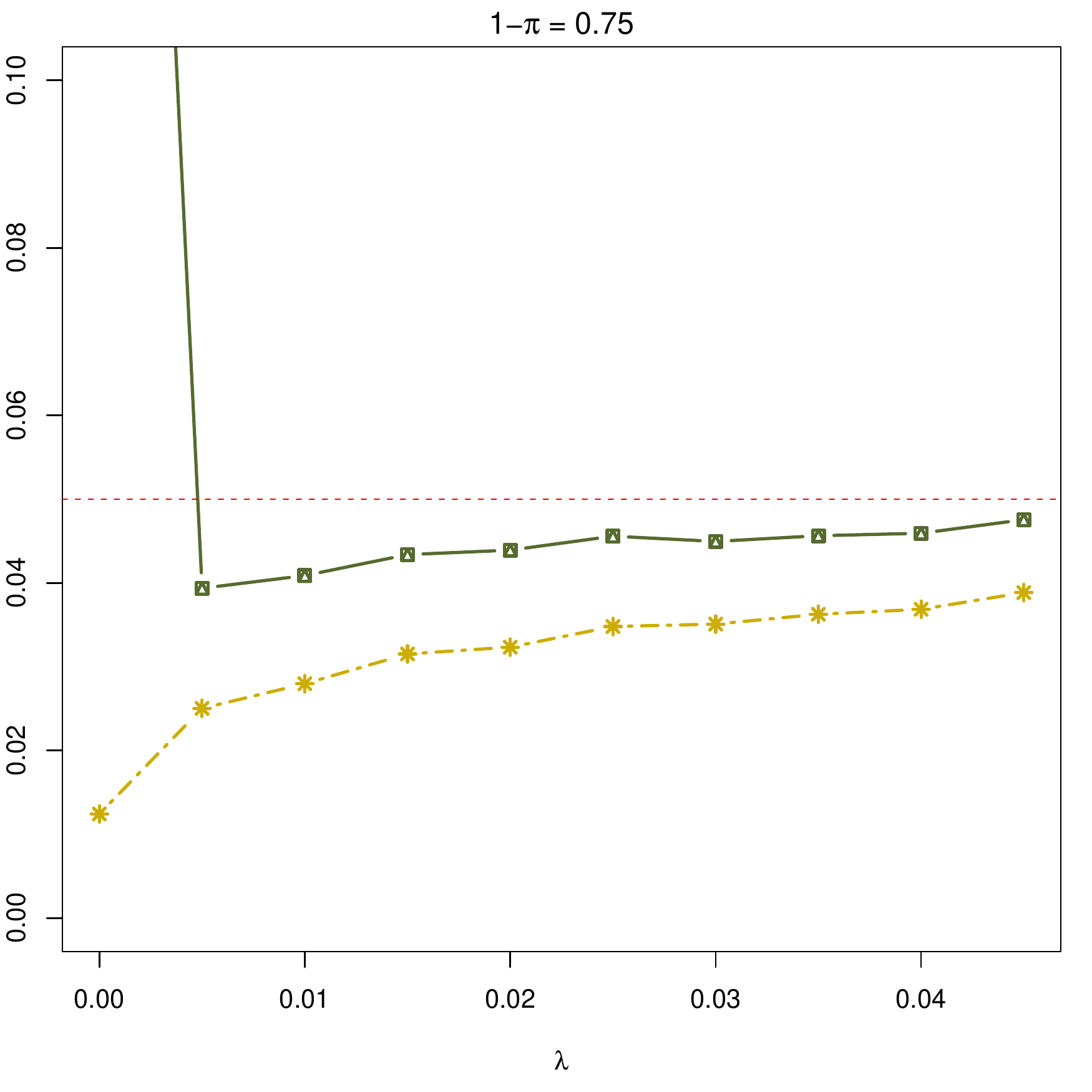}	
\includegraphics[width=0.22\textwidth]{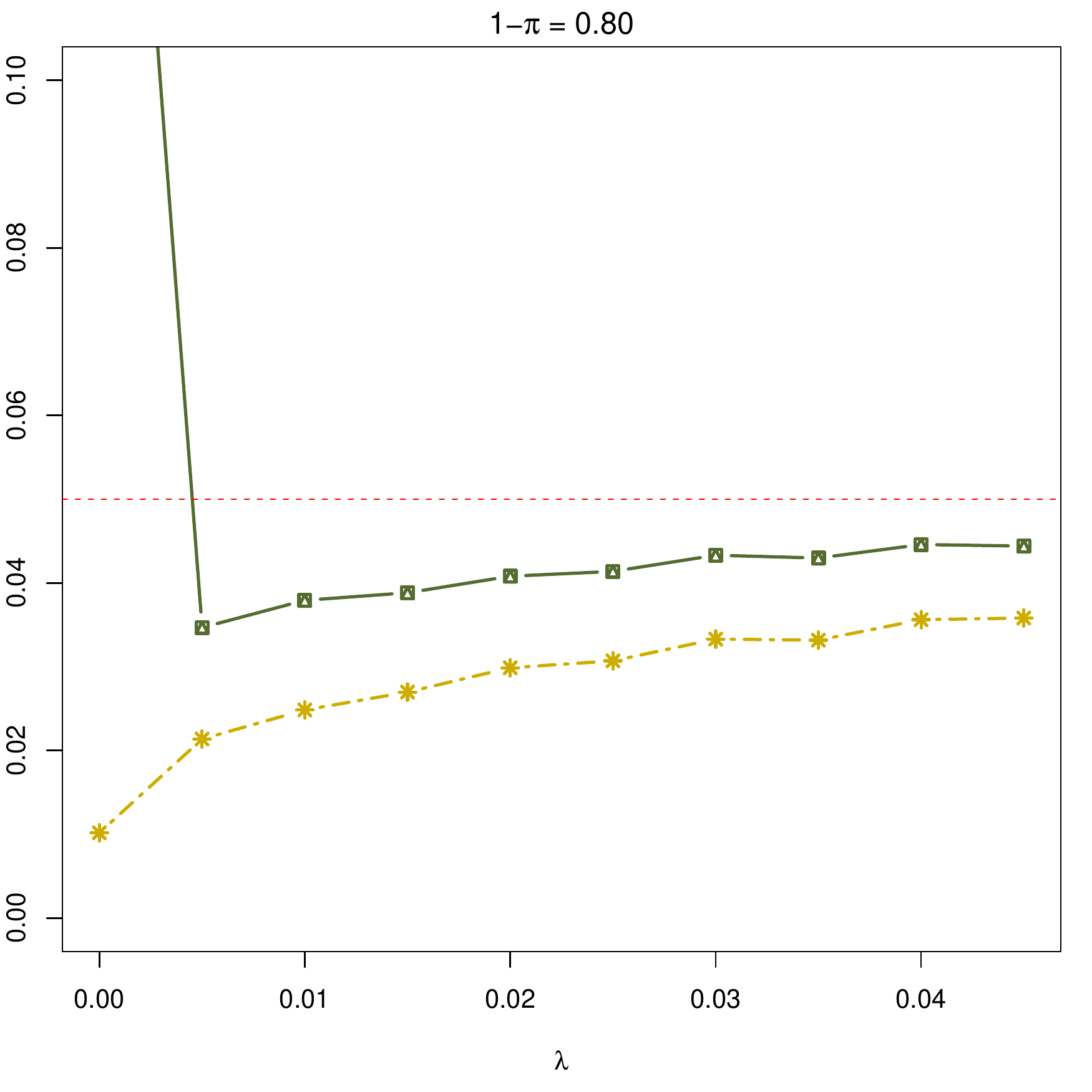}	
\includegraphics[width=0.22\textwidth]{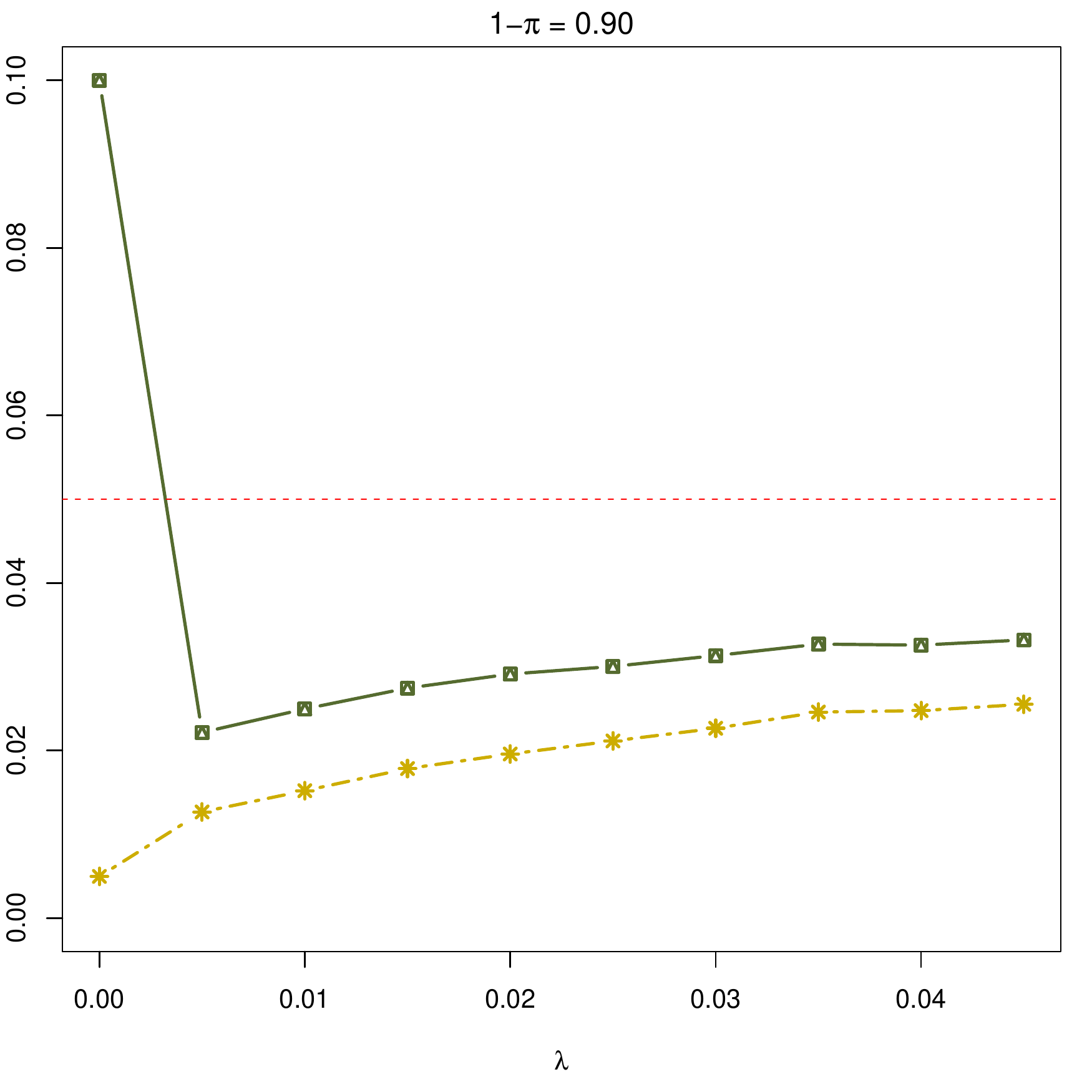}
\caption{FDR Comparisons}	
\label{fig3a}
\end{subfigure}\\

\begin{subfigure}[b]{\textwidth}
\centering
\includegraphics[width=0.22\textwidth]{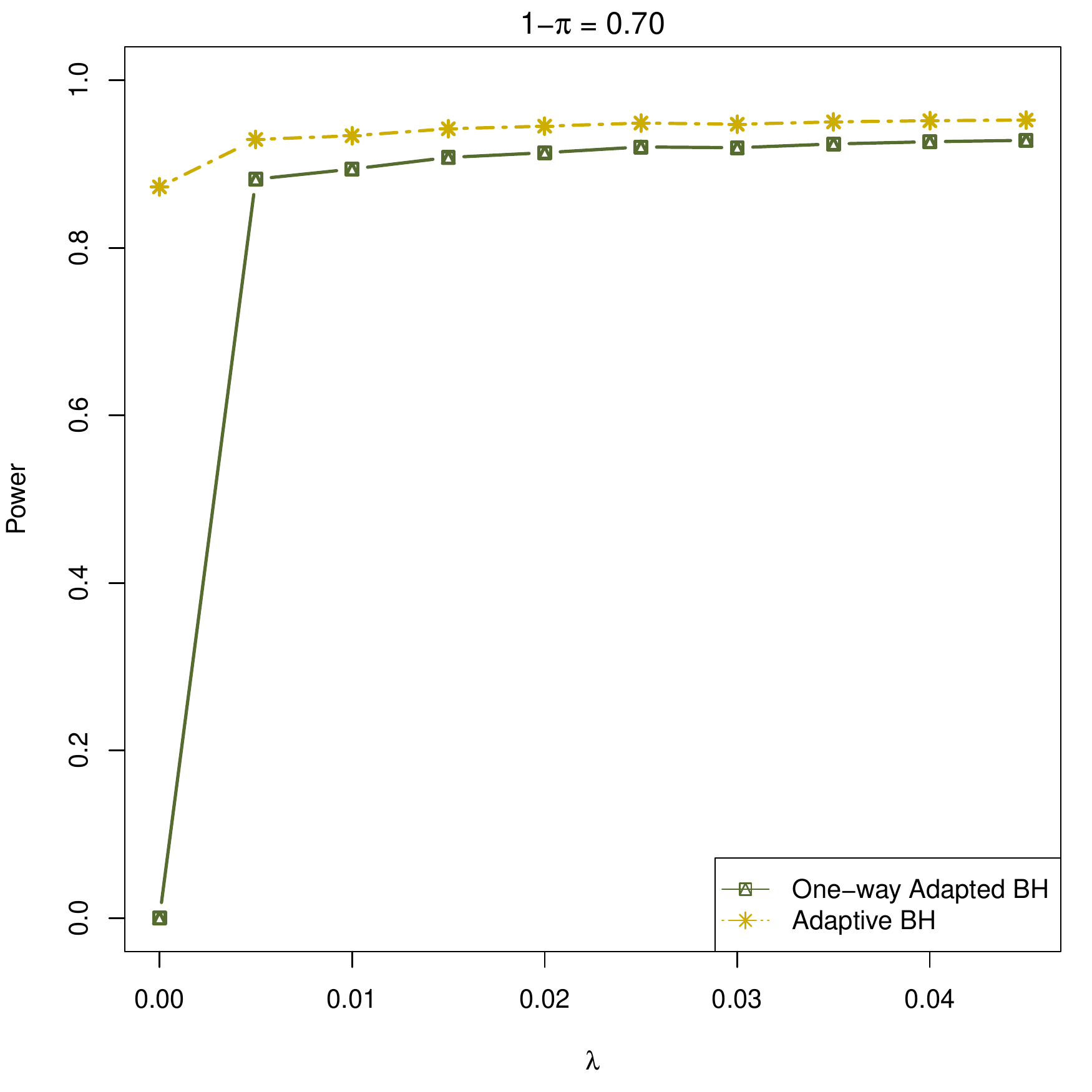}	
\includegraphics[width=0.22\textwidth]{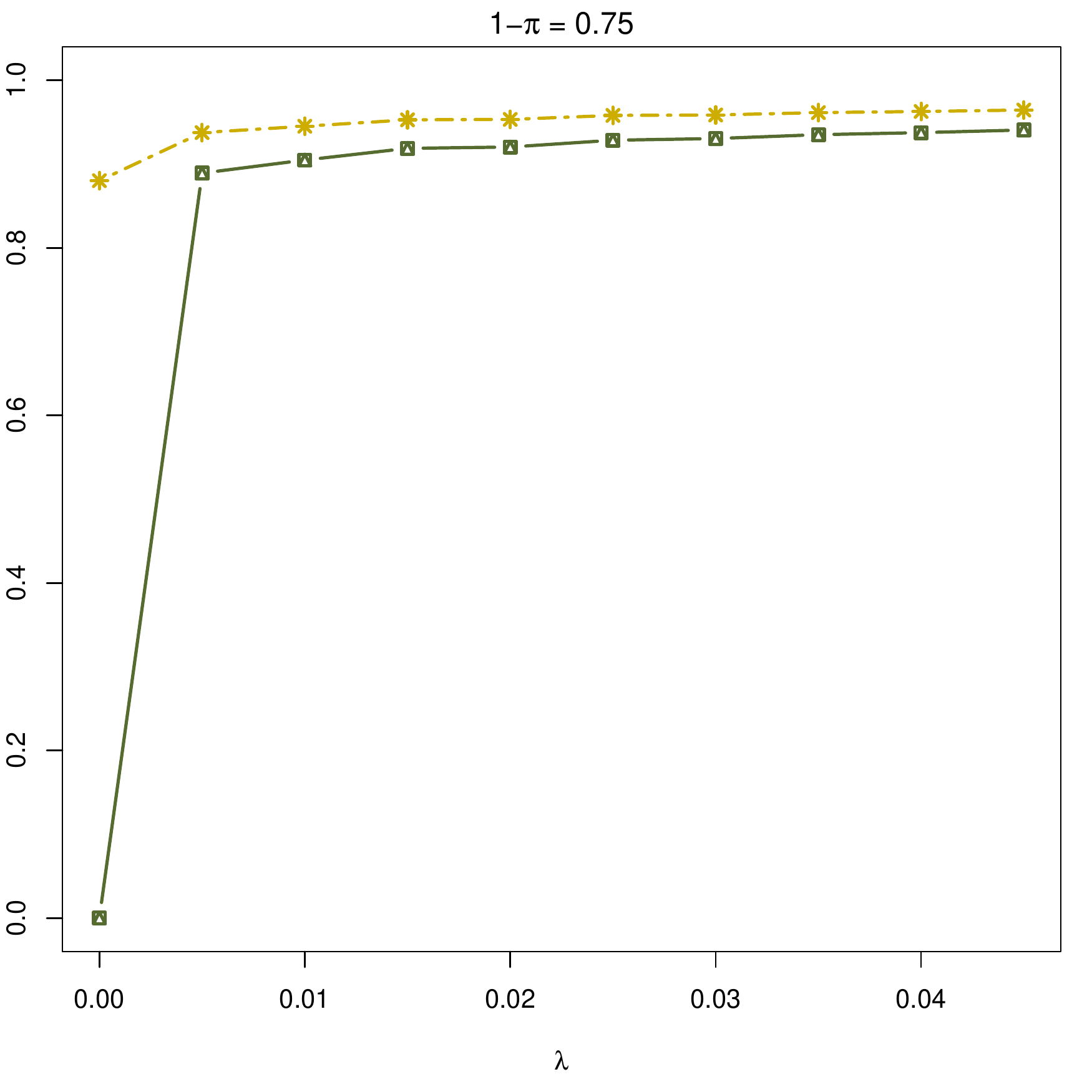}	
\includegraphics[width=0.22\textwidth]{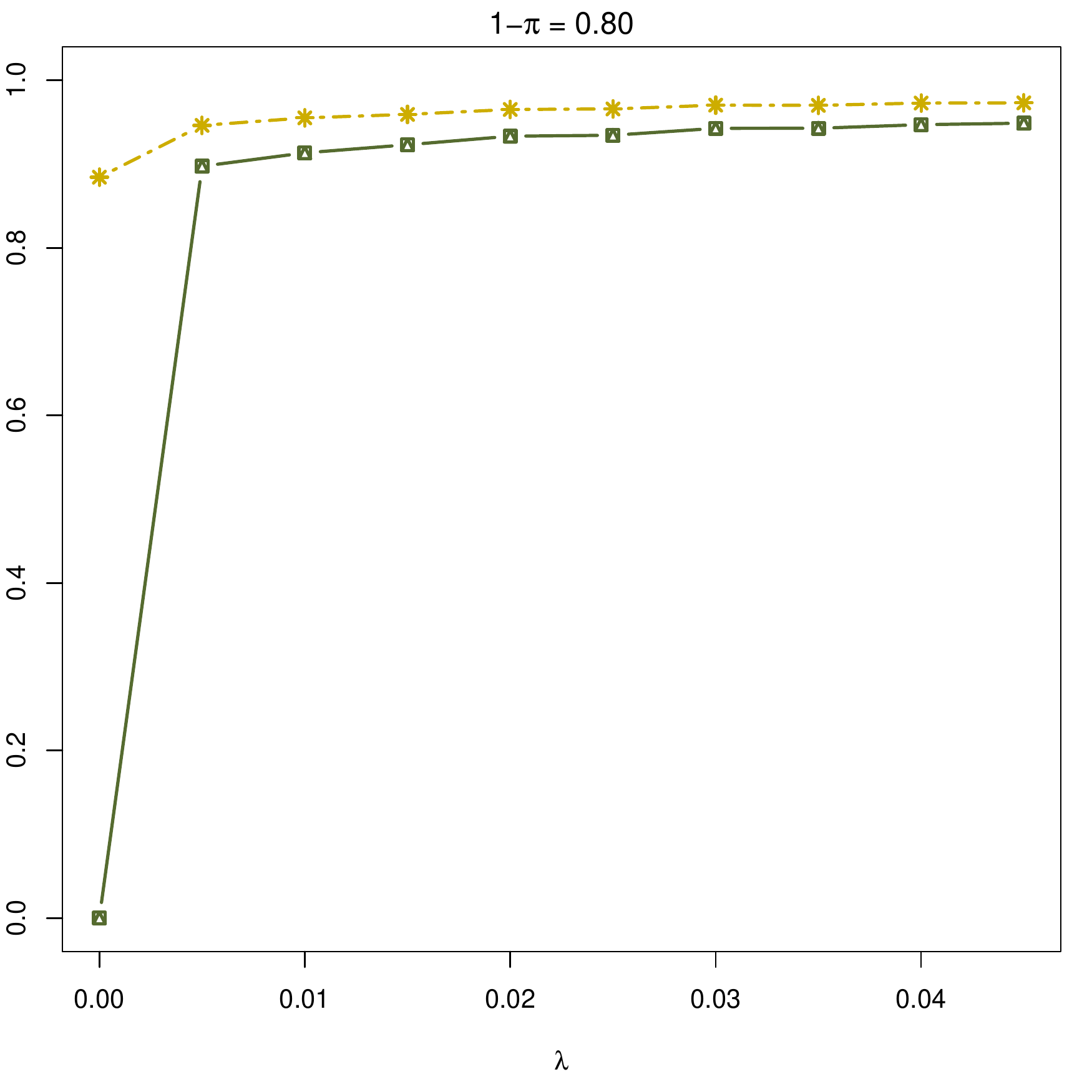}	
\includegraphics[width=0.22\textwidth]{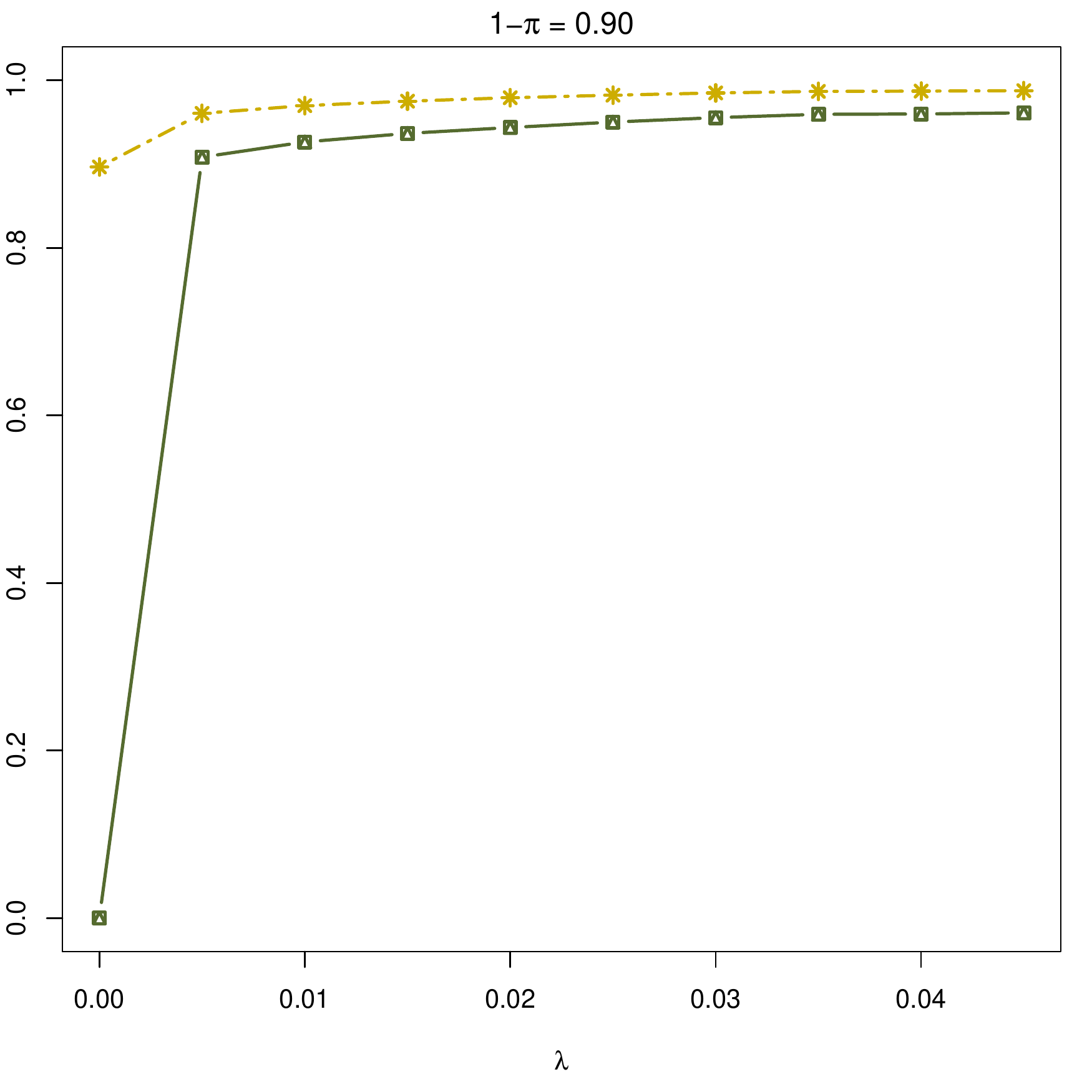}
\caption{Power Comparisons}
\label{fig3b}
\end{subfigure}
\caption{Comparison of the Data-Adaptive One-Way GBH with the naive Adaptive BH method under PRDS condition for varying choices of $0<\lambda<0.05(=\alpha)$ ($m=50, n = 100, \rho= 0.3, \pi_{\centerdot}=0, \pi$)}.\label{fig3}
\end{figure}

Figure \ref{fig3} displays the findings of the second type of investigation. As seen from it, our proposed procedure can potentially control FDR in scenarios where concentration of signals is high and for certain choices of $\lambda$, preferably $< \alpha$. A few such scenarios with varying density of signals uniformly distributed in all groups (i.e., $\pi_{\centerdot}=0$) and choices of $\lambda < \alpha$ have been shown in this figure.


\subsection{Two-Way Classified Hypotheses - One Hypothesis Per Cell} This section presents results associated with our simulation study that focused on investigating the performances of our proposed (i) Oracle Two-Way GBH$_1$ procedure (Theorem \ref{th3}) against the usual single-group BH procedure and the p-filter algorithm (\cite{2015arXiv151203397F}, \cite{2017arXiv170306222R}) in their oracle forms, and (ii) Data-Adaptive Two-Way GBH$_1$ procedure (Theorem \ref{th4}) against the Naive Data-Adaptive BH in terms of FDR control and power under normal distributional settings.

\subsubsection{Simulation Setting}

The simulation setting here is a natural extension of that in the above section. More specifically, it consists of the following steps:

\begin{itemize}

\item [1.] Generate $\boldsymbol{\Theta}_{mn}$ as an $m \times n$ random matrix of i.i.d. Ber$(1-\pi_{rc})$, $\boldsymbol{\theta}_m$ as a random vector of $m$ i.i.d. Ber$(1-\pi_{r})$, and $\boldsymbol{\theta}_n$ as a random vector of $n$ i.i.d. Ber$(1-\pi_c)$; \

\item [2.] Obtain
\begin{align}\label{e23}
& \boldsymbol{\Theta} = \boldsymbol{\Theta}_{mn}\star\boldsymbol{\theta}_m\boldsymbol{1}_n^T\star\boldsymbol{1}_m\boldsymbol{\theta}_n^T,
\end{align}
(with $A\star B$ denoting  the Hadamard product between matrices $A$ and $B$, and $\boldsymbol{1}_a$ representing the $a$-dimensional vector of $1$'s), \

\item [3.] Given $\boldsymbol{\Theta}$, generate a random $m \times n$ matrix $\mathbf{X}=((X_{gh}))$ as follows:
    \begin{eqnarray*}\label{e4.5}
\mathbf{X} & = & \mu\boldsymbol{\Theta} + \sqrt{(1-\rho_r)(1-\rho_c)}\mathbf{Z}_{mn} + \sqrt{(1-\rho_r)\rho_c}\mathbf{Z}_m\boldsymbol{1}_n^T + \nonumber \\  & & \qquad \sqrt{\rho_r(1-\rho_c)}\boldsymbol{1}_m\mathbf{Z}_n^T + \sqrt{\rho_r\rho_c}Z_0\boldsymbol{1}_m\boldsymbol{1}_n^T,
\end{eqnarray*} having generated $\mathbf{Z}_{mn}$ as $m \times n$ random matrix, $\mathbf{Z}_{m}$ as $m$-dimensional random vector, and $\mathbf{Z}_{n}$ as $n$-dimensional random vector, each comprising i.i.d. $N(0,1)$, and $Z_0$ as an additional $N(0,1)$ random variable.\

\item [4.] Apply each procedure at FDR level $\alpha=0.05$ for testing $H_{gh}: \text{E}(X_{gh}) = 0$ against $K_{gh}: \text{E}(X_{gh}) > 0$, simultaneously for all $g=1, \ldots, m, h=1, \ldots, n$, in terms of the corresponding p-values, and note the proportions of false rejections among all rejections and correct rejections among all false nulls.

\item [5.] Repeat Steps 1-4 200 times to simulate the values of FDR and power for each procedure by averaging out the corresponding proportions obtained in Step 4.\

\end{itemize}

\begin {remark} \rm Note that
\begin{align*} \label{es1}
\mathbf{vec}({\mathbf{X}}) \sim \text{N}_{m n}(\mathbf{vec}(\mu \boldsymbol{\Theta}), \Sigma_c \otimes \Sigma_r),
\end{align*}
where $\Sigma_r = (1 -\rho_r)I_n + \rho_r\boldsymbol{1}_n\boldsymbol{1}_n^T, \; \rho_{r} \in [0, 1)$, and $\Sigma_c = (1 -\rho_c)I_m + \rho_c\boldsymbol{1}_m\boldsymbol{1}_m^T, \; \rho_{c} \in [0,1).$ Thus, the test statistics are allowed to have different types of dependence structure by appropriately setting the value of $\rho_r$ and/or $\rho_c$ at $0$.

Also, as seen from equation (\ref{e23}), the hidden state of each row and each columns in terms of being significant or not has been factored into that of the hypothesis lying at their intersection. This enables us to incorporate the true effect of the underlying two-way classification structure into our simulation study. Specifically, we can regulate the density of signals in the entire matrix using the following
\begin{eqnarray*}
\pi_0 & = & 1 - (1 - \pi_{rc})(1 - \pi_r)(1 - \pi_c),
\end{eqnarray*} representing the proportion of true nulls in the entire set of mn hypotheses in terms of the proportions of rows ($1-\pi_{r}$) and columns ($1-\pi_{c}$) containing signals.  \end {remark}

\subsubsection{Simulation Findings} We fixed $m = 50$, $n = 100$, $\mu = 0$ for true null hypotheses, and $=3$ for true signals.

\vskip 10 pt
\noindent {\it Comparison of Oracle Procedures:} Here, we wanted to make two types of investigation of Oracle Two-Way GBH$_1$'s performance under independence as well as under PRDS compared to the other oracle procedures being considered - one in terms of identifying signals and the other in terms of FDR control and power. The findings of these are displayed in Figures 4-6.

\begin{figure}
\centering
$\begin{array}{rlrl}
\begin{subfigure}[b]{0.2\textwidth}
\includegraphics[width=\textwidth]{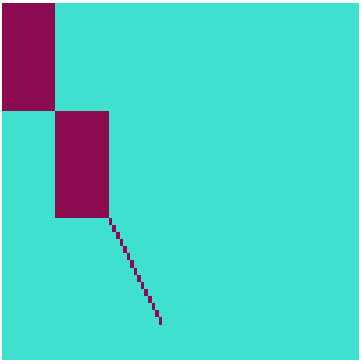}
\caption{Signals}
\label{fig4a}
\end{subfigure}&
\begin{subfigure}[b]{0.2\textwidth}
\includegraphics[width=\textwidth]{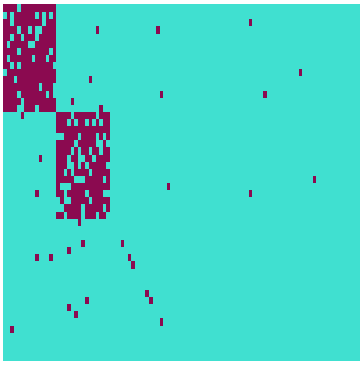}
\caption{Two-way GBH}
\label{fig4b}
\end{subfigure}&
\begin{subfigure}[b]{0.2\textwidth}
\includegraphics[width=\textwidth]{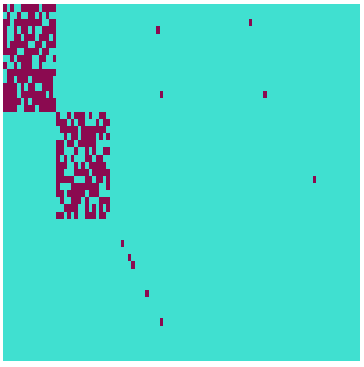}
\caption{pfilter}
\label{fig4c}
\end{subfigure}&
\begin{subfigure}[b]{0.2\textwidth}
\includegraphics[width=\textwidth]{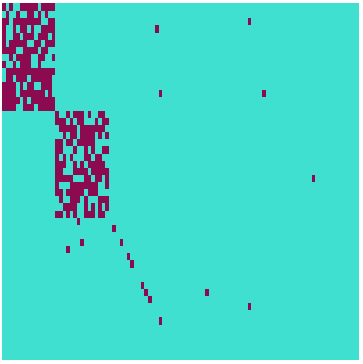}
\caption{BH}
\label{fig4d}
\end{subfigure}\\
&
\begin{subfigure}[b]{0.2\textwidth}
\includegraphics[width=\textwidth]{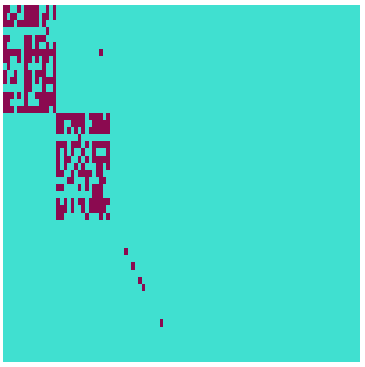}
\caption{Two-way GBH}
\label{fig4e}
\end{subfigure}&
\begin{subfigure}[b]{0.2\textwidth}
\includegraphics[width=\textwidth]{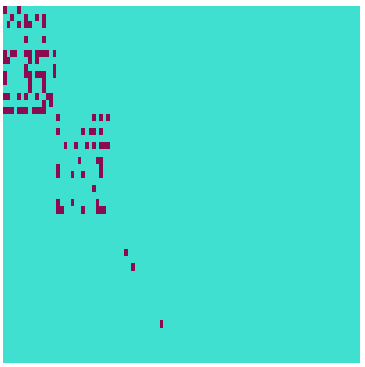}
\caption{pfilter}
\label{fig4f}
\end{subfigure}
&\begin{subfigure}[b]{0.2\textwidth}
\includegraphics[width=\textwidth]{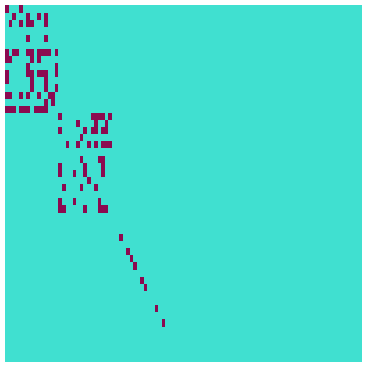}
\caption{BH}
\label{fig4g}
\end{subfigure}
\end{array}$
\caption{Comparison of the proposed oracle Two-way GBH with one hypothesis per cell, with other methods for one trial. (a) shows the layout of the significant hypotheses. (b), (c) and (d) show the performances of the proposed two-way GBH, the p-filter process and the BH procedure if the hypotheses are independent. (e), (f) and (g) show the performances of these methods for the same setup, respectively, if there is positive dependence among the hypotheses. We choose $\rho_r = 0.3$ and $\rho_c = 0.4$ for the case of dependent hypotheses.}\label{fig4}
\end{figure}

For the first type of investigation, in the $50 \times 100$ matrix, we arranged the significant hypotheses in two $15 \times 15$ blocks and along the diagonal of another $15\times 15$ block, as shown in Figure (\ref{fig4a}). This arrangement helps to analyze the performance of a multiple testing procedure when the signals are dense (in the two blocks) as well as when they are sparse (along the diagonal).
The performance of each method based on one trial is shown in the remaining plots in Figure (\ref{fig4}), with that being shown in Figures \ref{fig4b}-\ref{fig4d} for the independence case and in Figures \ref{fig4e} -\ref{fig4g} in the PRDS case (when $\rho_r = 0.3$ and $\rho_c = 0.4$). In either case, the proposed Oracle Two-Way GBH$_1$ procedure is seen to be successful in identifying maximum number of clustered signals, and almost equally efficient as the BH procedure when the signals are sparse. The performances of the p-filter algorithm and the BH procedure are comparable. The BH better identifies sparse signals, although under independence. It makes marginally higher number of false rejections than the p-filter process.

\begin{figure}[H]
\centering

\begin{subfigure}[b]{\textwidth}
\centering
\includegraphics[width=0.22\textwidth]{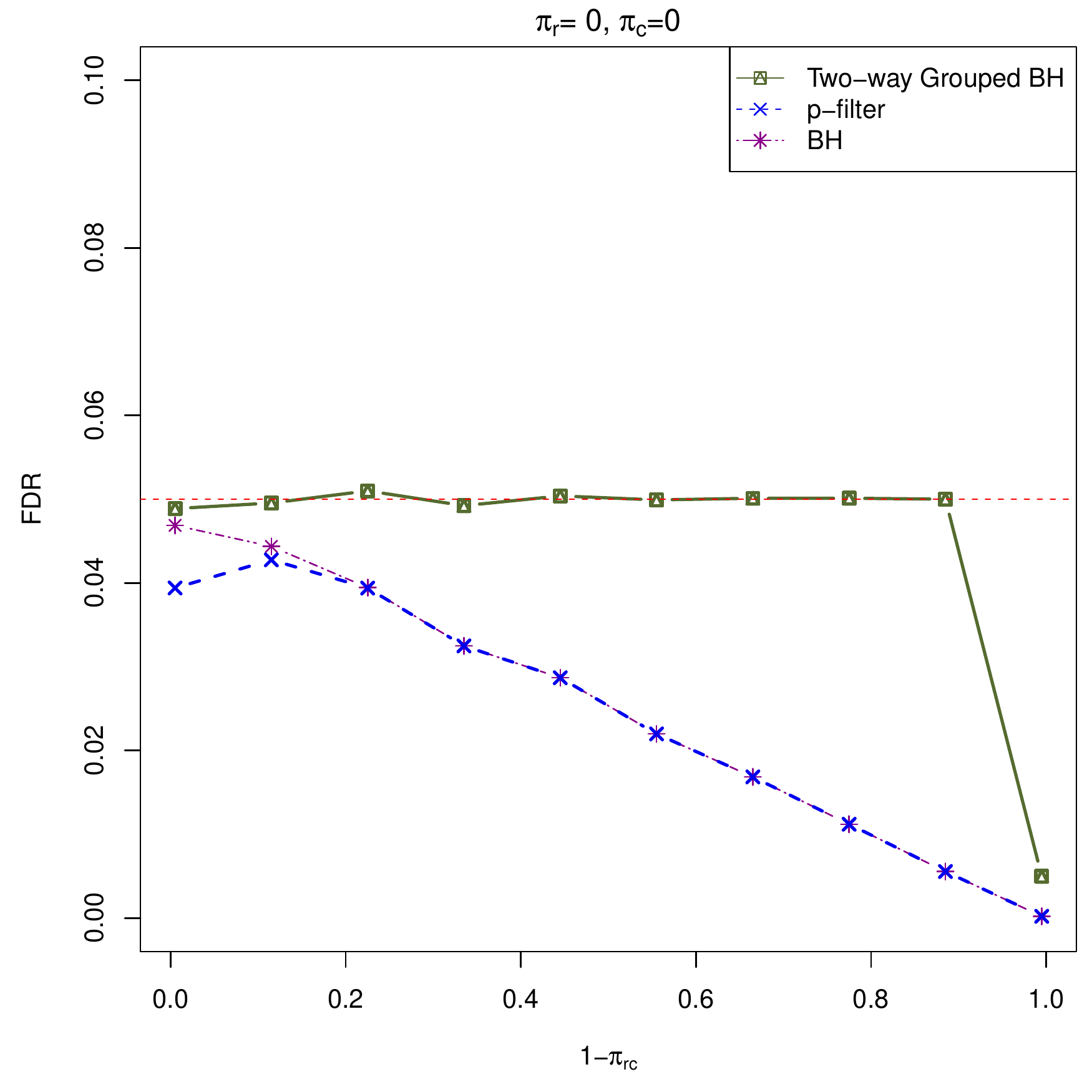}	
\includegraphics[width=0.22\textwidth]{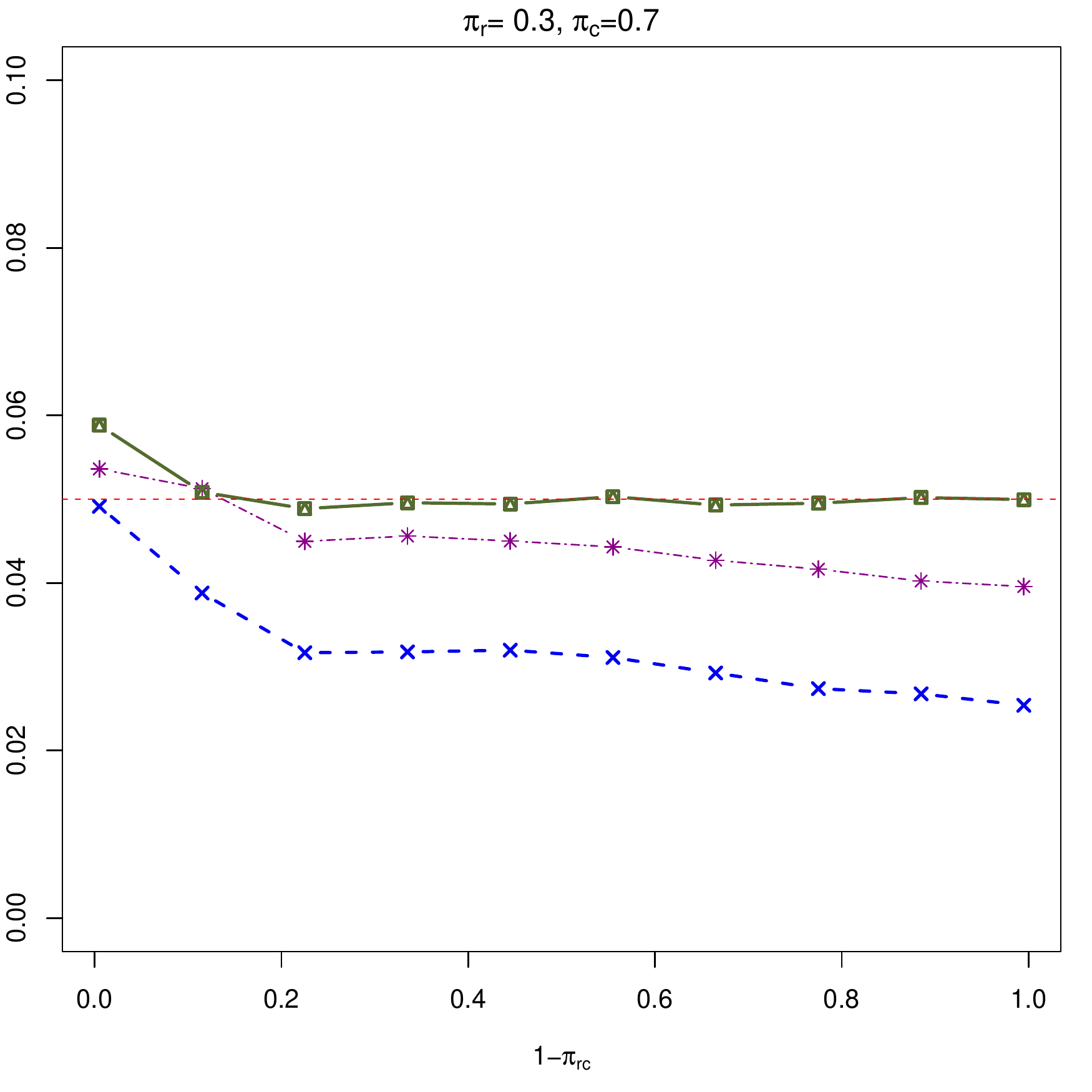}	
\includegraphics[width=0.22\textwidth]{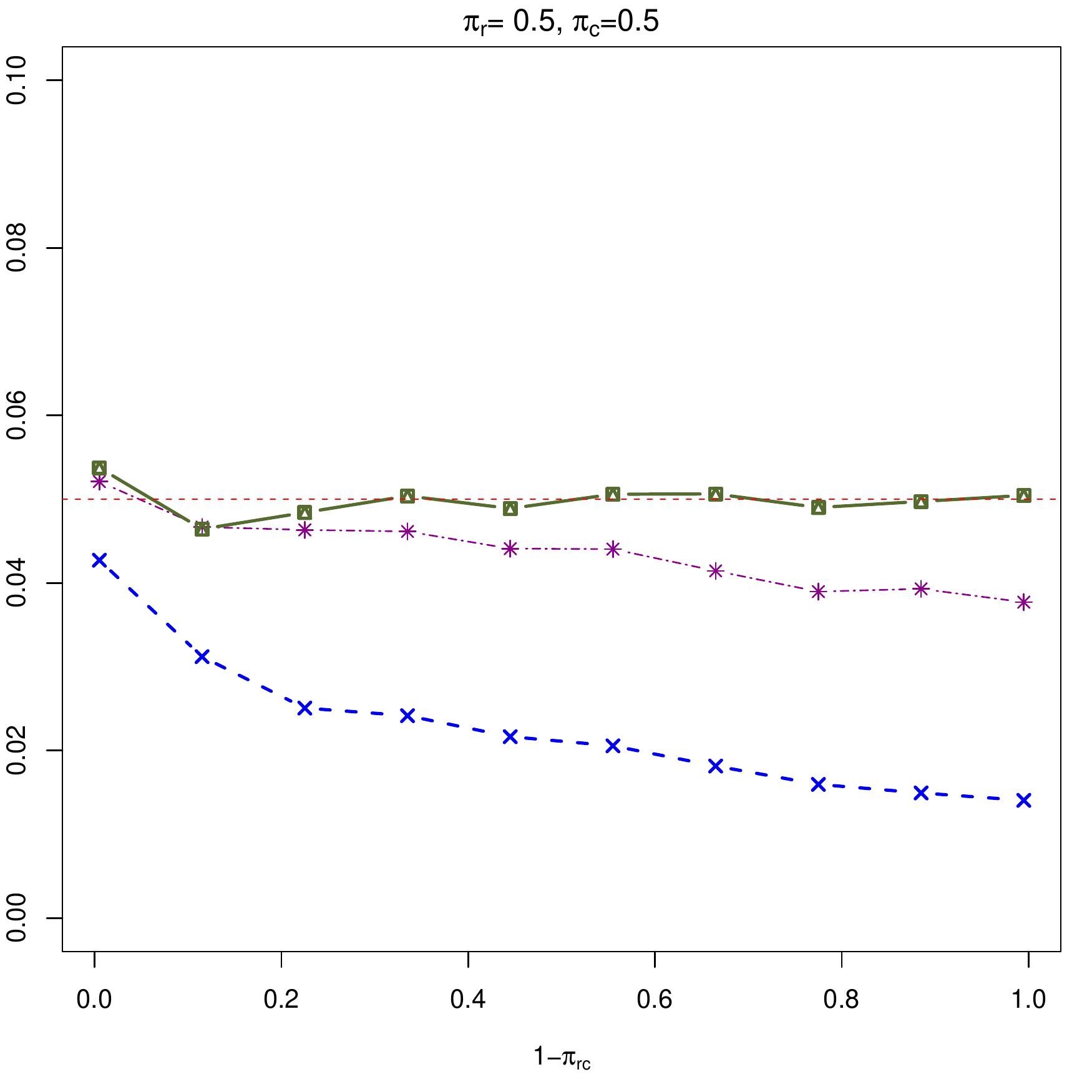}	
\includegraphics[width=0.22\textwidth]{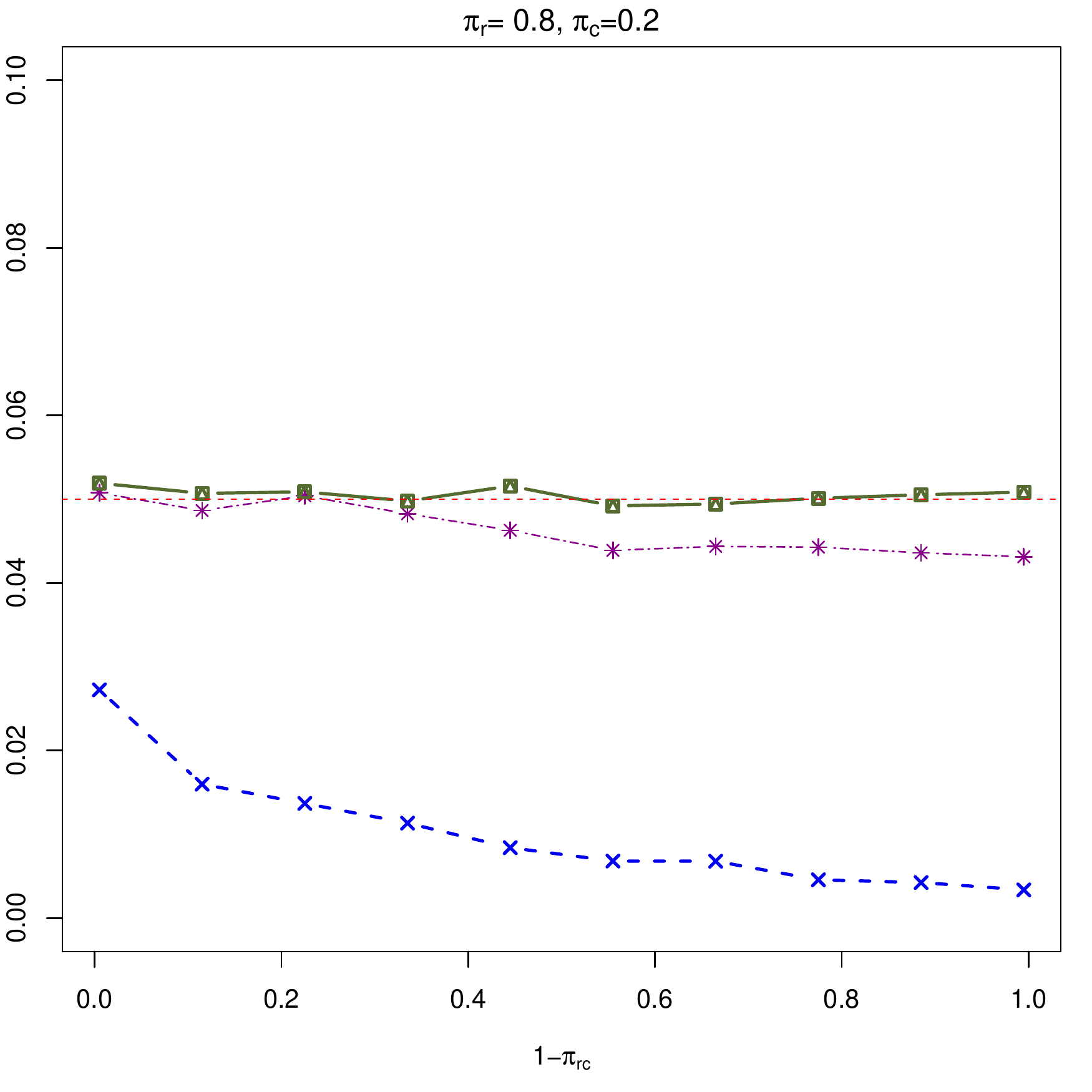}
\caption{FDR Comparisons}	
\label{fig5a}
\end{subfigure}\\

\begin{subfigure}[b]{\textwidth}
\centering
\includegraphics[width=0.22\textwidth]{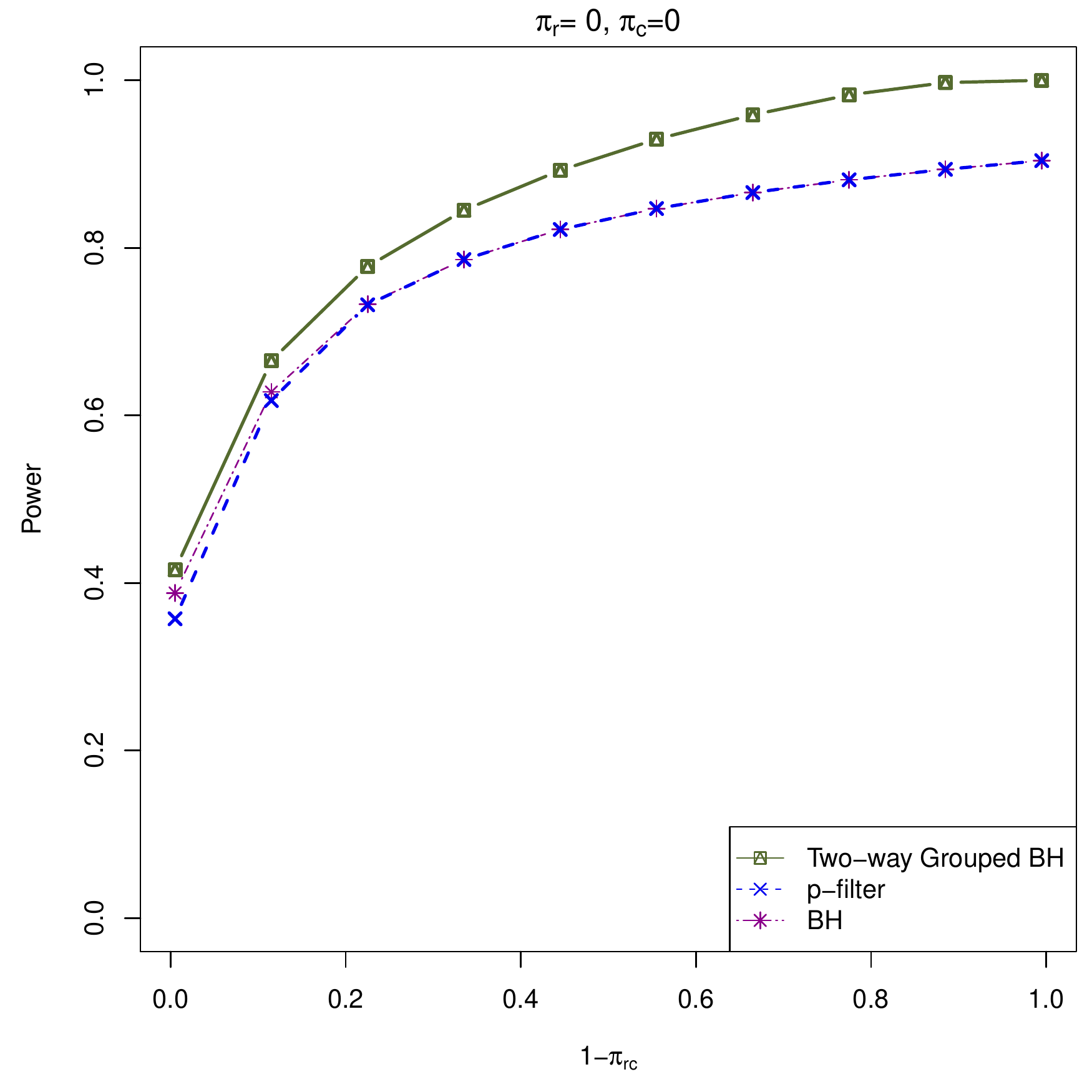}
\includegraphics[width=0.22\textwidth]{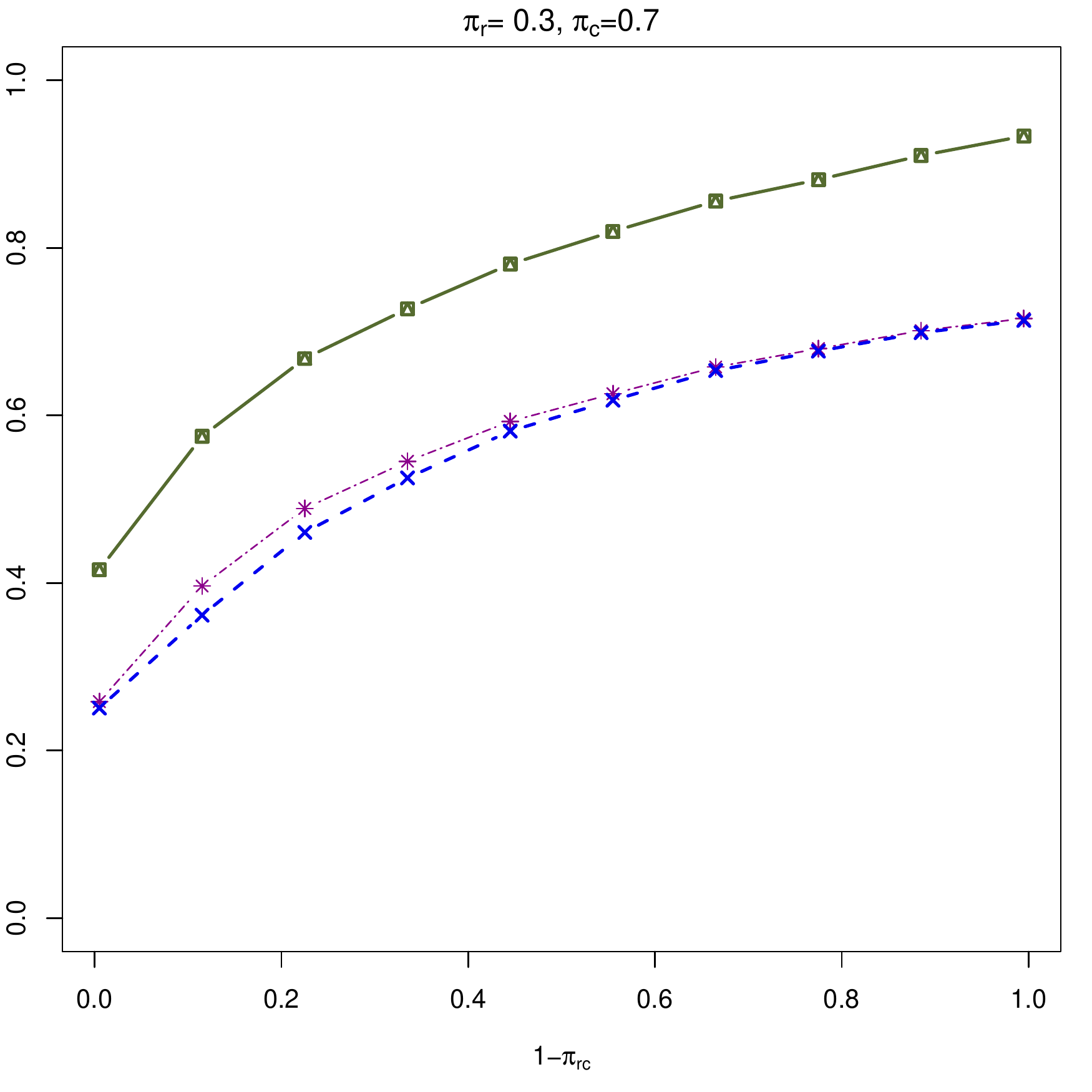}
\includegraphics[width=0.22\textwidth]{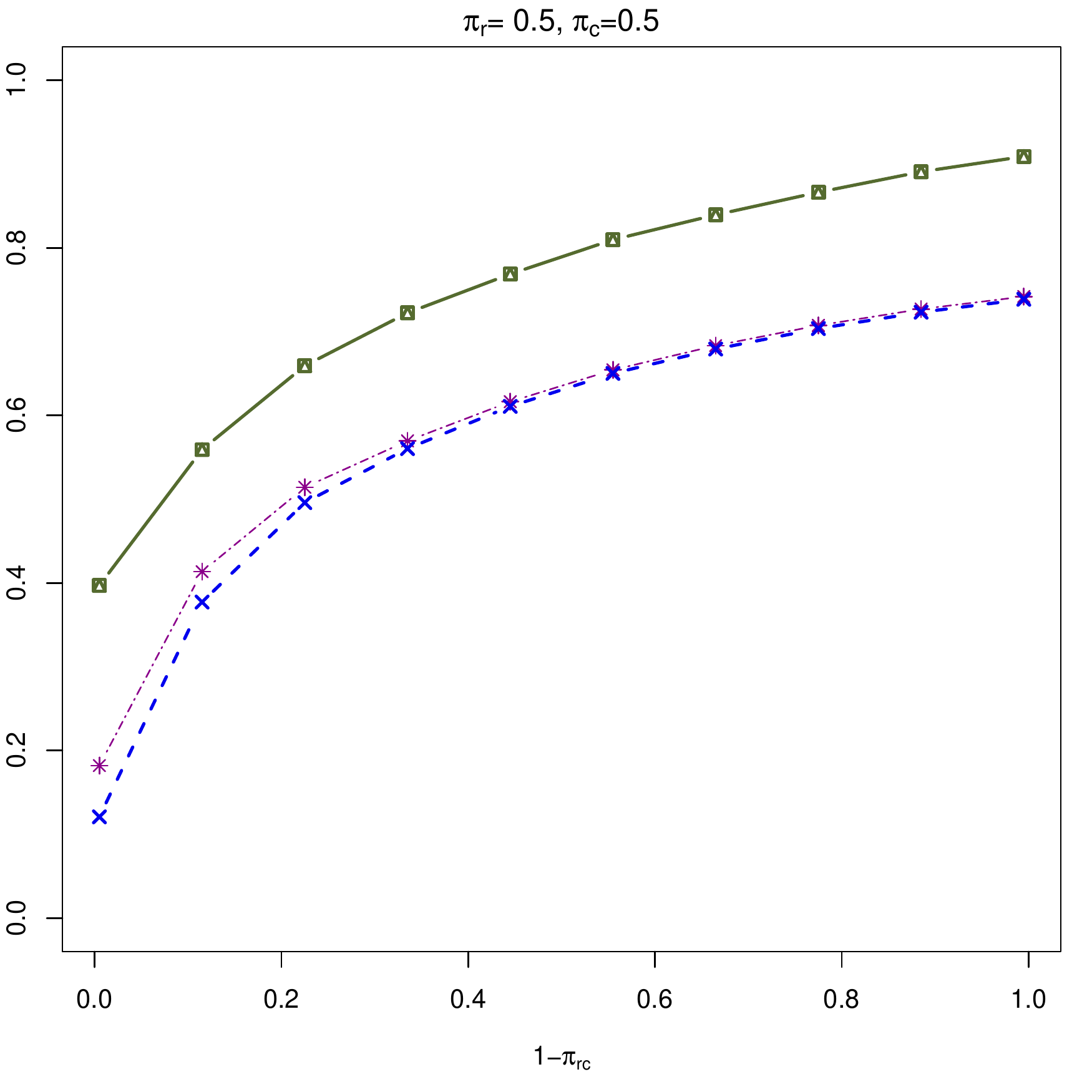}
\includegraphics[width=0.22\textwidth]{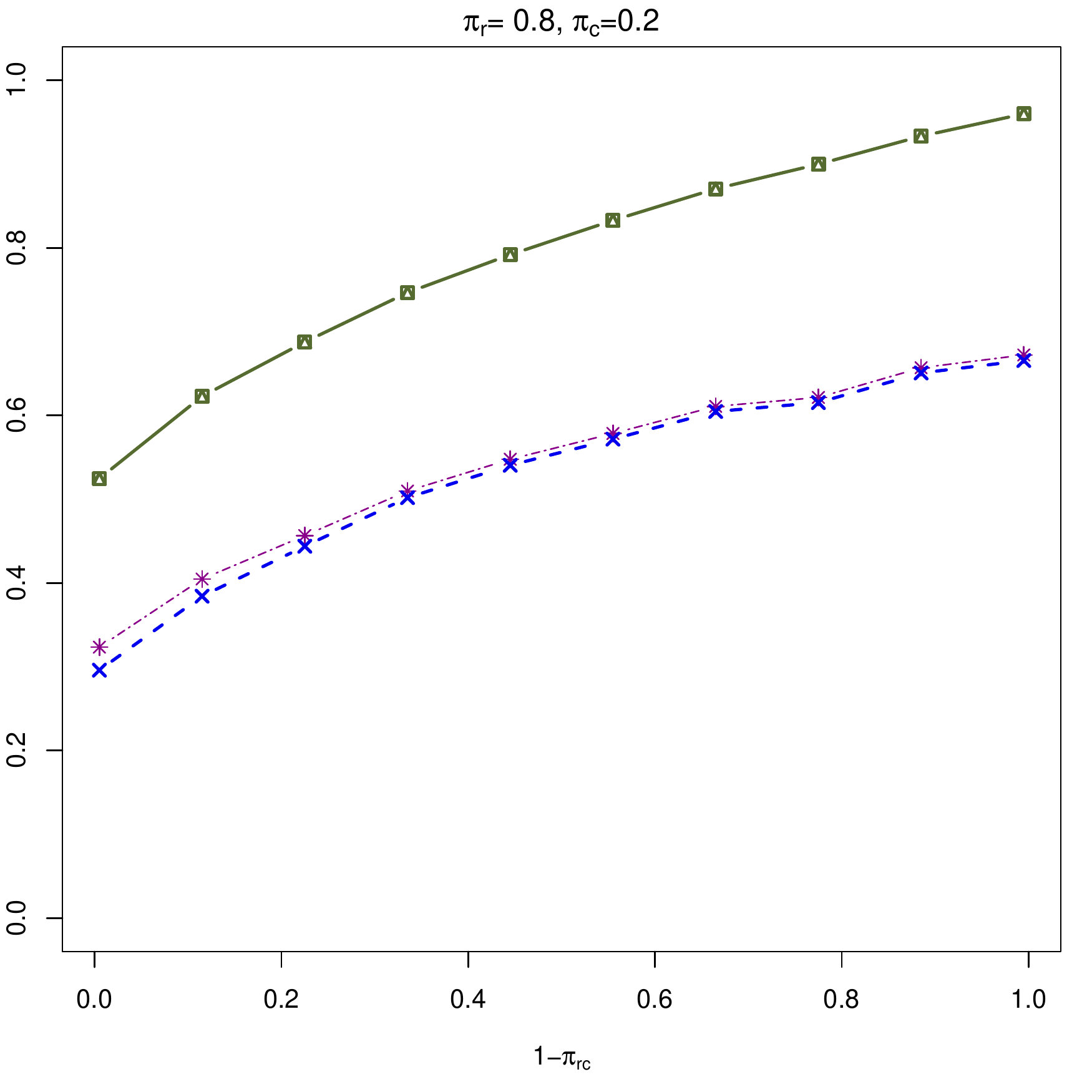}
\caption{Power Comparisons}
\label{fig5b}
\end{subfigure}
\caption{Comparison of the oracle Two-way GBH$_1$ procedure with other methods, under independence. Set of parameters used is $(m=50, n = 100, \rho_r= 0, \rho_c=0, \pi_r, \pi_c, \pi_{rc})$.}\label{fig5}
\end{figure}

\begin{figure}
\centering

\begin{subfigure}[b]{\textwidth}
\centering
\includegraphics[width=0.22\textwidth]{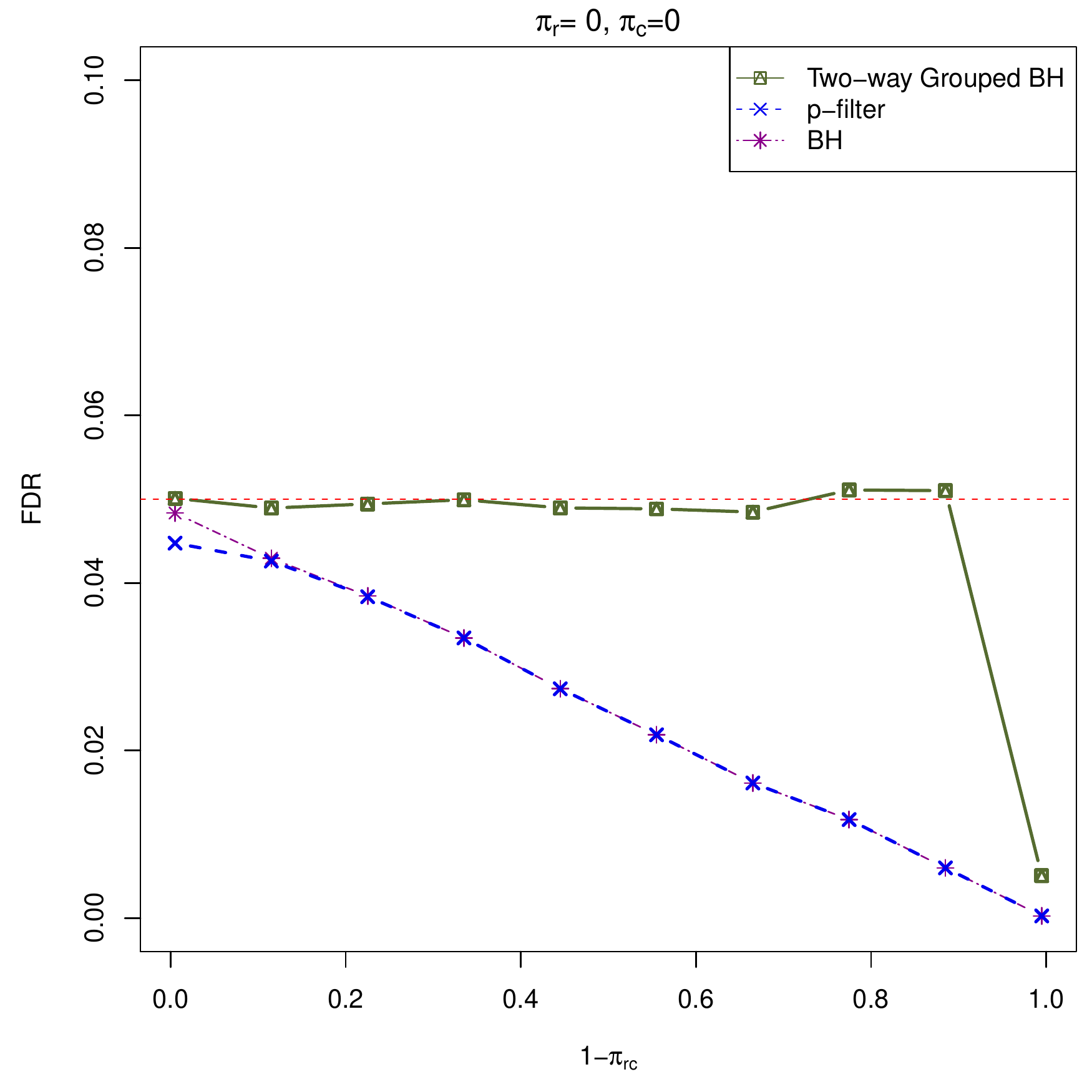}	
\includegraphics[width=0.22\textwidth]{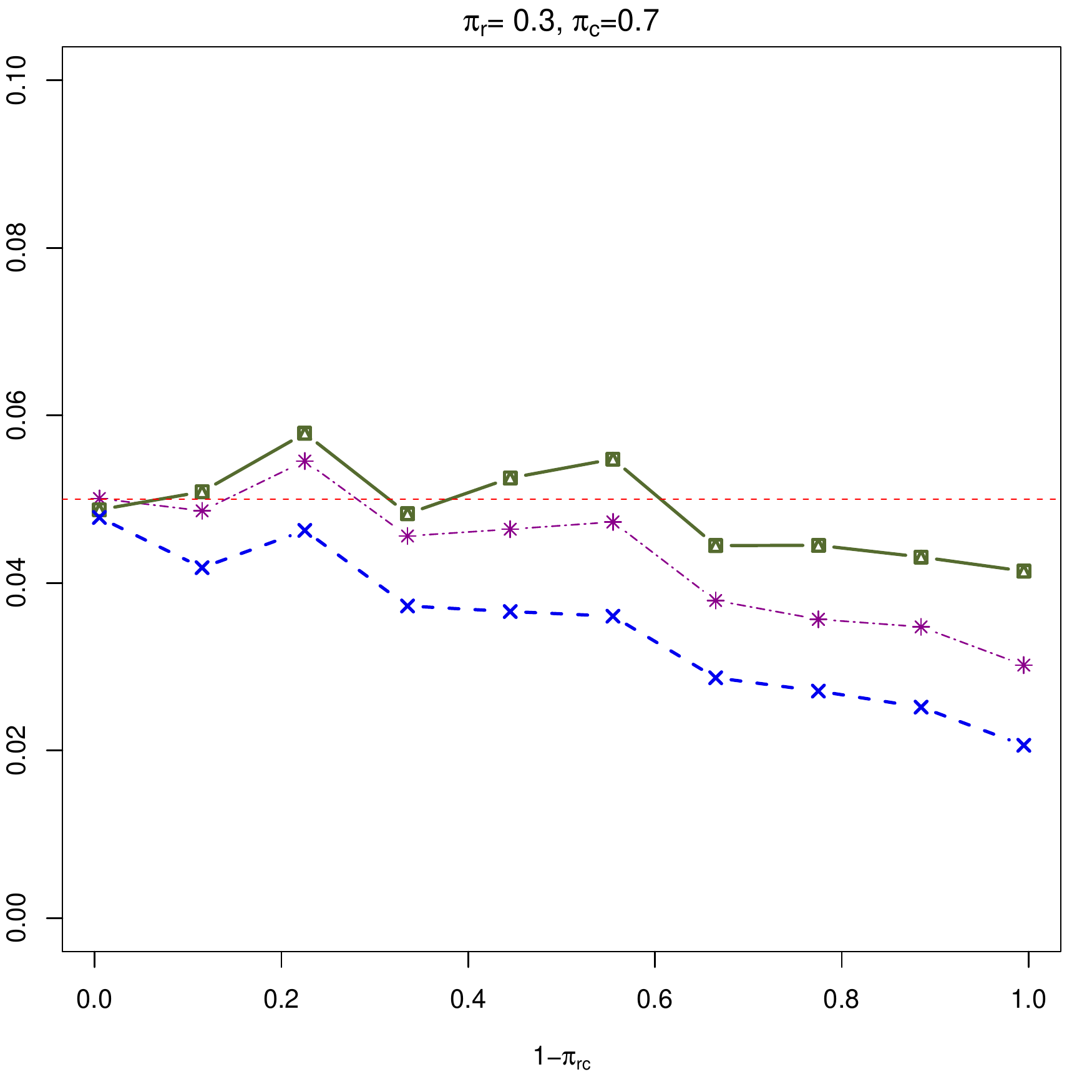}	
\includegraphics[width=0.22\textwidth]{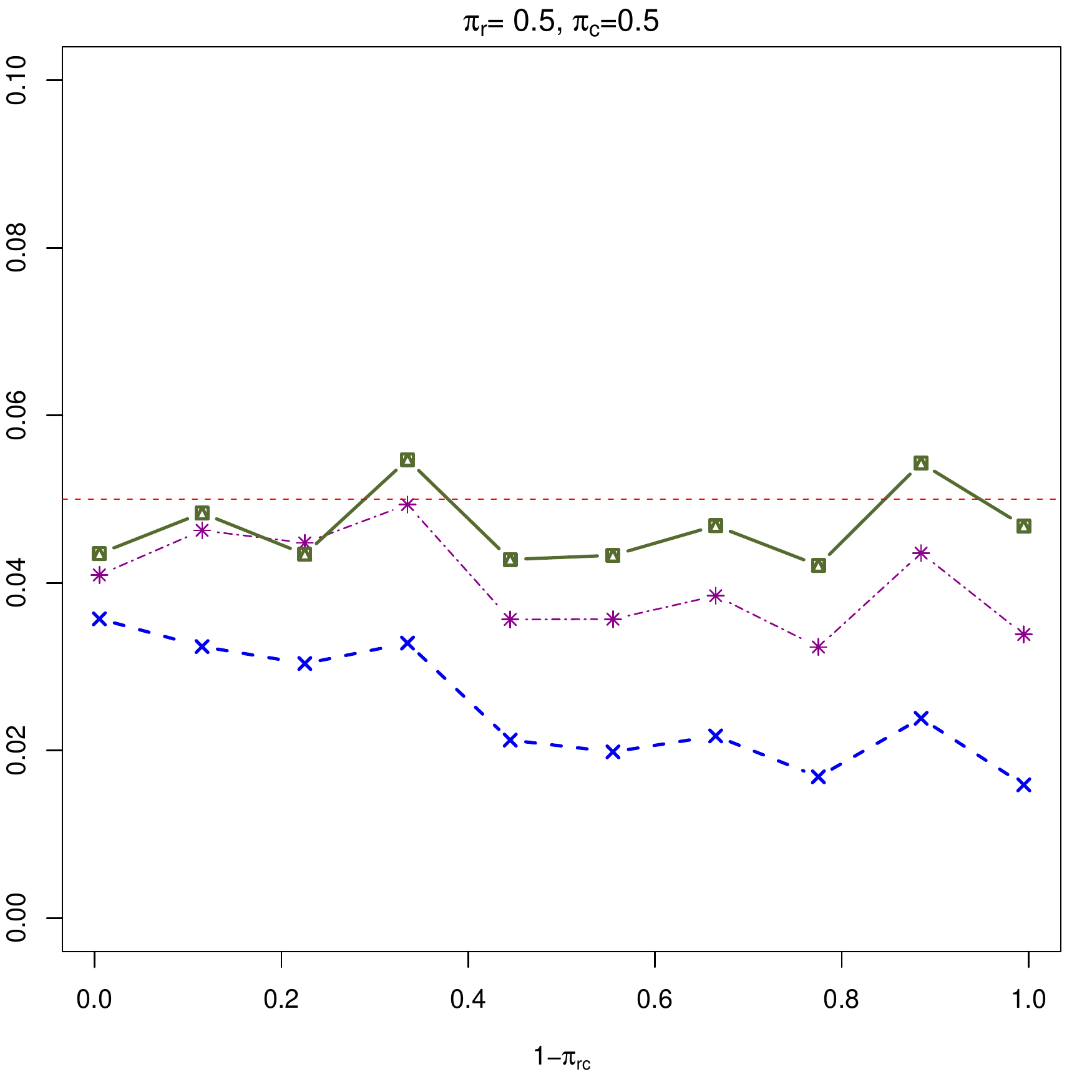}	
\includegraphics[width=0.22\textwidth]{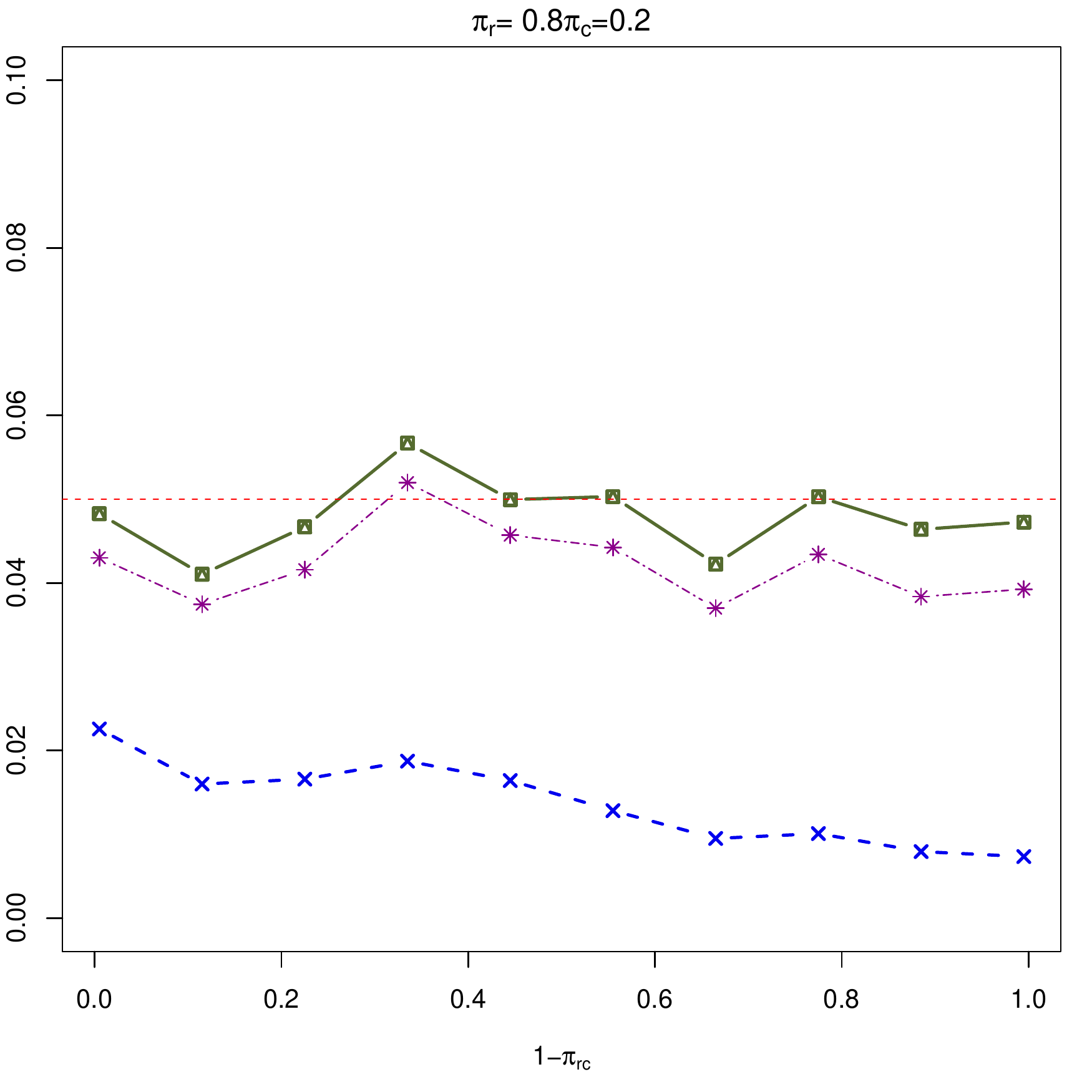}
\caption{FDR Comparisons}	
\label{fig6a}
\end{subfigure}\\

\begin{subfigure}[b]{\textwidth}
\centering
\includegraphics[width=0.22\textwidth]{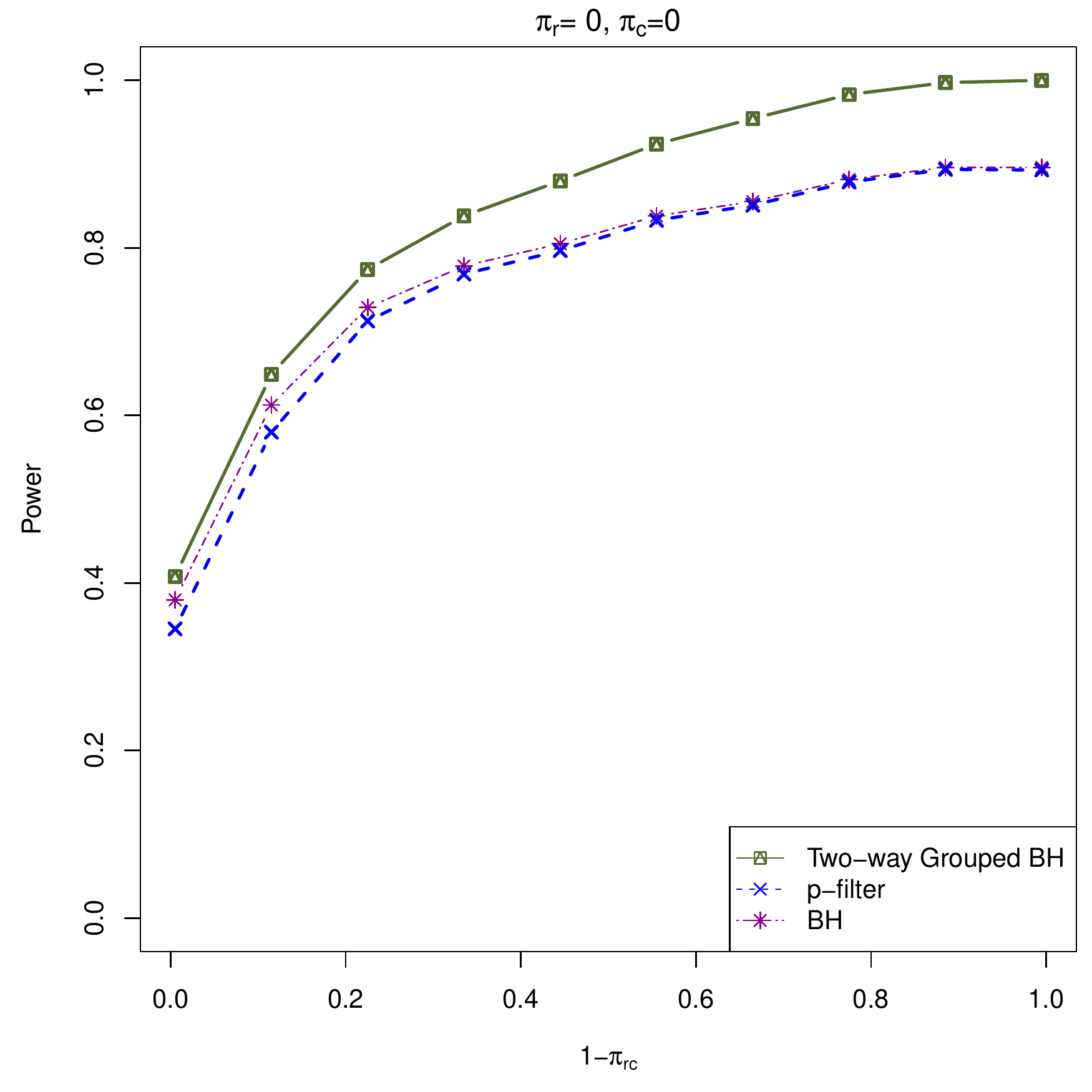}	
\includegraphics[width=0.22\textwidth]{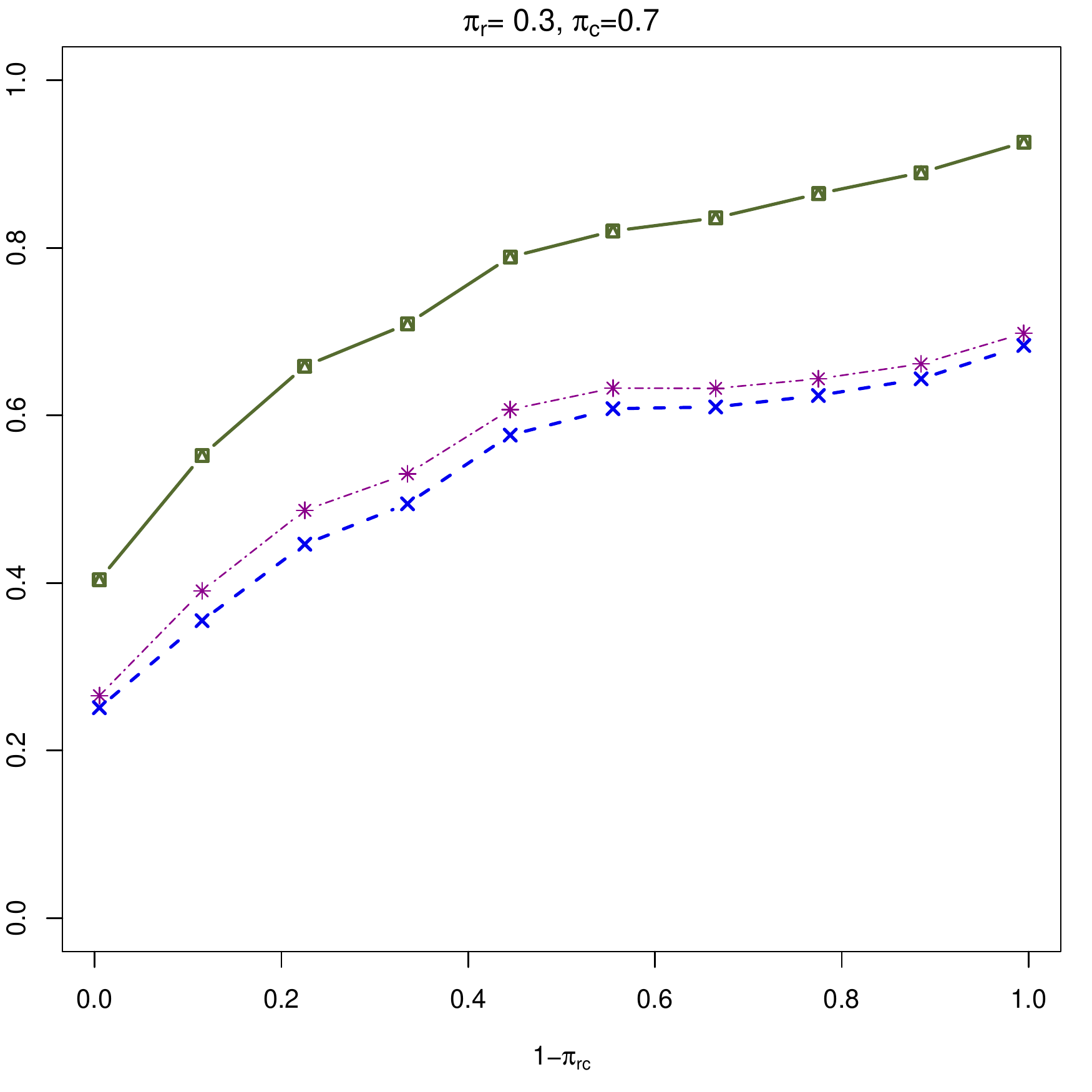}
\includegraphics[width=0.22\textwidth]{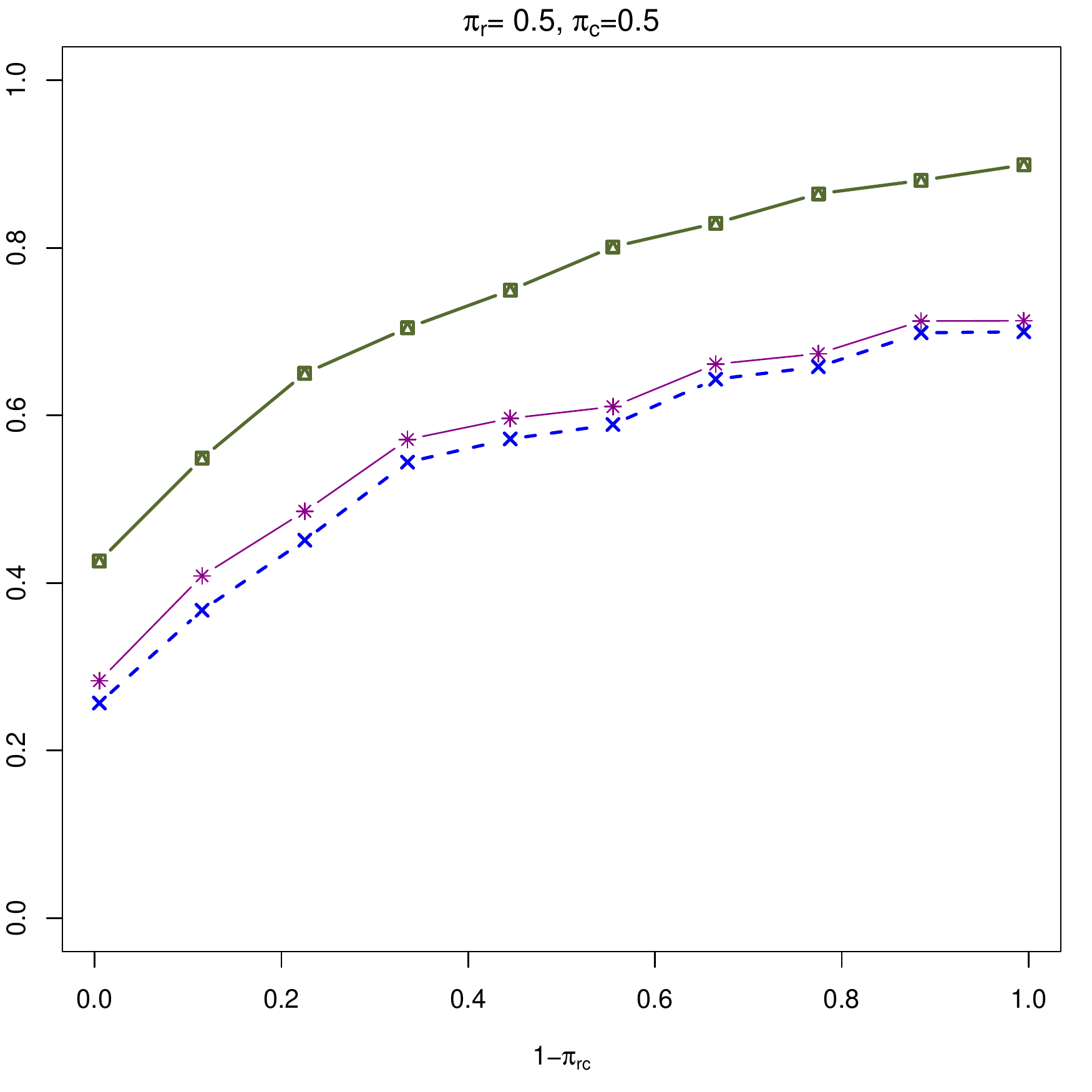}
\includegraphics[width=0.22\textwidth]{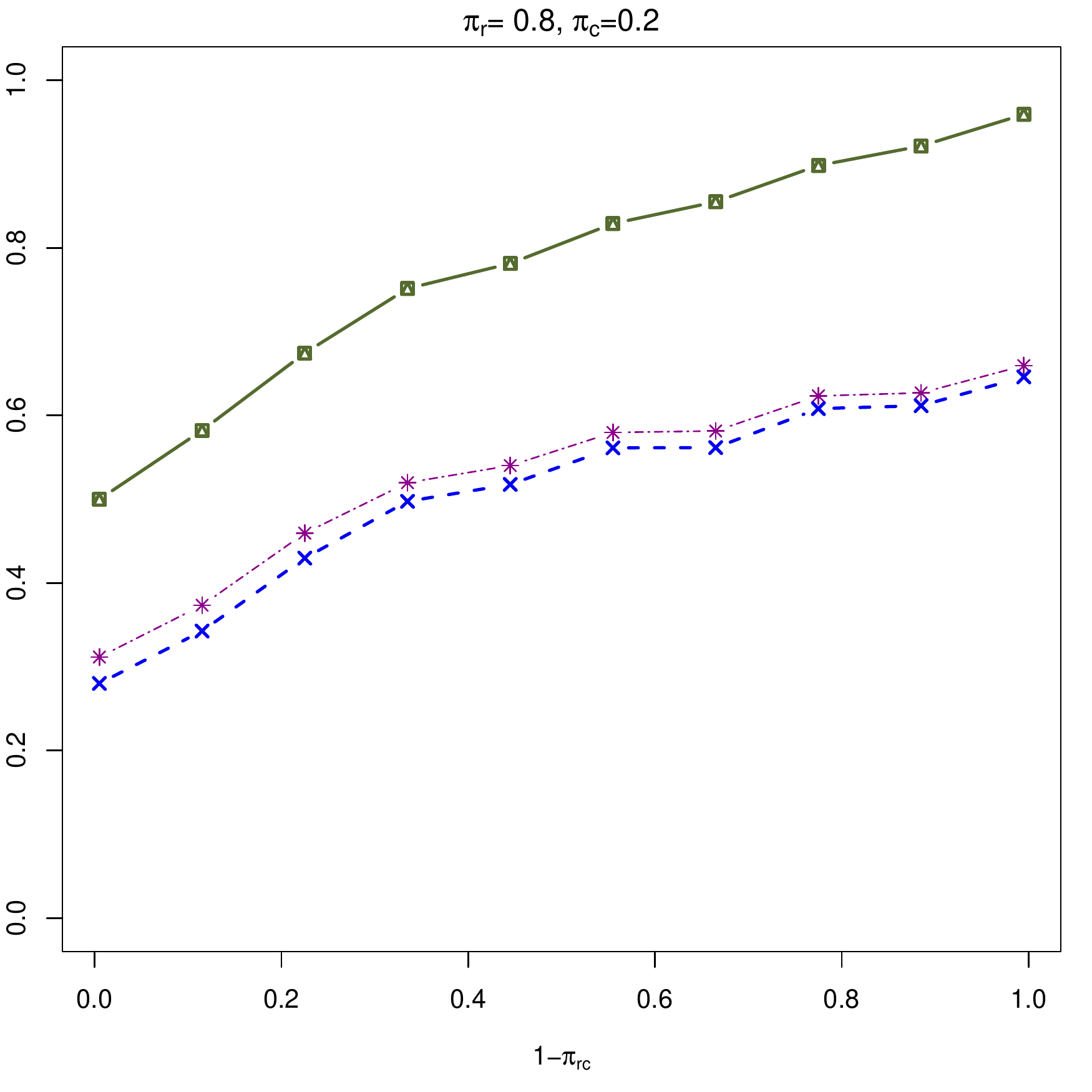}
\caption{Power Comparisons}
\label{fig6b}
\end{subfigure}

\caption{Comparison of the oracle Two-way GBH$_1$ procedure for hypotheses with PRDS property, with other methods. Set of parameters used is $(m=50, n = 100, \rho_r= 0.3, \rho_c=0.4, \pi_r, \pi_c, \pi_{rc})$.}\label{fig6}
\end{figure}

For the second type of investigation, we varied the density of true signals in the $50 \times 100$ blocks. We chose different values for $\pi_r$ and $\pi_c$ to regulate the proportions of significant rows and columns. For each choice of $(\pi_r,\, \pi_c)$, we varied $\pi_{rc}$ between $0$ and $1$ to determine the density of signals in the significant rows and columns.
We evaluated the performance of each of the three methods in terms of simulated FDR and power at each level of $(1 - \pi_{rc})$. The results are displayed in Figure \ref{fig5} for the independent case and in Figure \ref{fig6} for the PRDS case. Our proposed Oracle Two-Way GBH$_1$ procedure performs better than either p-filter algorithm or the BH procedure in terms of both FDR control and power. Performances of the p-filter process and the BH procedure are comparable. As the density of true signals increases, the proposed method maintains control on FDR at level $\alpha$ and is more powerful than the other two procedures.

\noindent {\it Comparison of Data-Adaptive Procedures}: Here, our focus had been to investigate the following two questions regarding performance of our proposed Data Adaptive Two-Way GBH$_1$ in Theorem \ref{th4} compared to its natural competitor, which is Naive Data-Adaptive BH: (i) How well it performs under independence when both are theoretically known to control FDR? (ii) If it can it possibly control FDR under PRDS in view of the fact that such  control is yet to be theoretically proved for both of these procedures.

\begin{figure}
\centering

\begin{subfigure}[b]{\textwidth}
\centering
\includegraphics[width=0.22\textwidth]{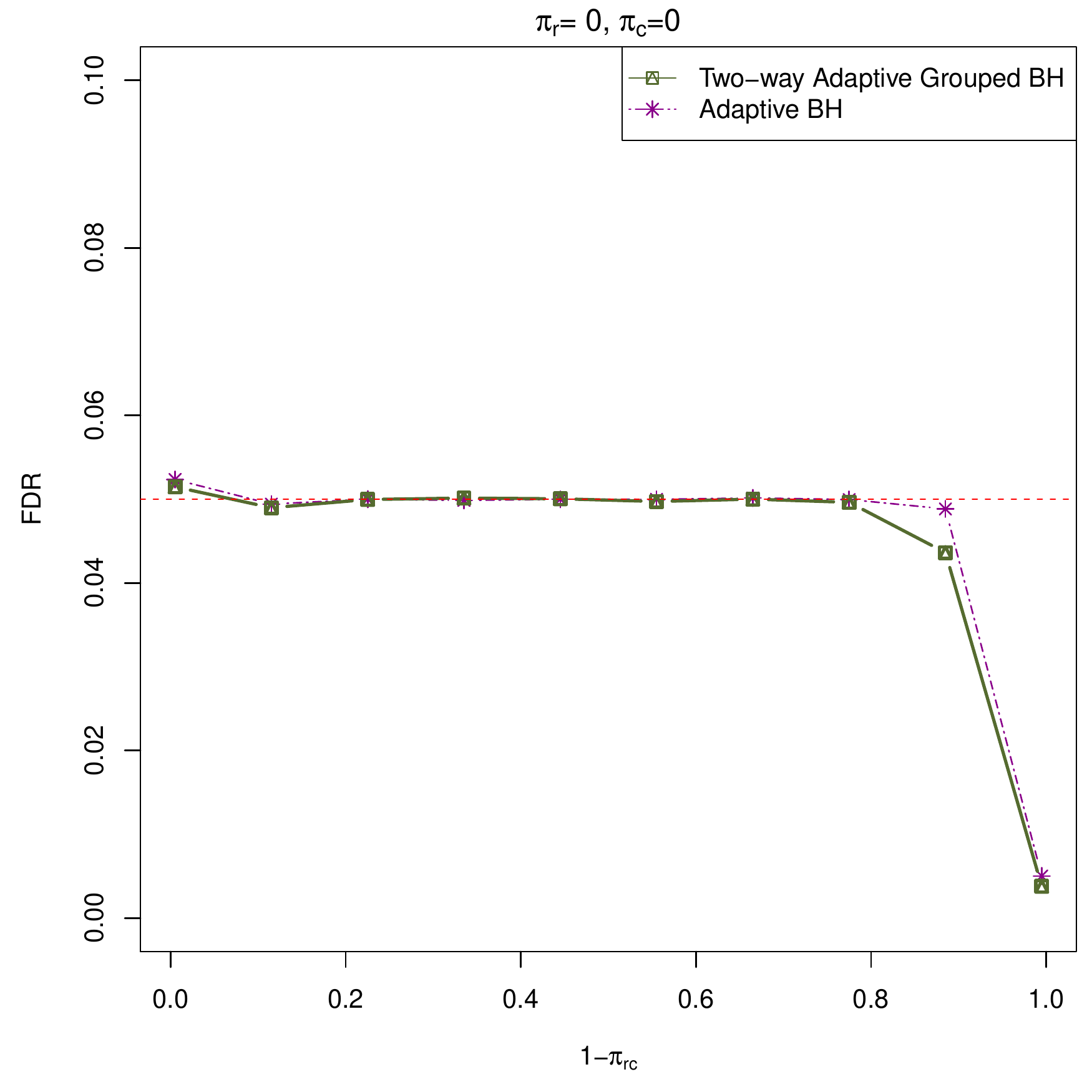}	
\includegraphics[width=0.22\textwidth]{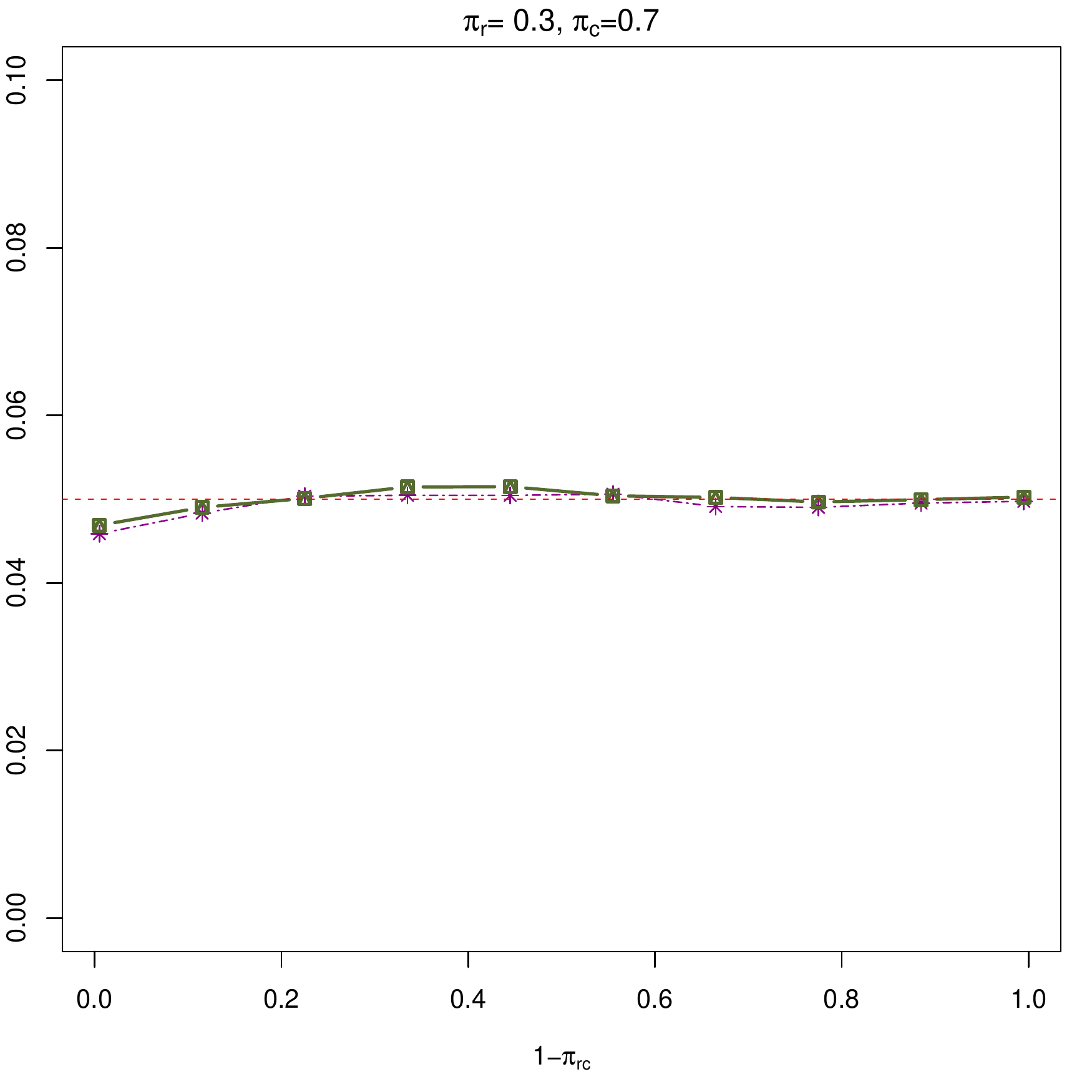}	
\includegraphics[width=0.22\textwidth]{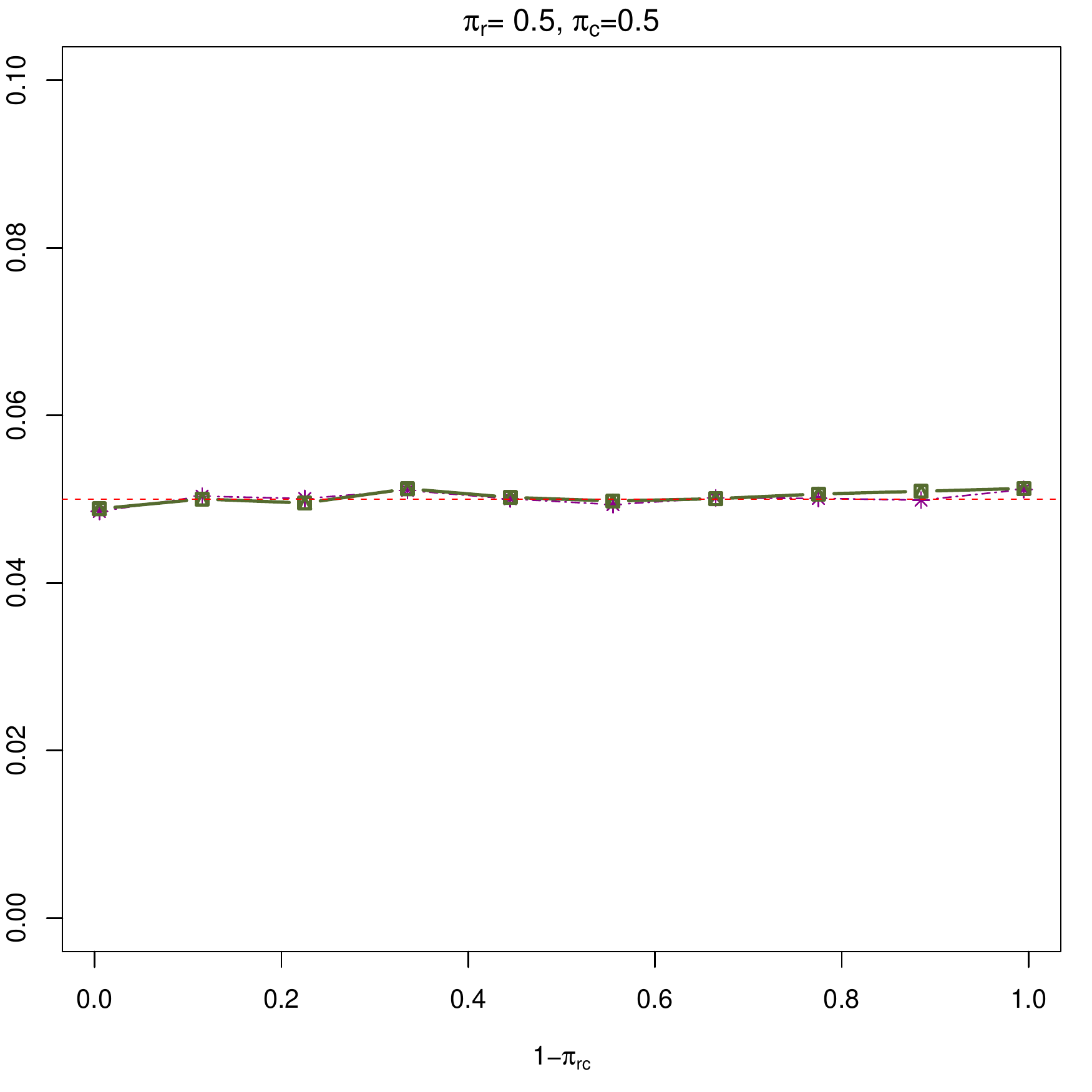}	
\includegraphics[width=0.22\textwidth]{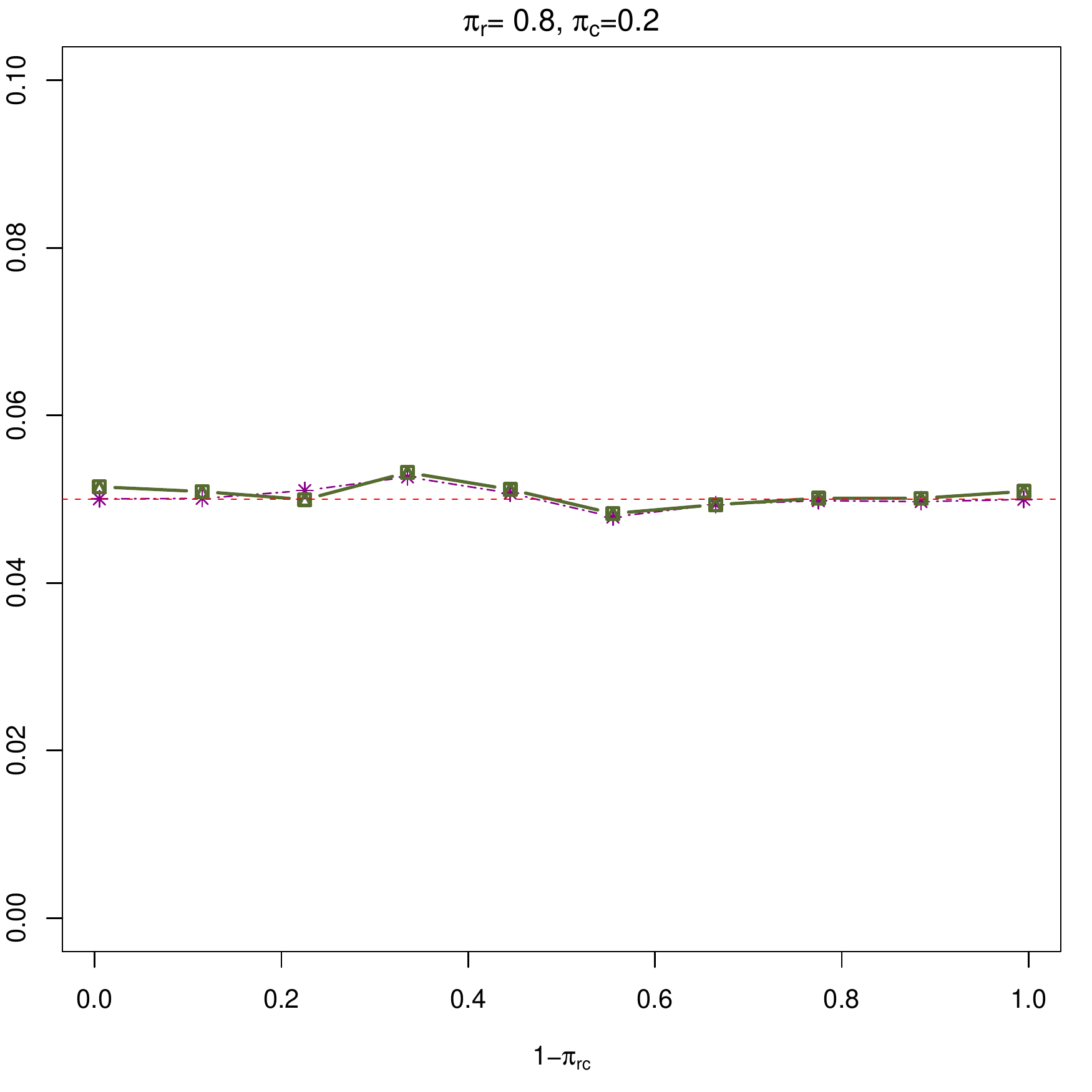}
\caption{FDR Comparisons}	
\label{fig7a}
\end{subfigure}\\

\begin{subfigure}[b]{\textwidth}
\centering
\includegraphics[width=0.22\textwidth]{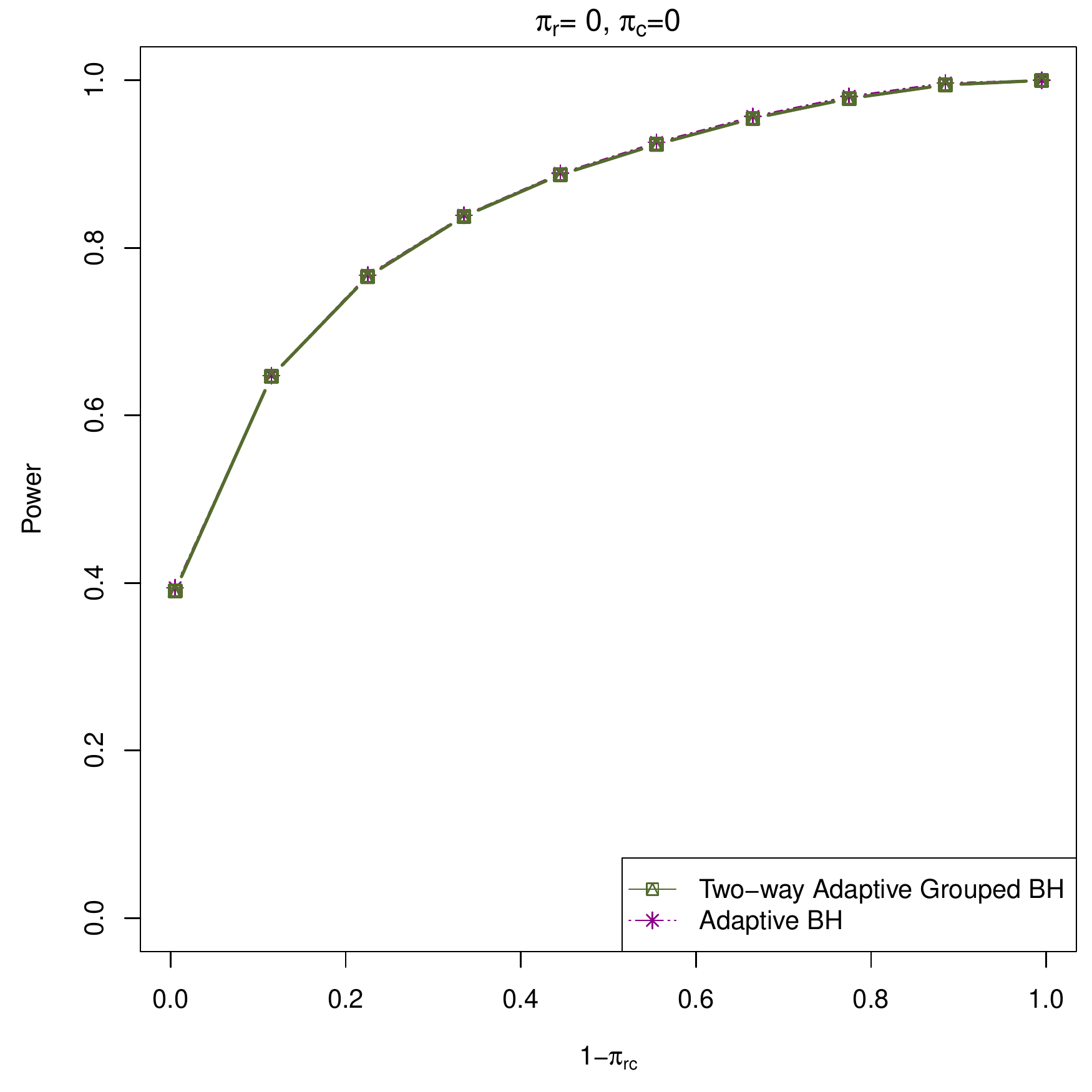}	
\includegraphics[width=0.22\textwidth]{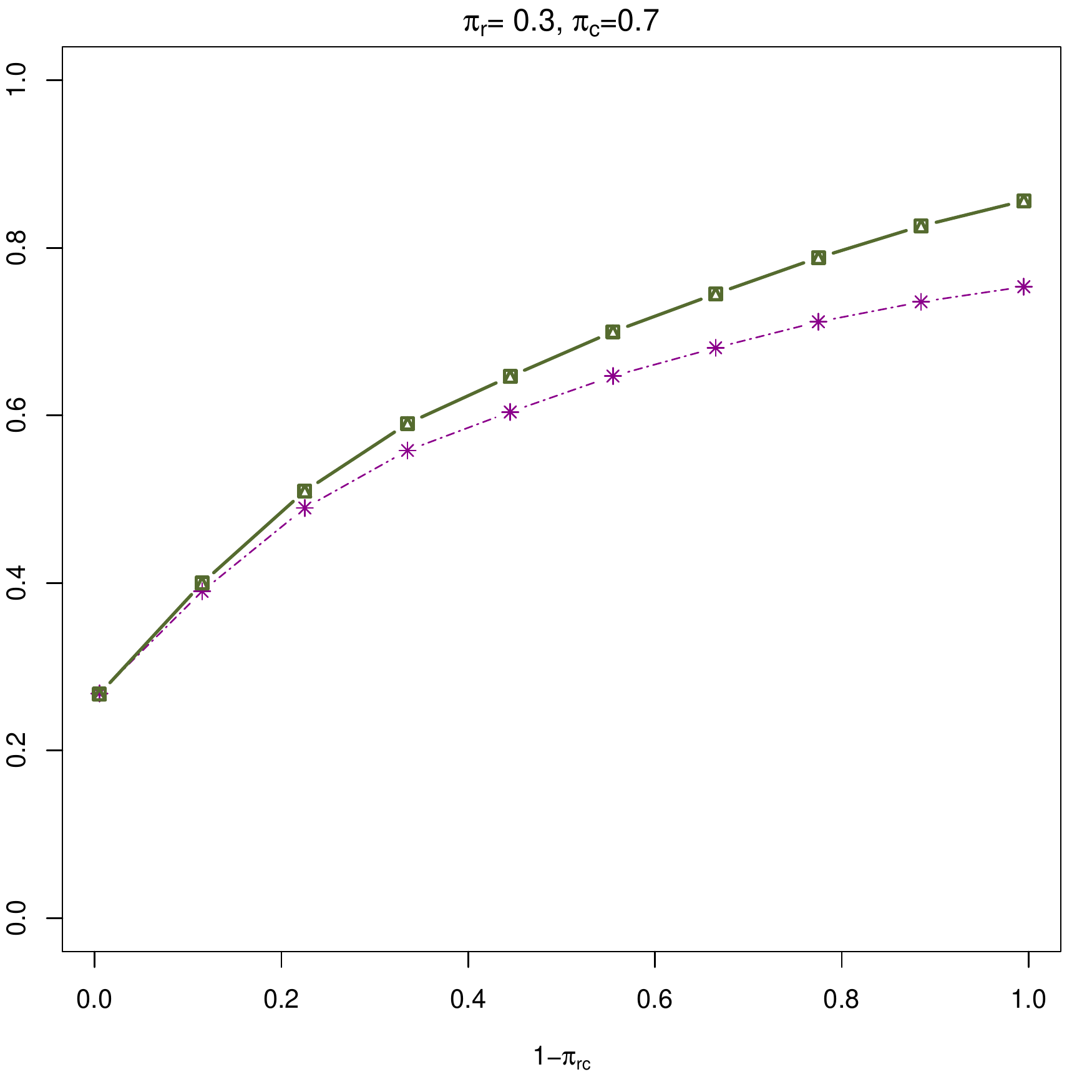}
\includegraphics[width=0.22\textwidth]{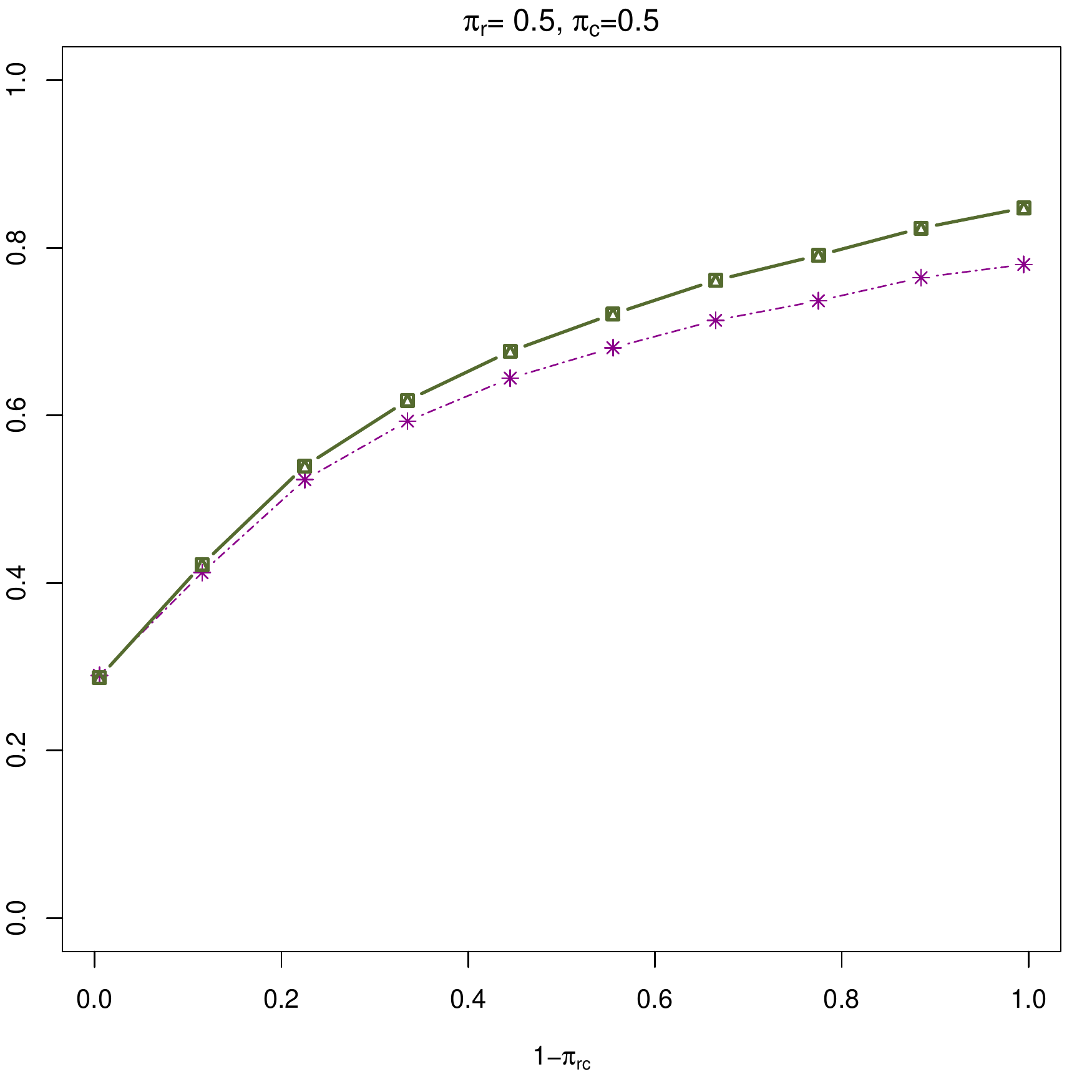}
\includegraphics[width=0.22\textwidth]{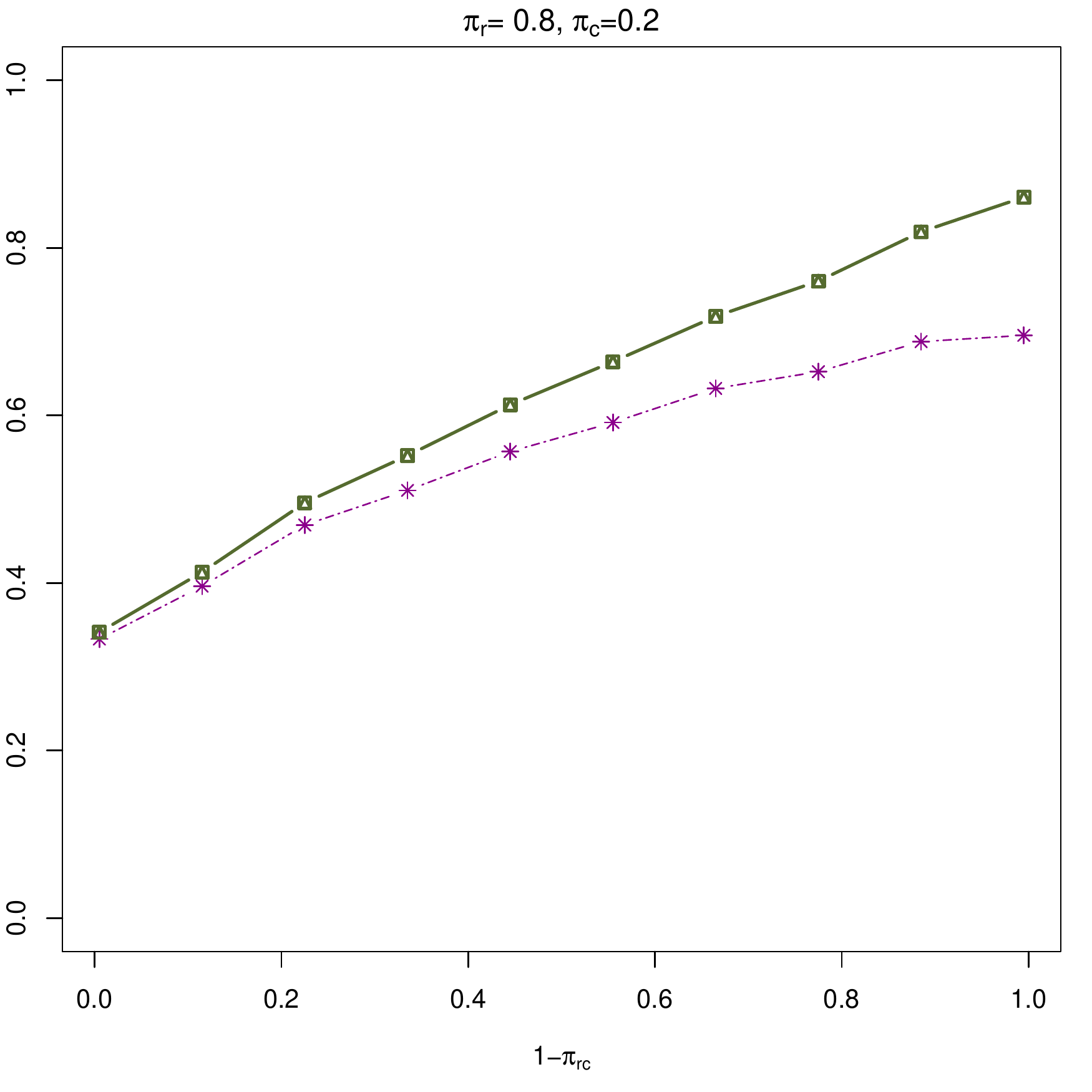}
\caption{Power Comparisons}
\label{fig7b}
\end{subfigure}

\caption{Comparison of the data-adaptive Two-way GBH$_1$ procedure with the naive Adaptive BH method, under independence. Set of parameters used is $(m = 50, n = 100, \rho_r =0, \rho_c = 0, \pi_r, \pi_c, \pi_{rc})$}\label{fig7}
\end{figure}

\begin{figure}
\centering

\begin{subfigure}[b]{\textwidth}
\centering
\includegraphics[width=0.22\textwidth]{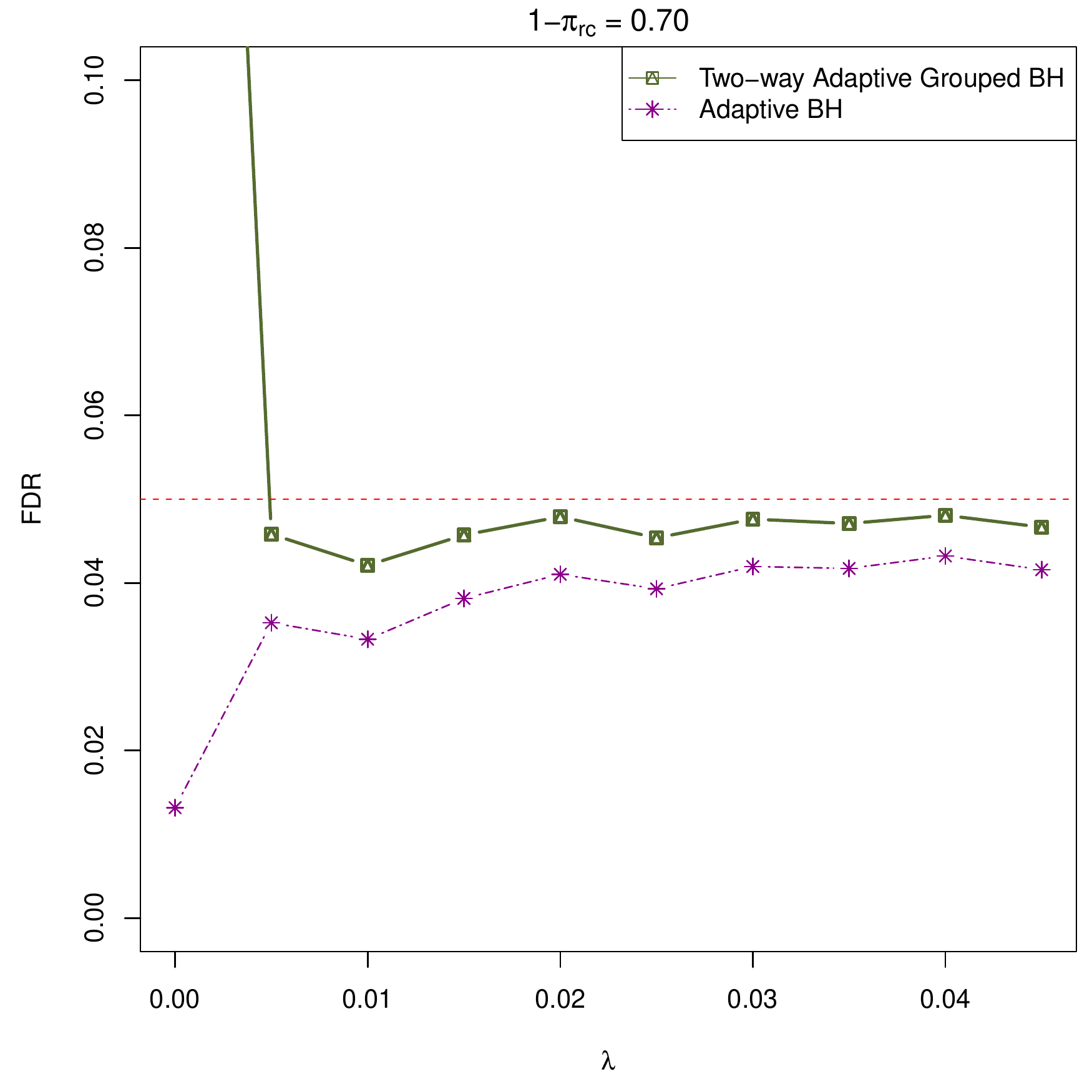}	
\includegraphics[width=0.22\textwidth]{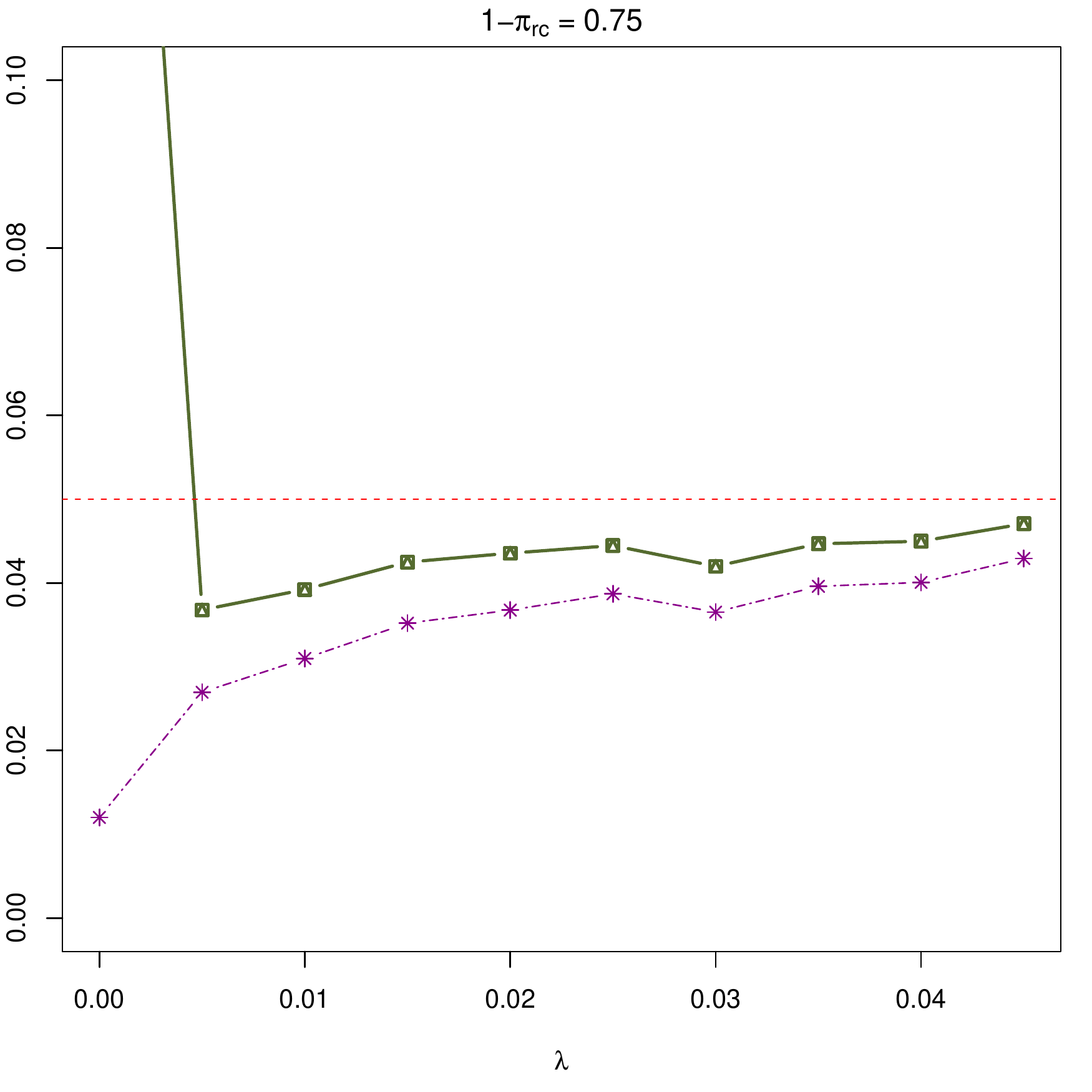}	
\includegraphics[width=0.22\textwidth]{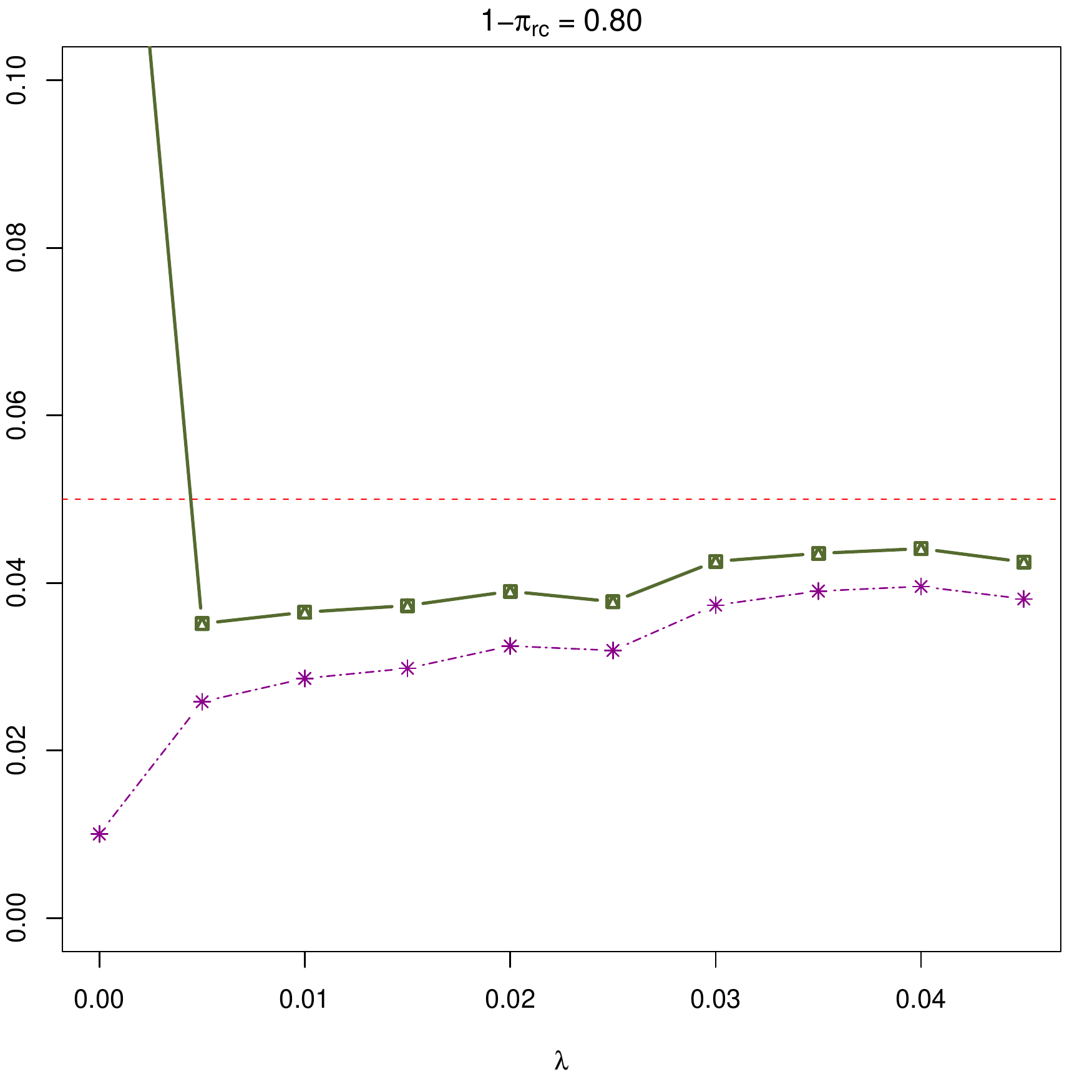}	
\includegraphics[width=0.22\textwidth]{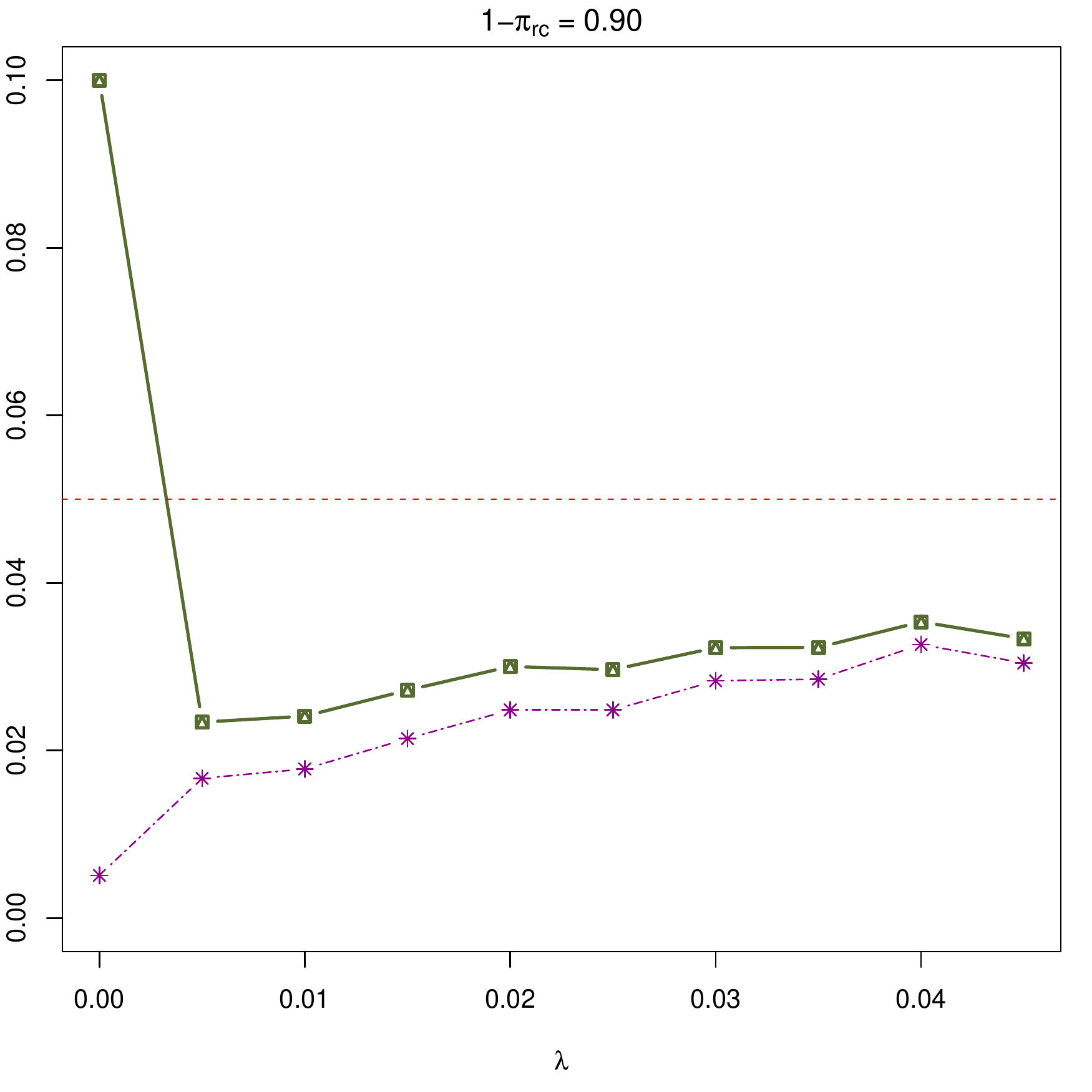}
\caption{FDR Comparisons}	
\label{fig8a}
\end{subfigure}\\

\begin{subfigure}[b]{\textwidth}
\centering
\includegraphics[width=0.22\textwidth]{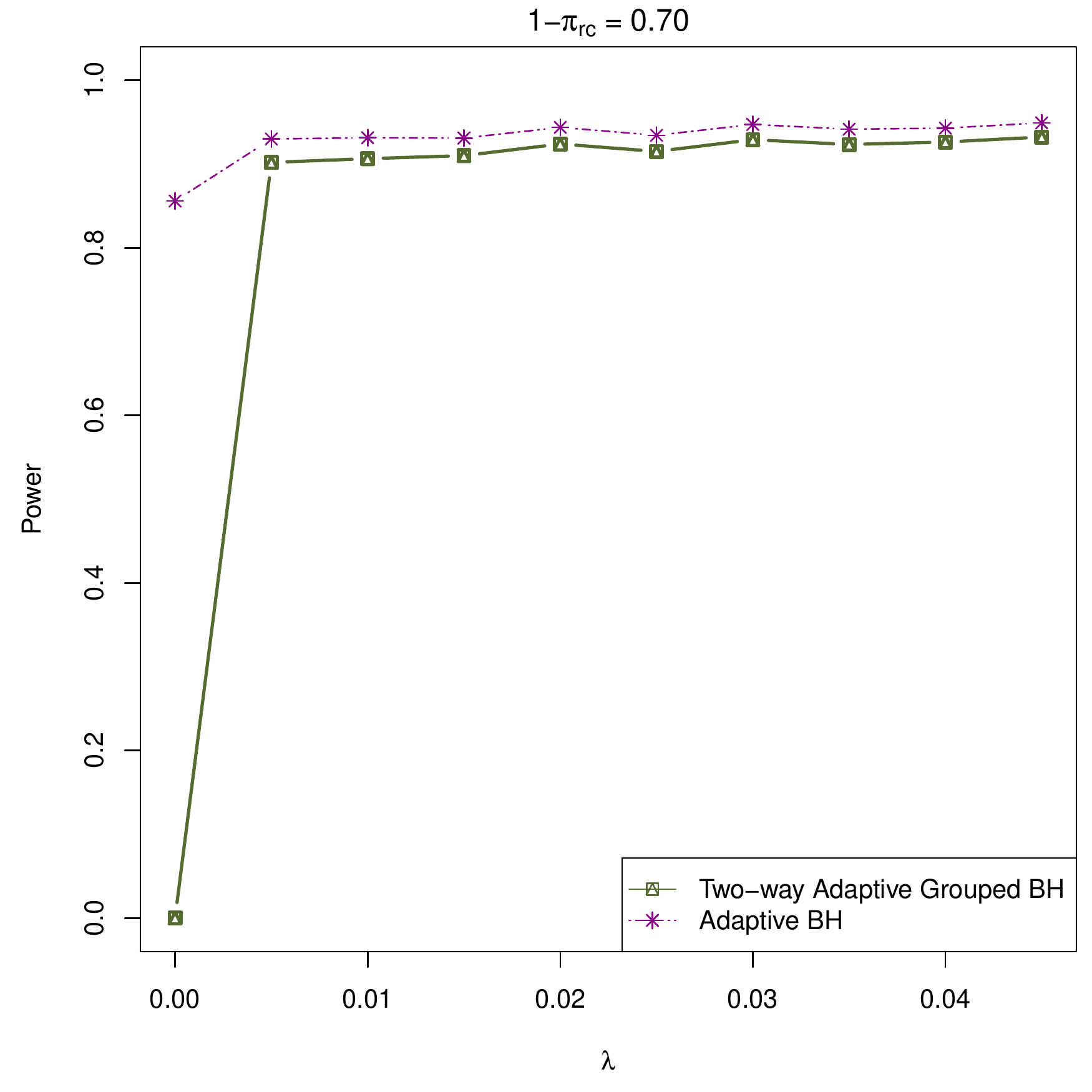}	
\includegraphics[width=0.22\textwidth]{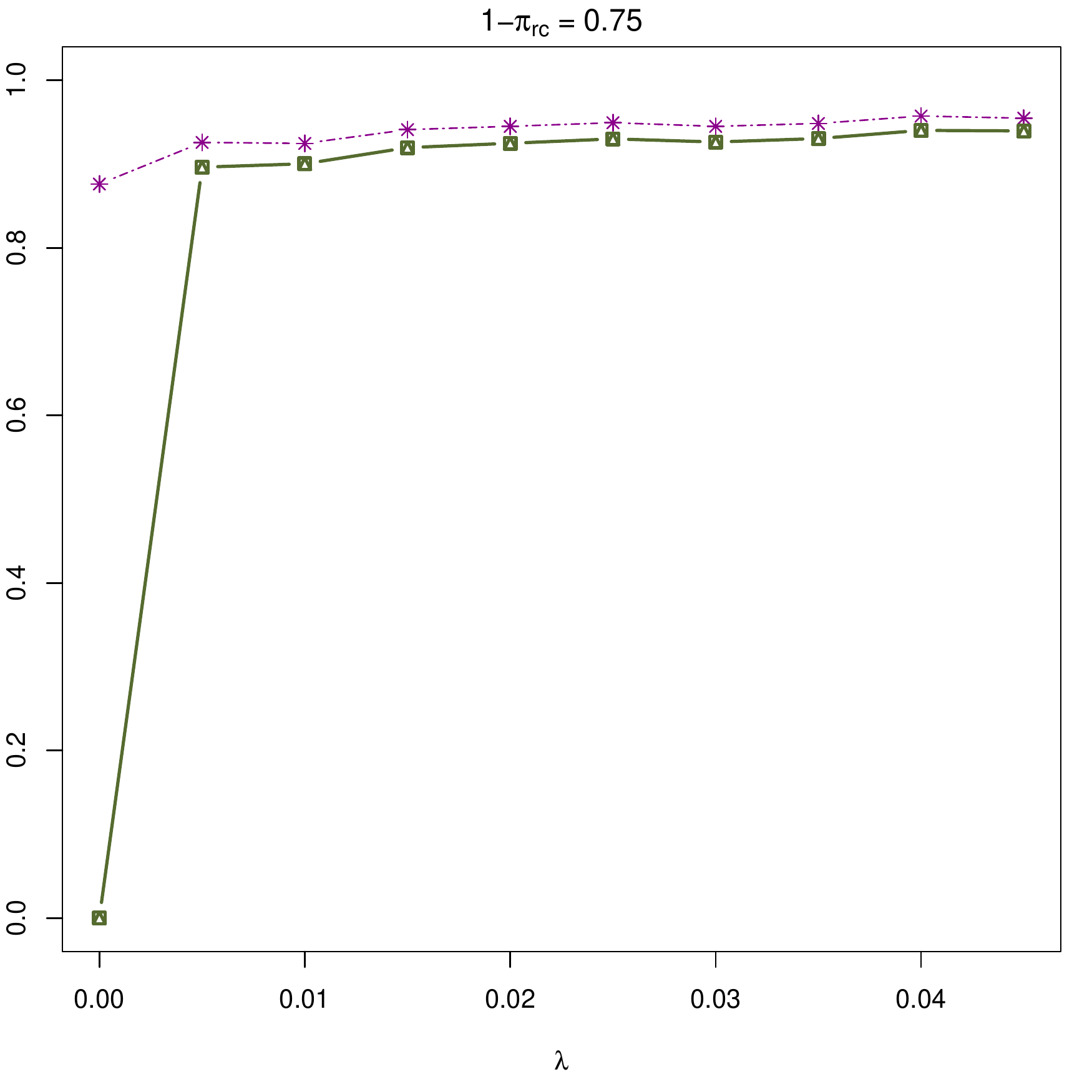}
\includegraphics[width=0.22\textwidth]{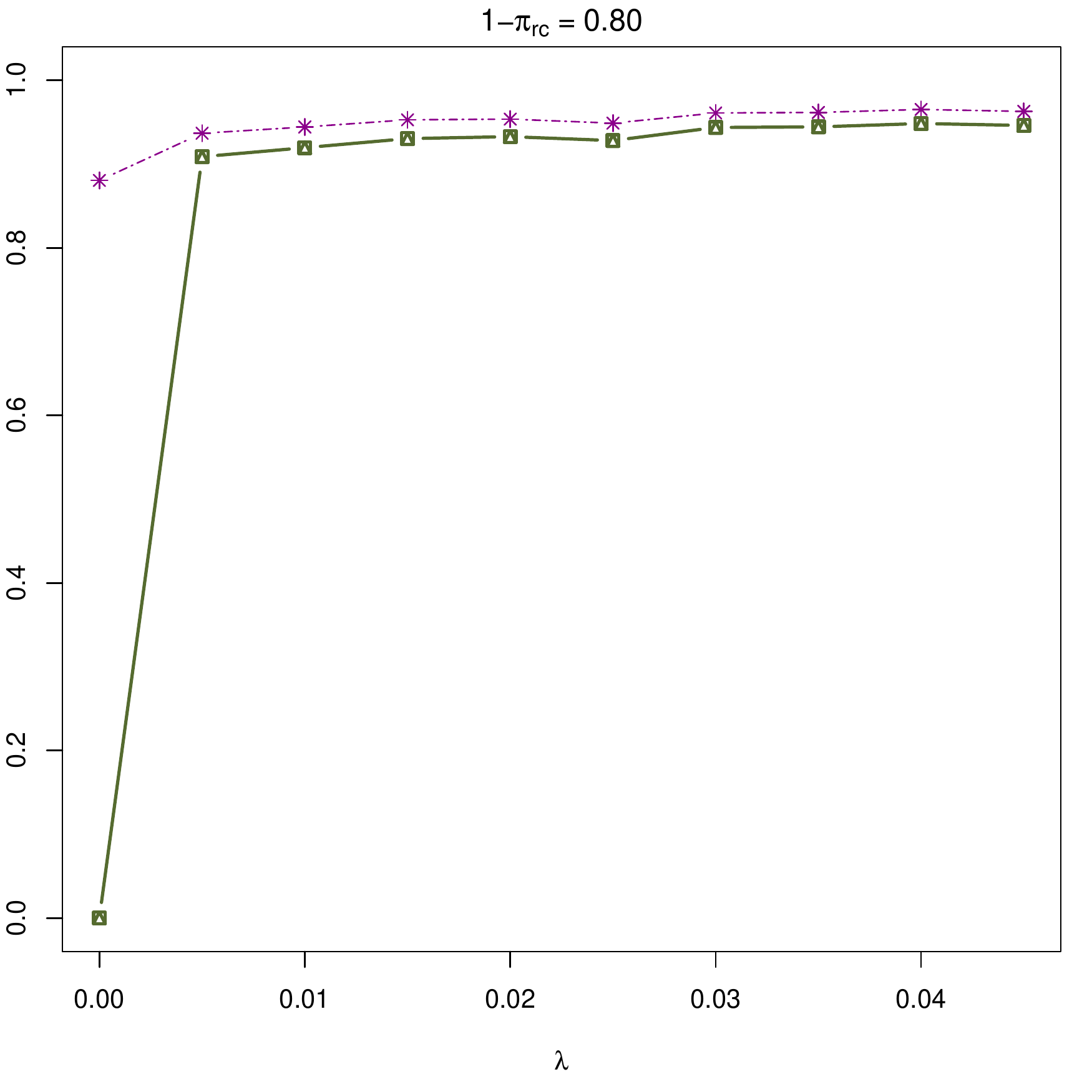}
\includegraphics[width=0.22\textwidth]{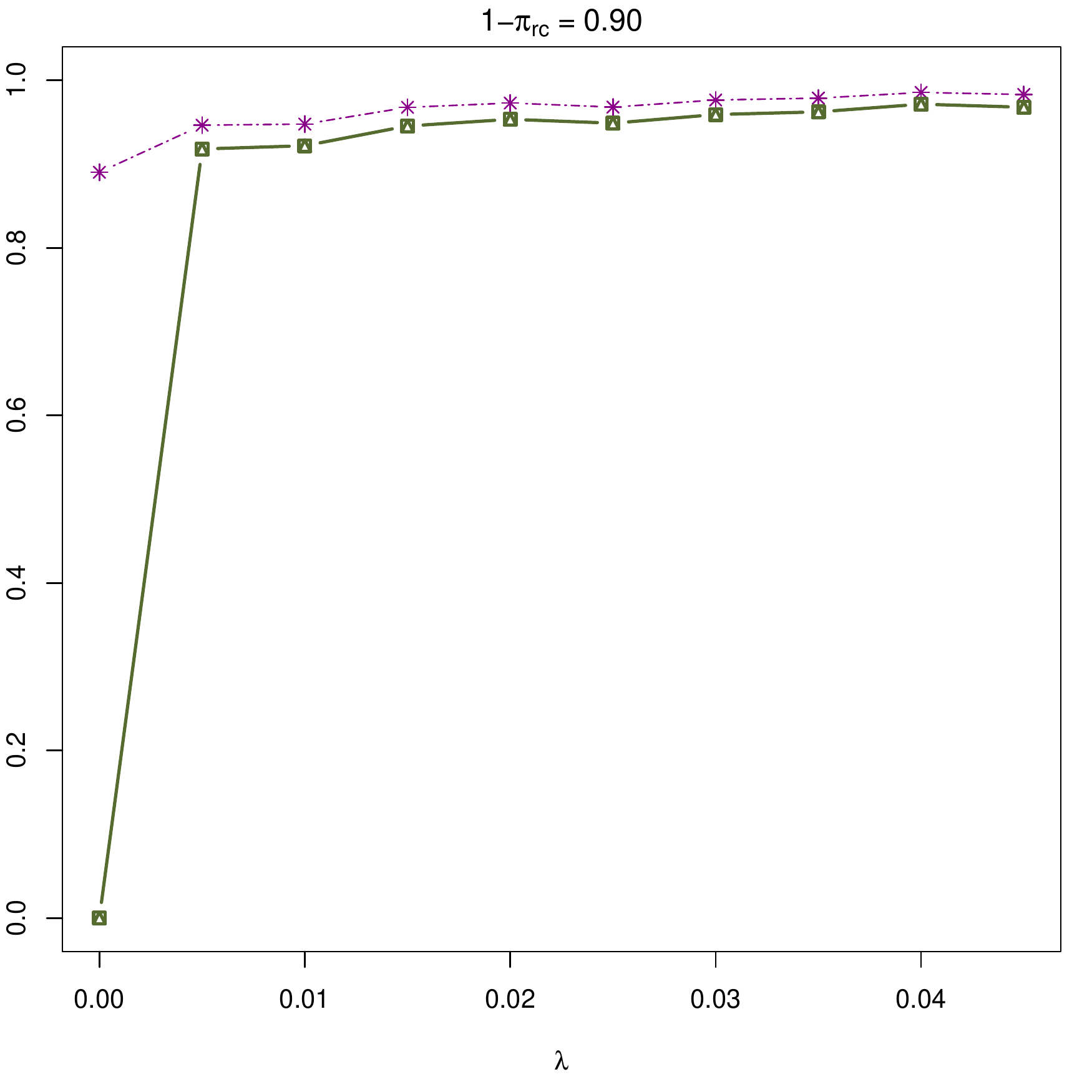}
\caption{Power Comparisons}
\label{fig8b}
\end{subfigure}

\caption{Comparison of the data-adaptive Two-way GBH$_1$ procedure for hypotheses with PRDS property, with the naive Adaptive BH procedure, for varying choices of $0< \lambda < 0.05 (=\alpha)$. Set of parameters used is $(m = 50, n = 100, \rho_r =0.3, \rho_c = 0.4, \pi_r=0, \pi_c=0, \pi_{rc})$}\label{fig8}
\end{figure}

Figures 7 and 8 display the findings of these investigation. Figure \ref{fig7}, which summarizes the results associated with answering question (i) (with $\lambda=0.5$), indicates that, though both these methods have comparable power when the signals are uniformly distributed in all rows and columns (i.e., when $\pi_r = \pi_c = 0$), our proposed method seems significantly more powerful in these situations. We chose $\rho_r = 0.3$ and $\rho_c = 0.4$ to answer question (ii), with  the related findings being summarized in Figure \ref{fig8}. It shows that the proposed Data-Adaptive Two-Way GBH$_1$ possibly can control FDR under PRDS when there is a high density of signals across all row and column groups; however, the choice of $\lambda$ seems crucial in such situations, and its values should be chosen in the range $(0, \alpha)$.

\subsection{Two-way Classified Hypotheses - Multiple hypotheses at each intersection} This section presents results from our simulation study carried out to investigate the performances of (i) Oracle Two-Way GBH$_{>1}$ procedure (Theorem \ref{th5}) against the single-group BH procedure in its oracle form and (ii) Data-Adaptive Two-Way GBH$_{>1}$ (Theorem \ref{th6}) against the naive Adaptive BH in terms of FDR control and power under normal distributional settings.

\subsubsection{Simulation Setting}

We considered the case where $n_{gh} = p$ for all $(g,h)$, so that our data generating process had to be designed to produce a random pair of third order tensors of dimension $m \times n \times p$, ($\mathbf{X}$, $\boldsymbol{\Theta}$), consisting of normally distributed test statistics and the Bernoulli hidden states of the corresponding hypotheses, respectively. The following were the steps in that process:

\begin{itemize}

\item [1.] Generate $\boldsymbol{\Theta}_{mnp}$ as an $m \times n \times p$ dimensional random tensor of i.i.d. Ber$(1-\pi_{rc})$, $\boldsymbol{\theta}_m$ as a random vector of $m$ i.i.d. Ber$(1-\pi_{r})$, and $\boldsymbol{\theta}_n$ as a random vector of $n$ i.i.d. Ber$(1-\pi_c)$. \
\item [2.] Obtain
\begin{align*}
& \boldsymbol{\theta} = \boldsymbol{\theta}_{mnp}\, \star \, (\boldsymbol{\theta}_m \circ \boldsymbol{1}_n \circ \boldsymbol{1}_p )\, \star \, (\boldsymbol{1}_m\circ \boldsymbol{\theta}_n \circ \boldsymbol{1}_p)
\end{align*}
(with $\boldsymbol{a}\circ \boldsymbol{b}$ denoting the outer product between the vectors $\boldsymbol{a}$ and $\boldsymbol{b}$);
\item [3.] Given $\boldsymbol{\Theta}$, generate $\mathbf{X}$ as an $m \times n \times p$ dimensional tensor having a tensor normal distribution given below using its vectored form:
\begin{align*}
\mathbf{vec}({\mathbf{X}}) \sim \text{N}_{mn p}(\mathbf{vec}(\mu \boldsymbol{\theta}), \Sigma_p \otimes \Sigma_c\otimes \Sigma_r),
\end{align*} where $\Sigma_r = (1 -\rho_r)I_n + \rho_r\boldsymbol{1}_n\boldsymbol{1}_n^T, \; \rho_{r} \in [0, 1)$, $\Sigma_c = (1 -\rho_c)I_m + \rho_c\boldsymbol{1}_m\boldsymbol{1}_m^T, \; \rho_{c} \in [0,1)$, and $\Sigma_p = (1 -\rho_p)I_p + \rho_p\boldsymbol{1}_p\boldsymbol{1}_p^T, \; \rho_{p} \in [0,1)$.
\end {itemize}

Let $X_{ghk}$ be the $k$th layer test statistic in the $(g,h)$ cell. They can have different types of positive dependence structures determined through appropriate choices of the correlation coefficients $\rho_r$ $\rho_c$ and $\rho_p$. If there is independence along any dimension of the tensor $\mathbf{X}$, the corresponding correlation coefficient is set to $0$. We considered the problem of testing $H_{gh}: \text{E}(X_{ghk}) = 0$ against $K_{ghk}: \text{E}(X_{ghk}) > 0$, simultaneously for all $g=1, \ldots, m, h=1, \ldots, n, k=1, \ldots, p$. So, the next two steps in our simulation study were the following:

\begin {itemize}
\item [4.] Apply each of the aforementioned procedures  at FDR level $\alpha=0.05$, and note down each of the the proportions of false rejections among all rejections and correct rejections among all false nulls. 

\item [5.] Repeat Steps 1-4 200 times to simulate the values of FDR and power for each procedure by averaging out the corresponding proportions noted in Step 4.

\end{itemize}

\subsubsection{Simulation Findings} We considered fixed $m = 50$, $n = 100$, $p=10$, and set $\mu$ at $0$ for true null hypotheses and at $=3$ for all true signals. The rest of the parameters are regulated to generate different situations and analyze the performance of our method in those settings. The combination of parameters ($\pi_r$, $\pi_c$) chosen are similar to those in the case of two-way classification with one hypothesis
per cell. For each combination of values for ($\pi_{r}$, $\pi_{c}$), $1- \pi_{rc}$ was varied between $0$ and $1$. Signals are sparse for smaller values of $1-\pi_{rc}$ and the density increases with its value.

\begin{figure}
	\centering
	
	\begin{subfigure}[b]{\textwidth}
		\centering
		\includegraphics[width=0.22\textwidth]{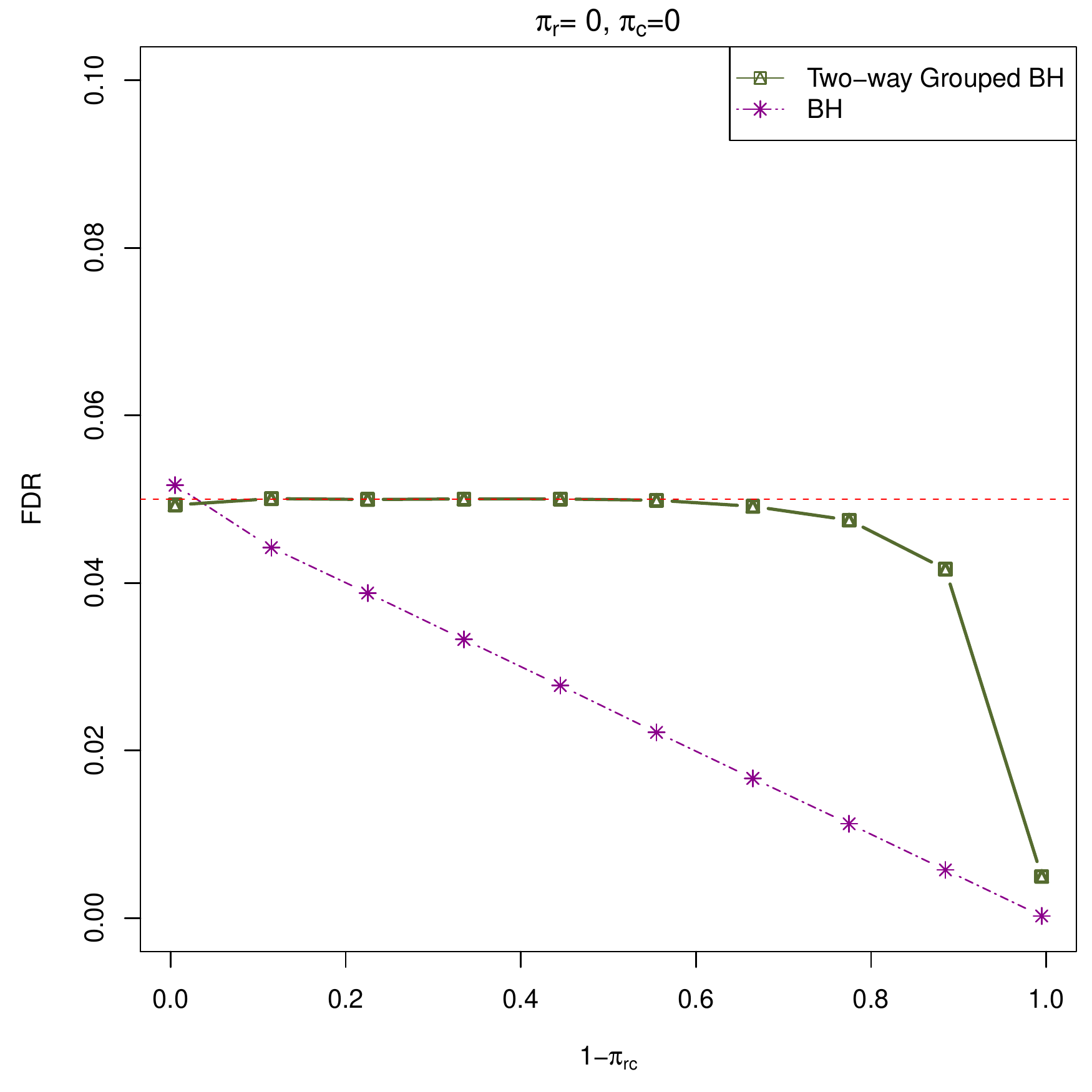}	
		\includegraphics[width=0.22\textwidth]{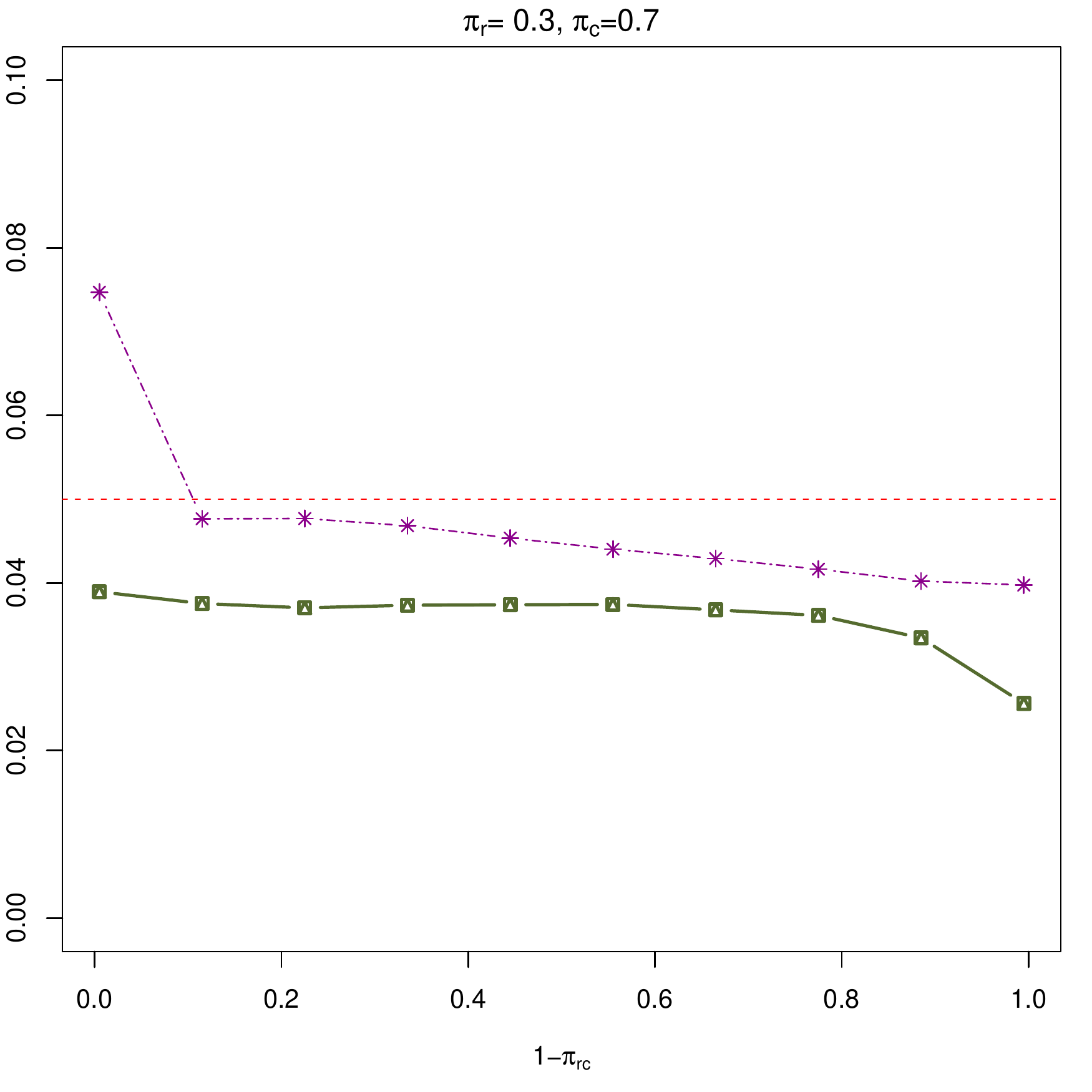}	
		\includegraphics[width=0.22\textwidth]{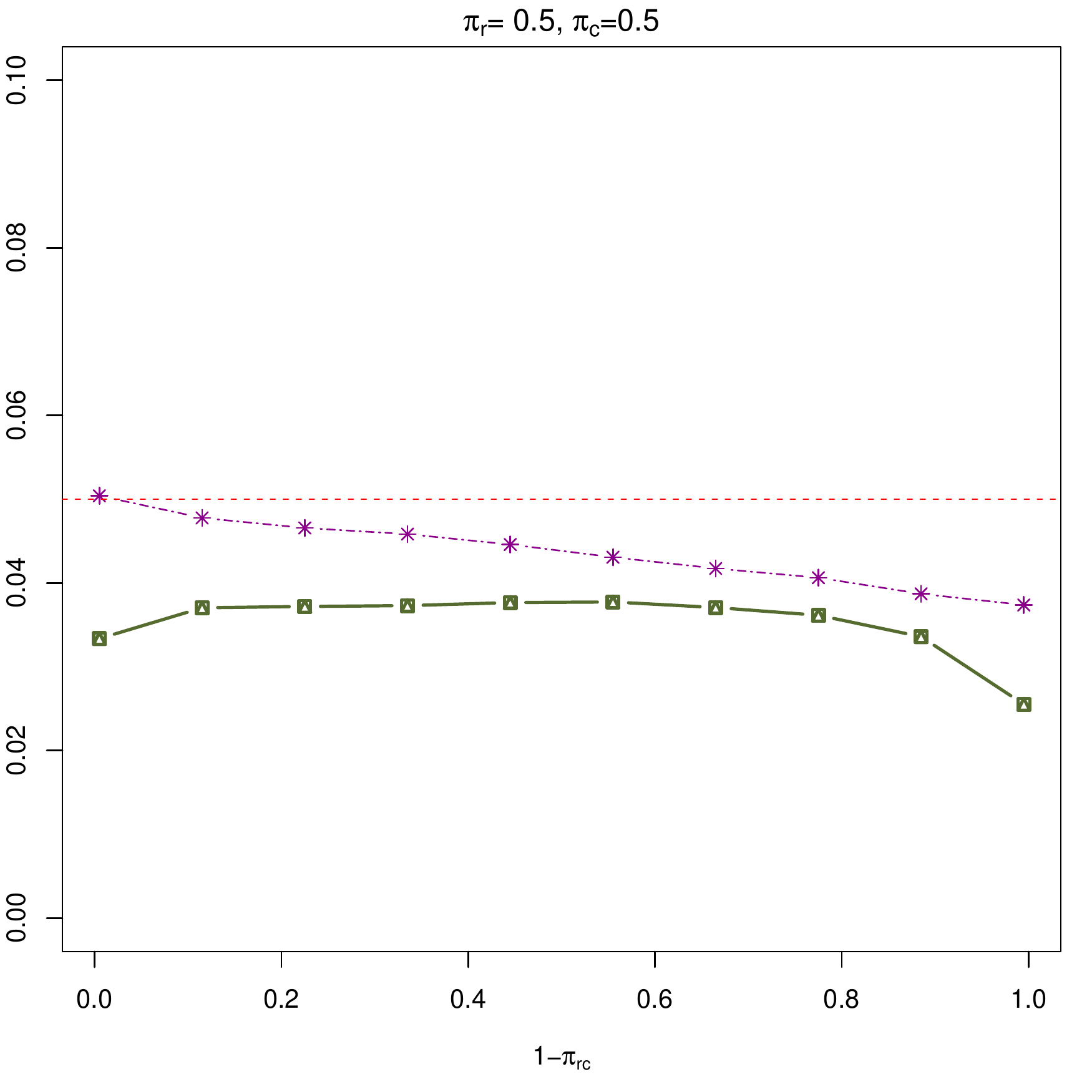}	
		\includegraphics[width=0.22\textwidth]{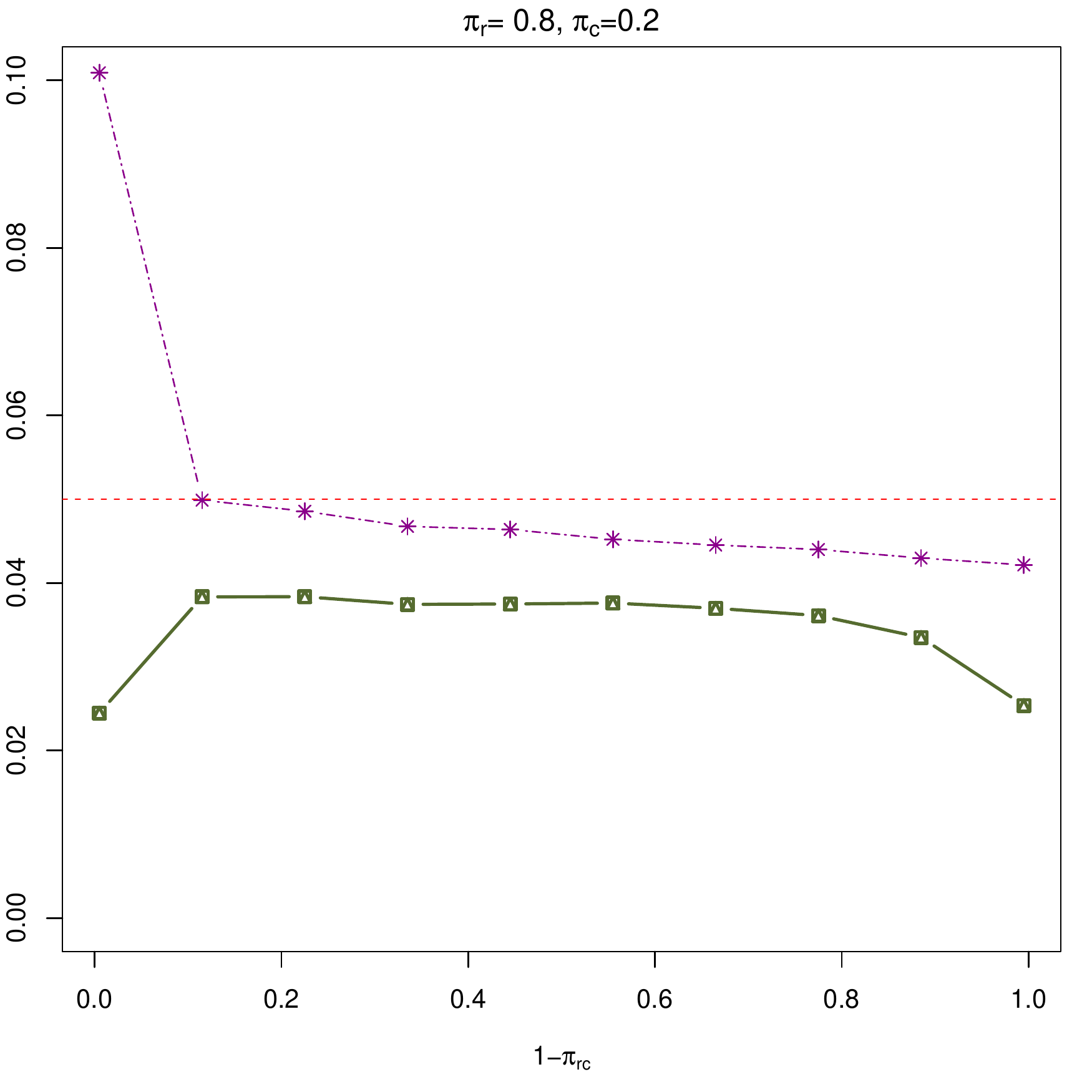}
		\caption{FDR Comparisons}	
		\label{fig9a}
	\end{subfigure}\\
	
	\begin{subfigure}[b]{\textwidth}
		\centering
		\includegraphics[width=0.22\textwidth]{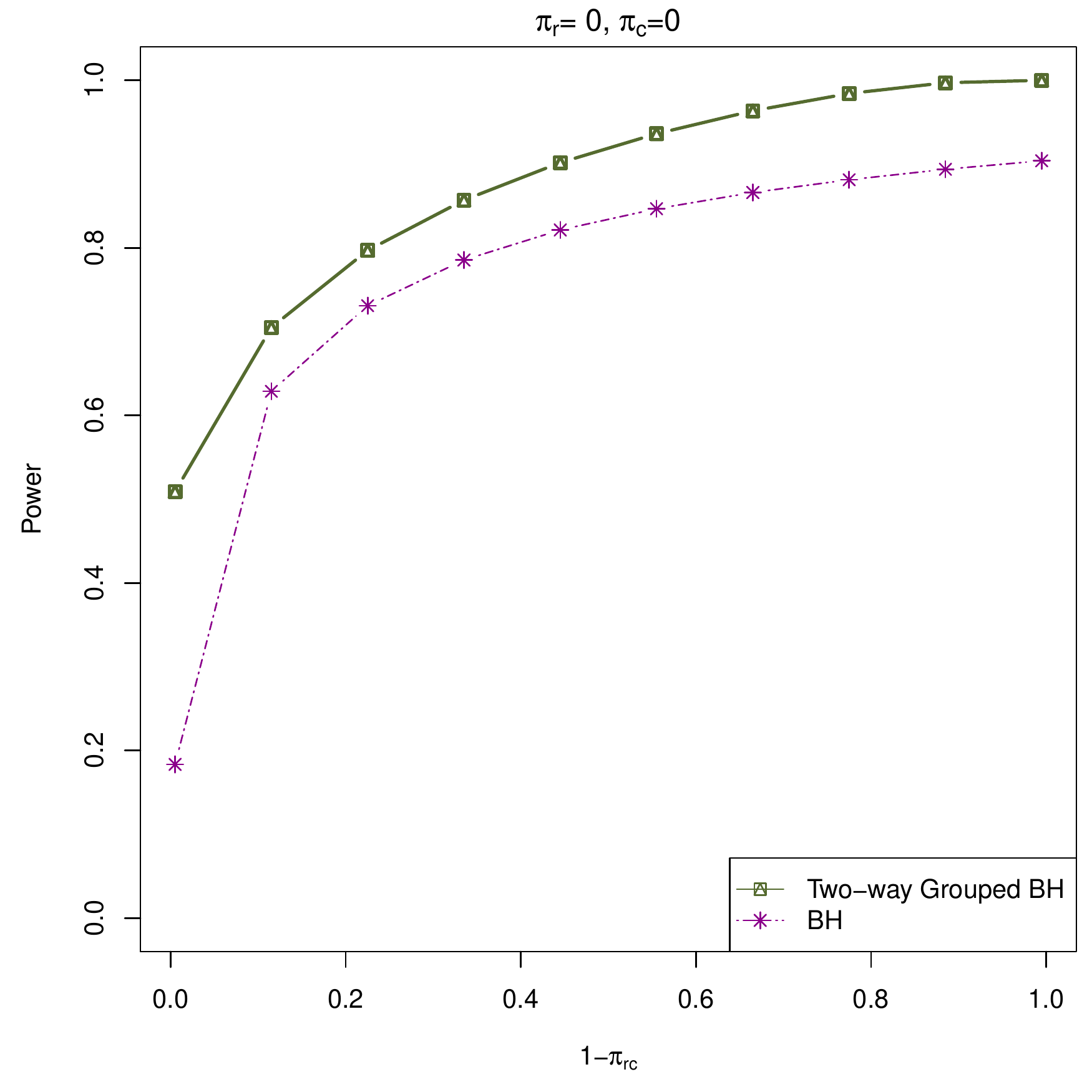}	
		\includegraphics[width=0.22\textwidth]{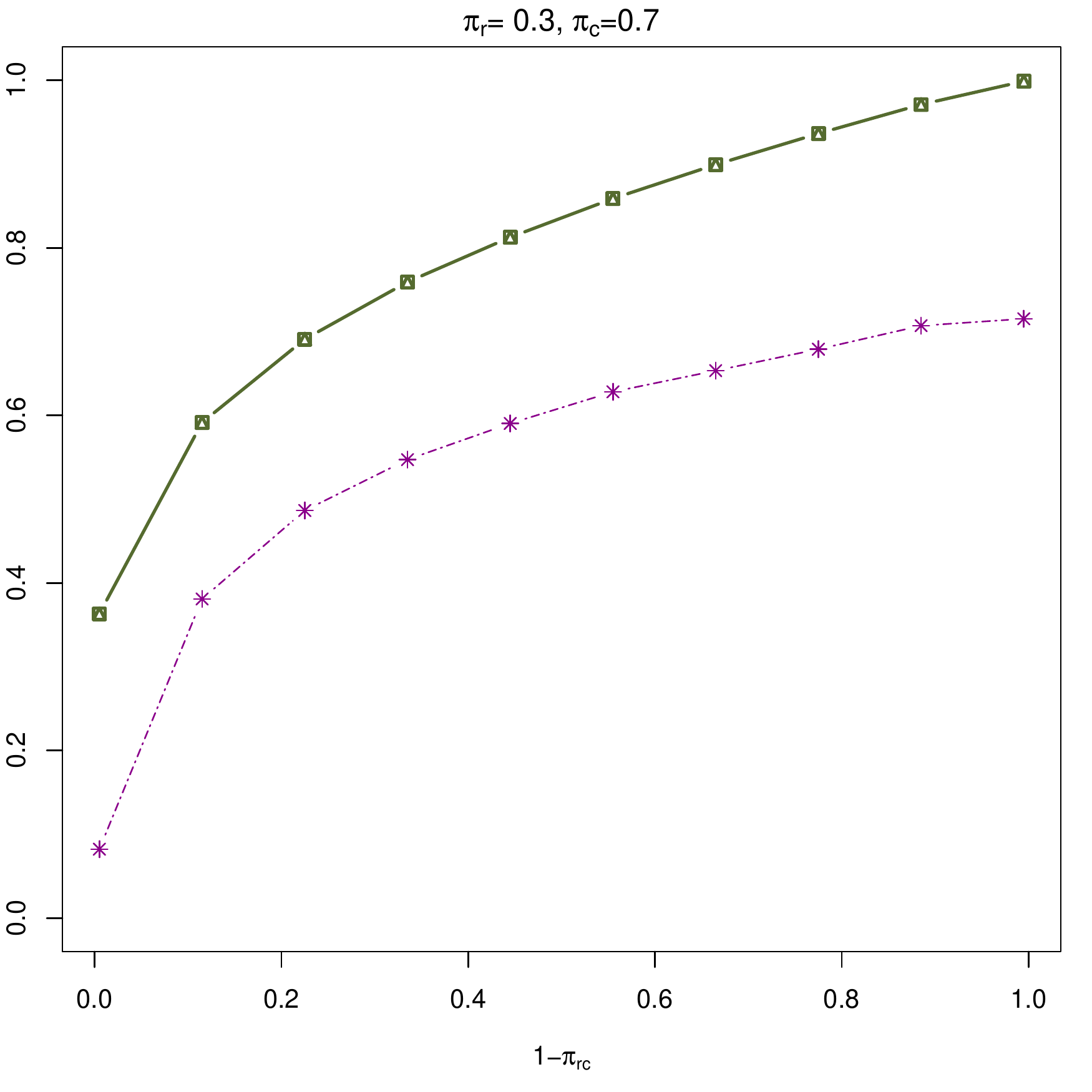}
		\includegraphics[width=0.22\textwidth]{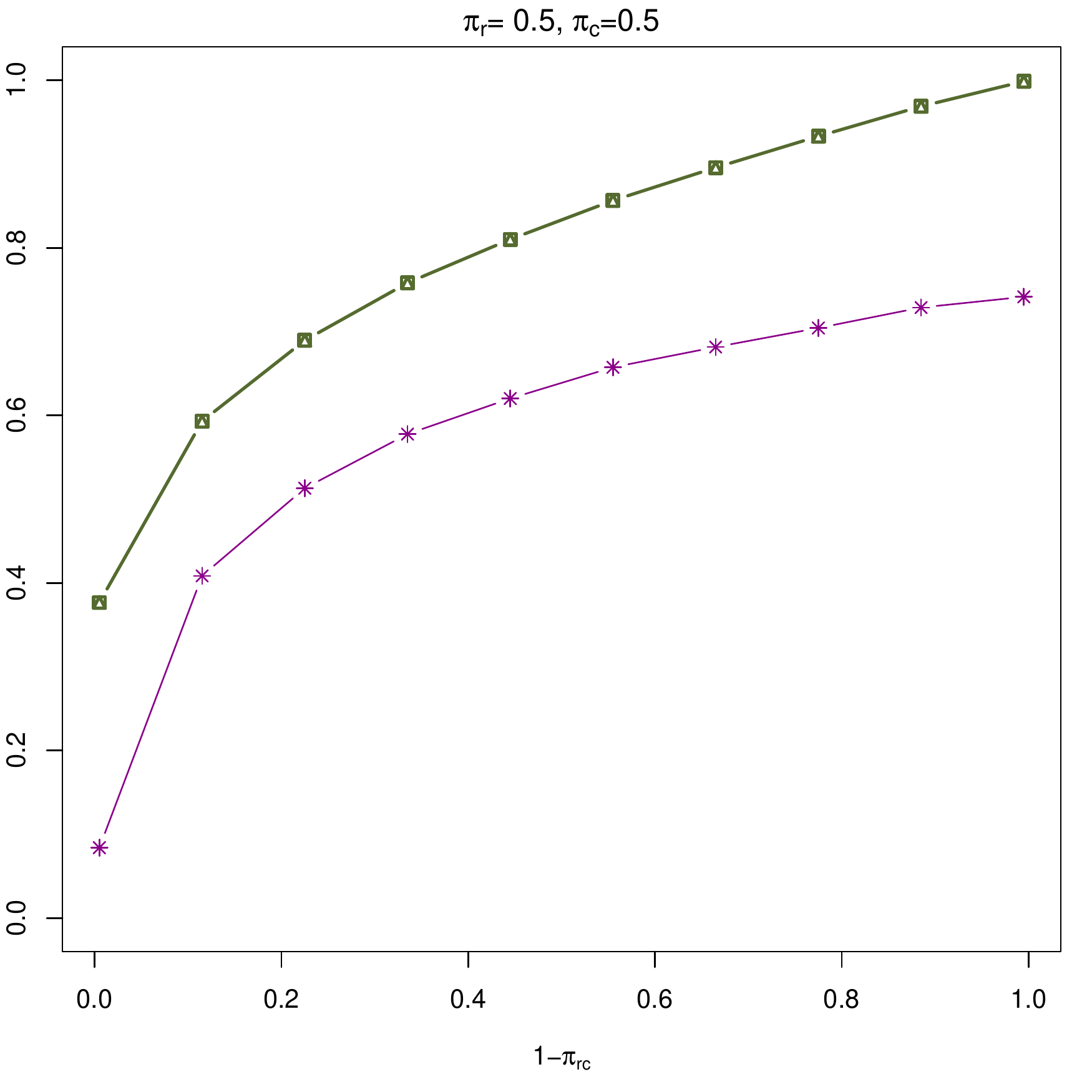}
		\includegraphics[width=0.22\textwidth]{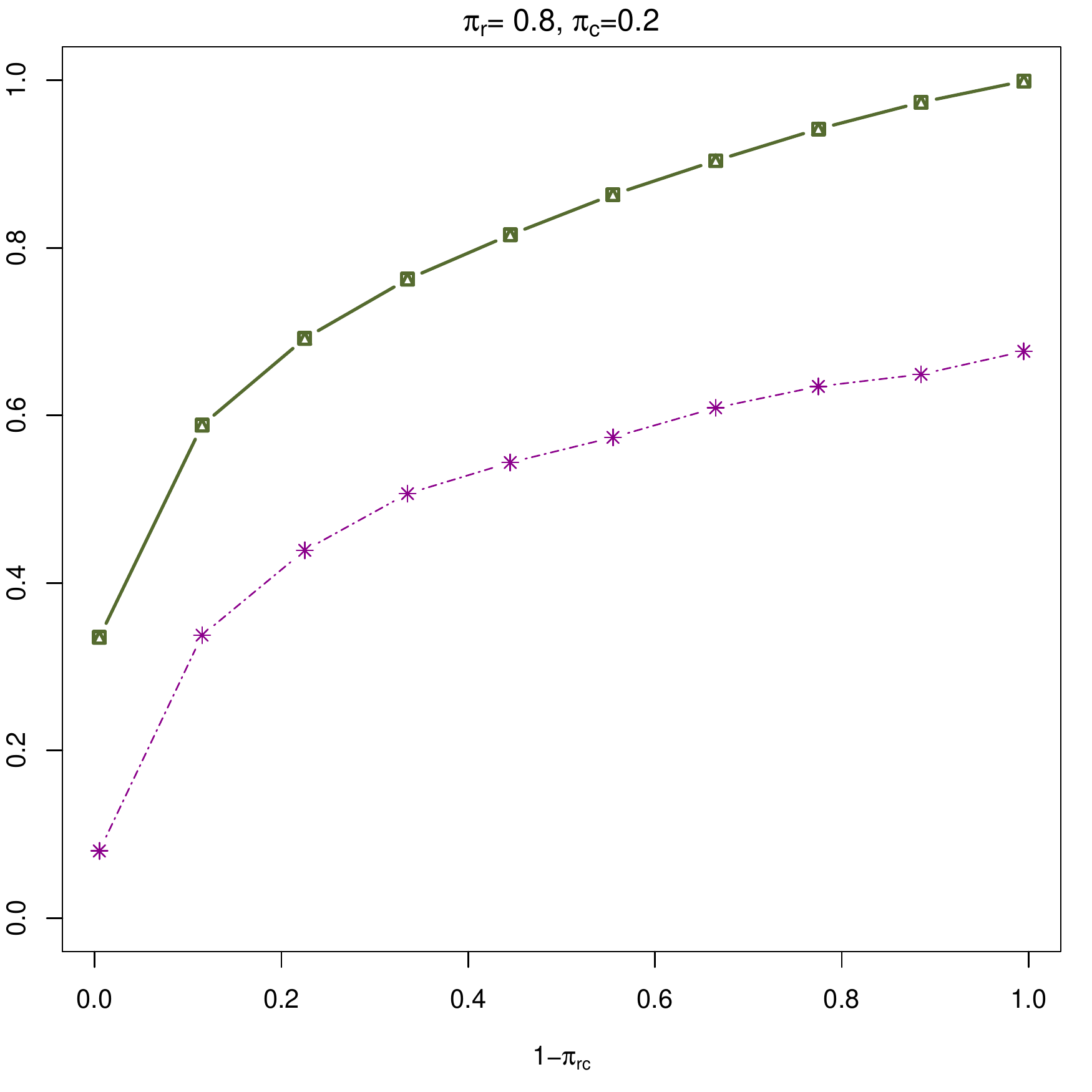}
		\caption{Power Comparisons}
		\label{fig9b}
	\end{subfigure}
	
	\caption{Comparison of the oracle Two-way GBH$_{>1}$ procedure with the BH method, applied to independent hypotheses. Set of parameters used is $(m = 50, n = 100, p=10, \rho_r =0, \rho_c = 0, \rho_p =0,\pi_r, \pi_c, \pi_{rc})$}\label{fig9}
\end{figure}

\begin{figure}
	\centering
	
	\begin{subfigure}[b]{\textwidth}
		\centering
		\includegraphics[width=0.22\textwidth]{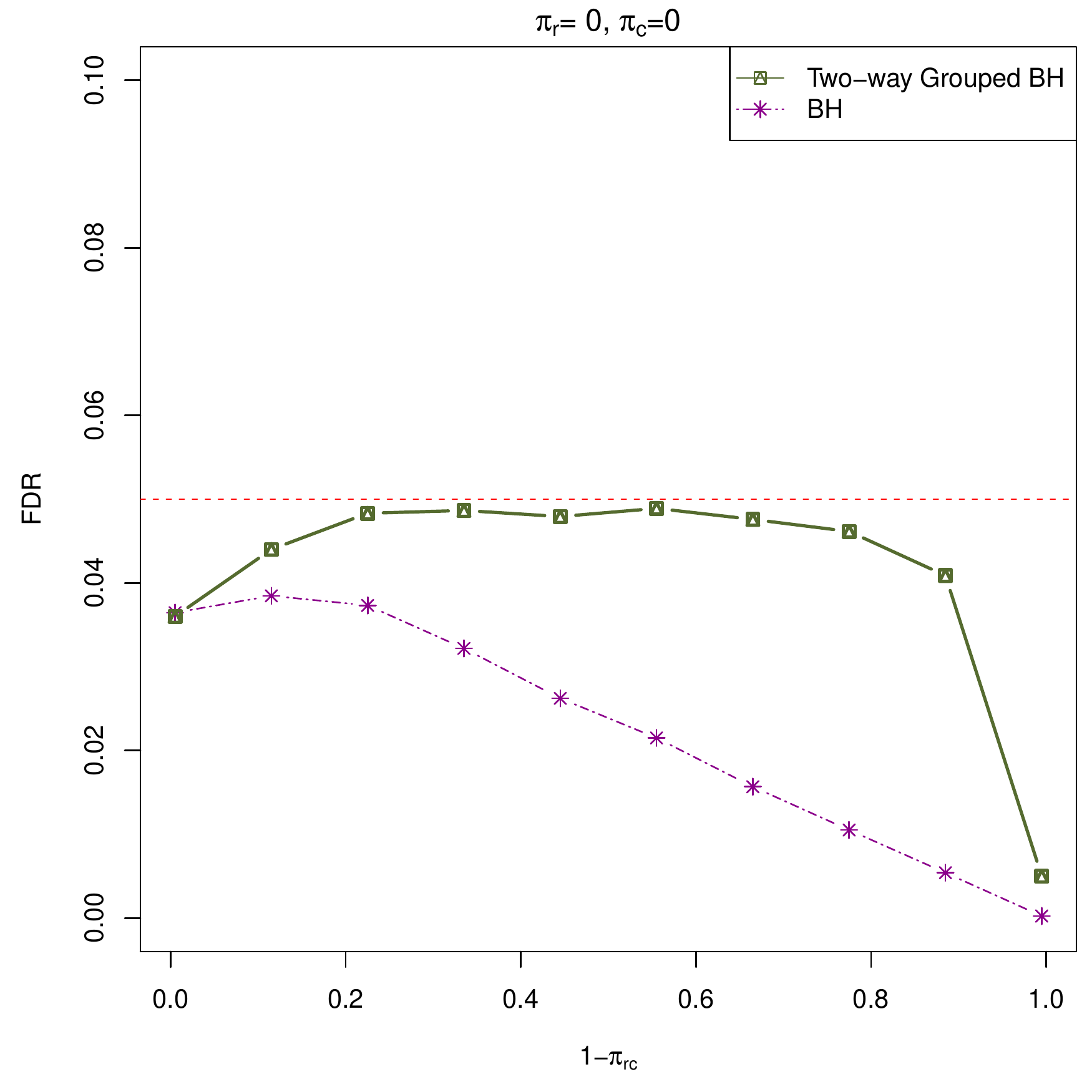}	
		\includegraphics[width=0.22\textwidth]{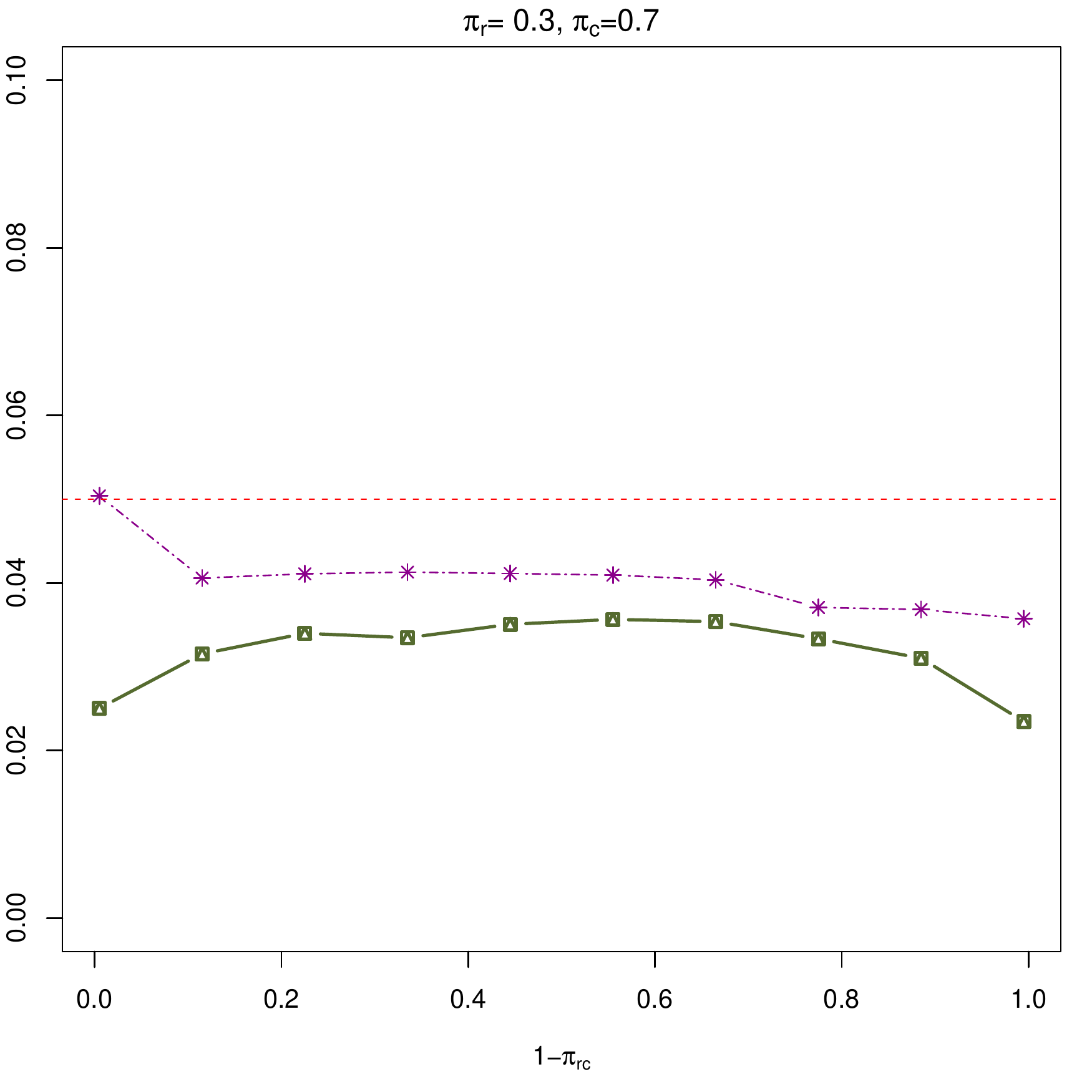}	
		\includegraphics[width=0.22\textwidth]{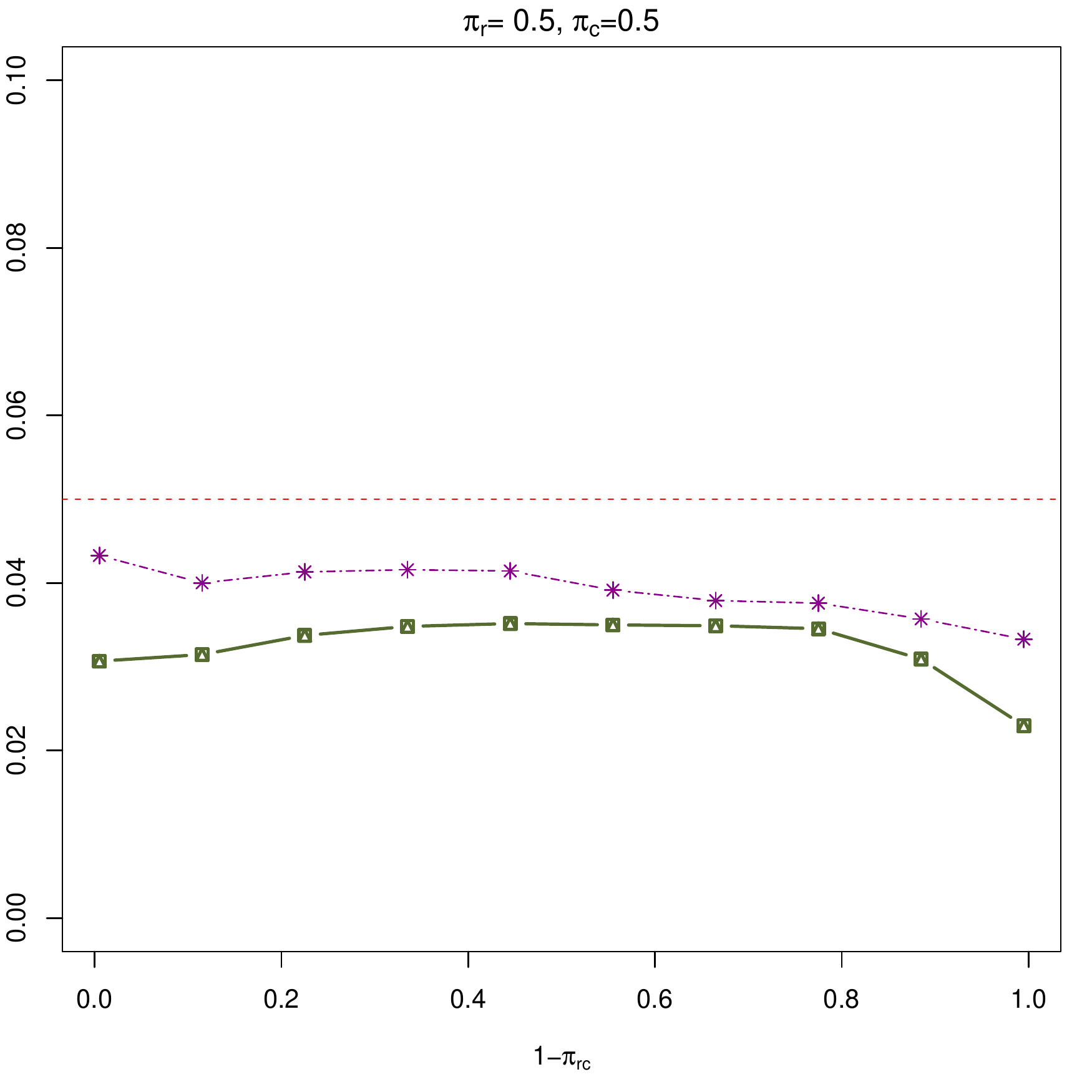}	
		\includegraphics[width=0.22\textwidth]{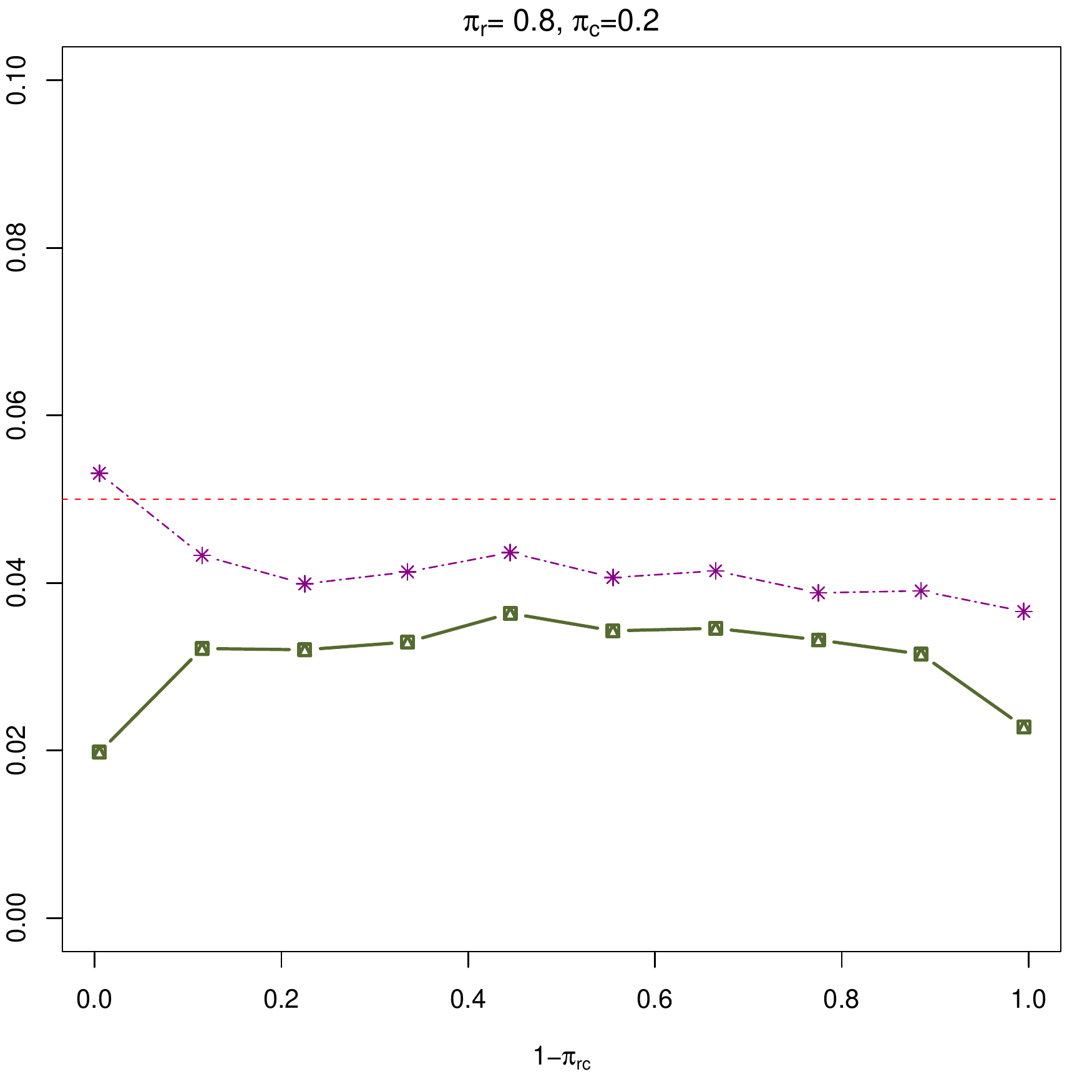}
		\caption{FDR Comparisons}	
		\label{fig10a}
	\end{subfigure}\\
	
	\begin{subfigure}[b]{\textwidth}
		\centering
		\includegraphics[width=0.22\textwidth]{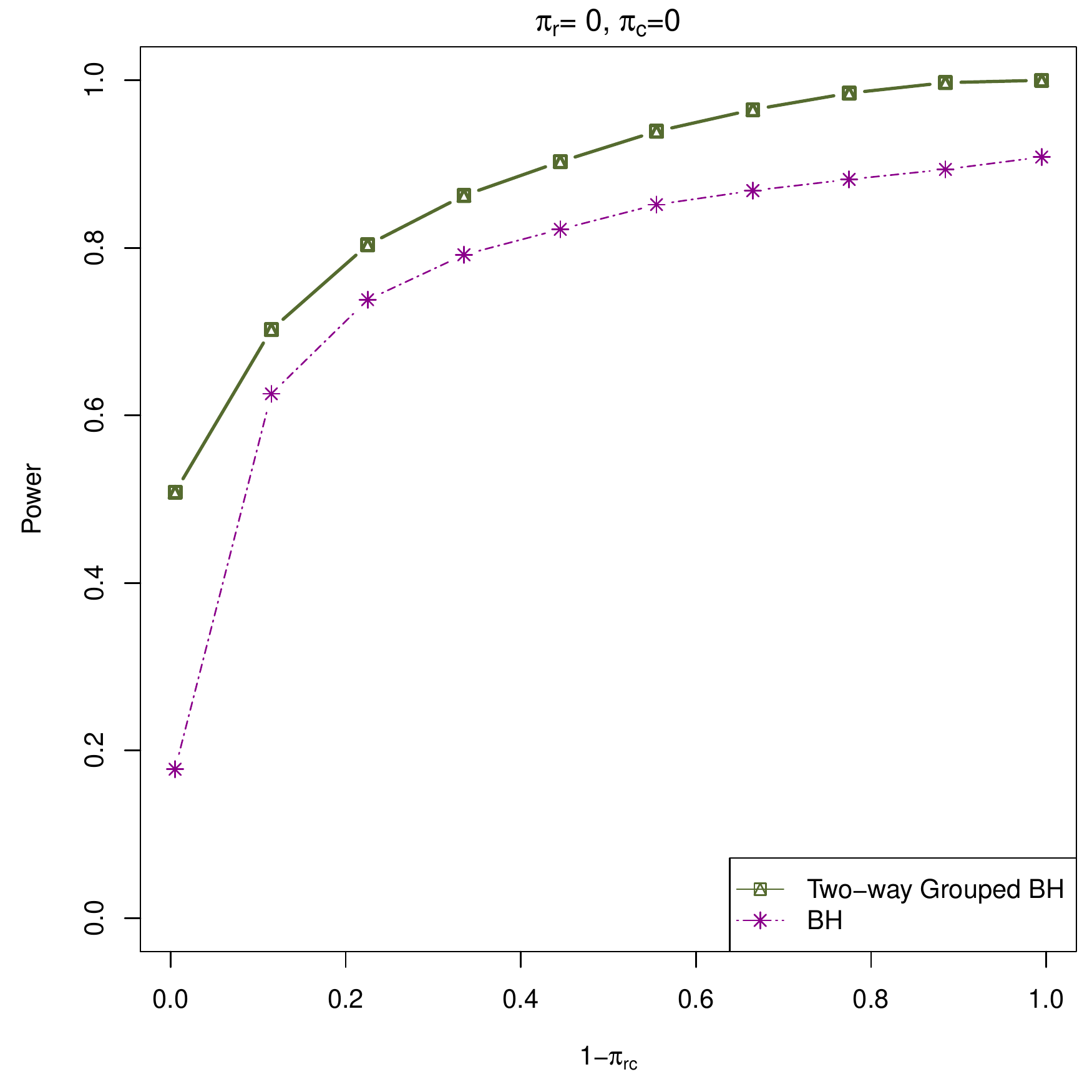}	
		\includegraphics[width=0.22\textwidth]{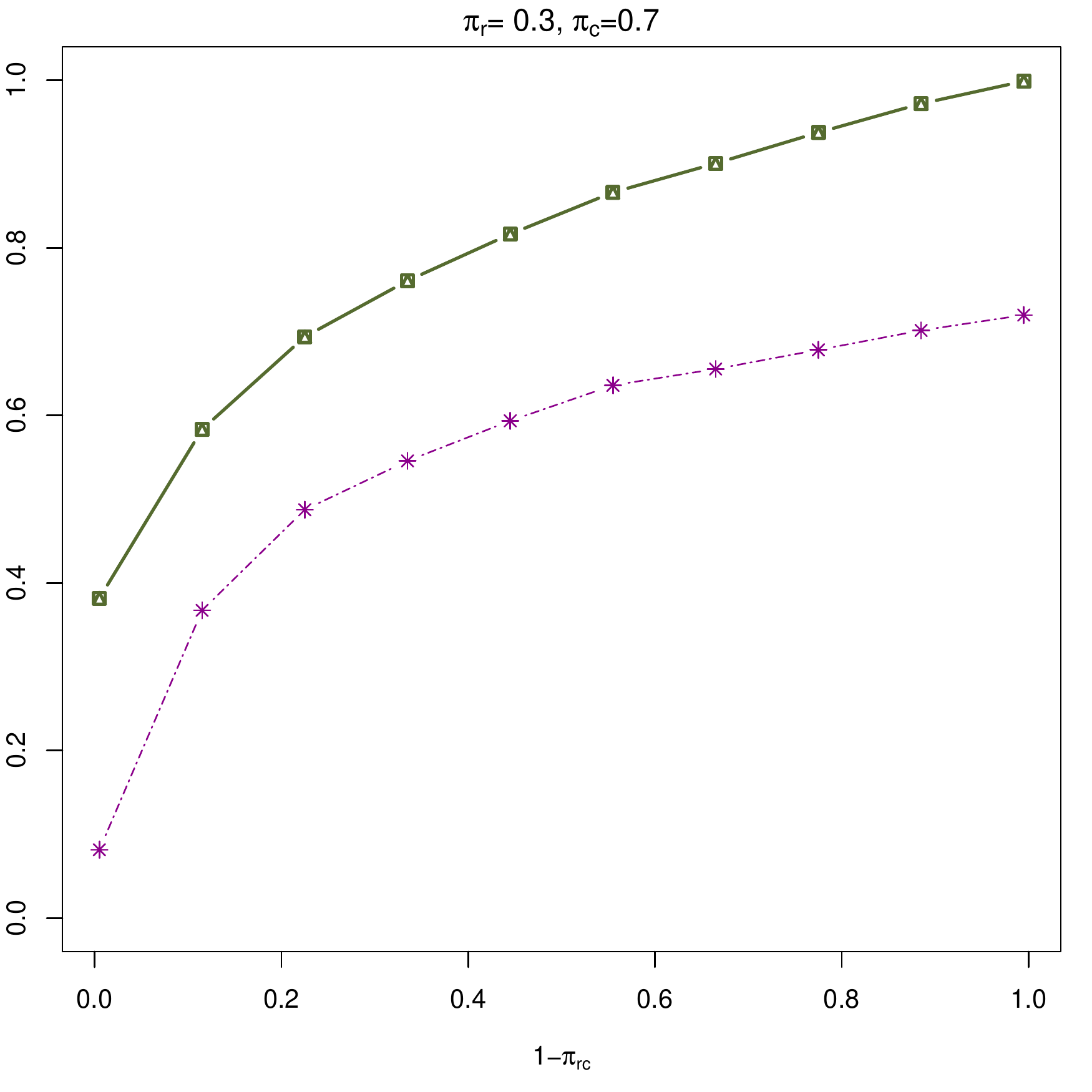}
		\includegraphics[width=0.22\textwidth]{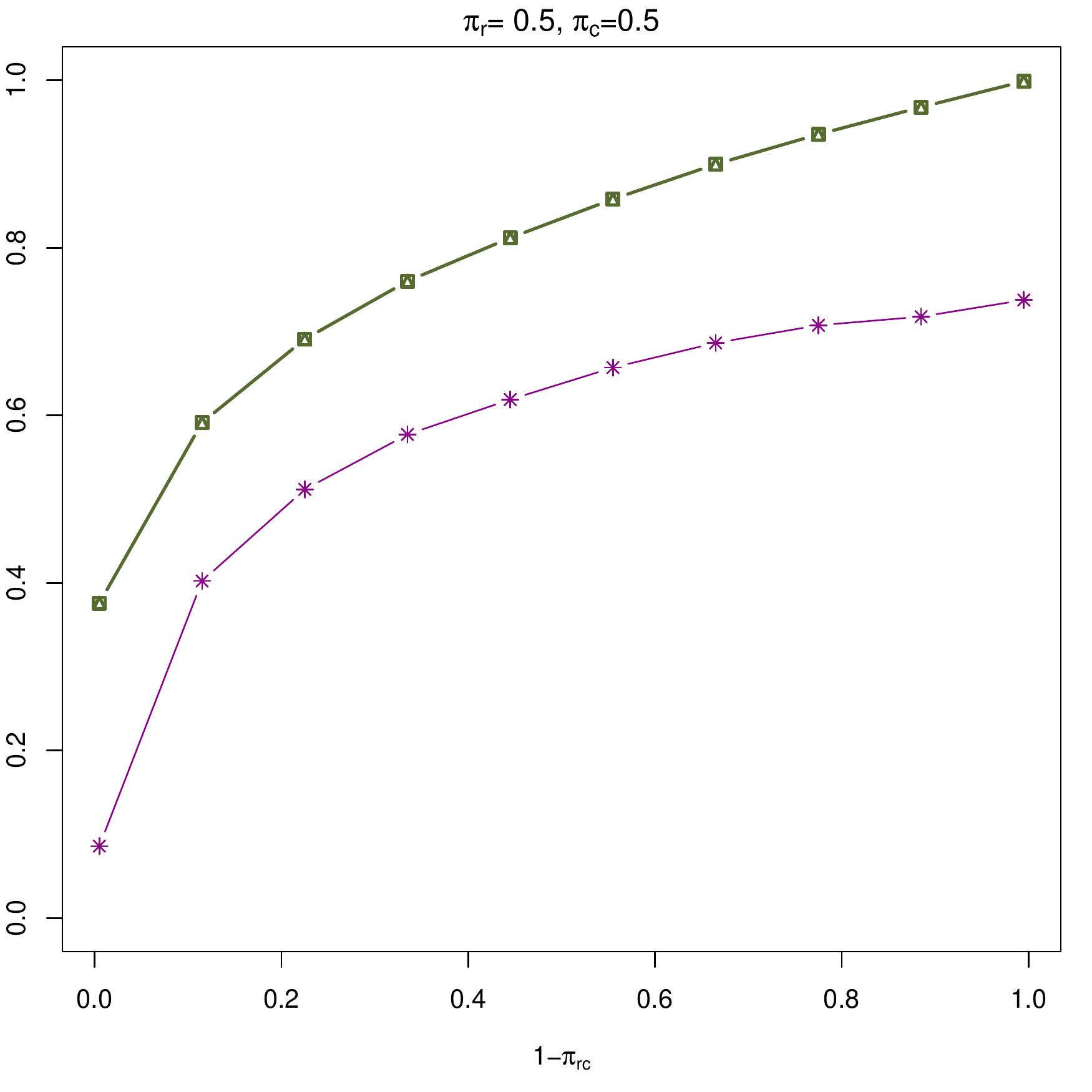}
		\includegraphics[width=0.22\textwidth]{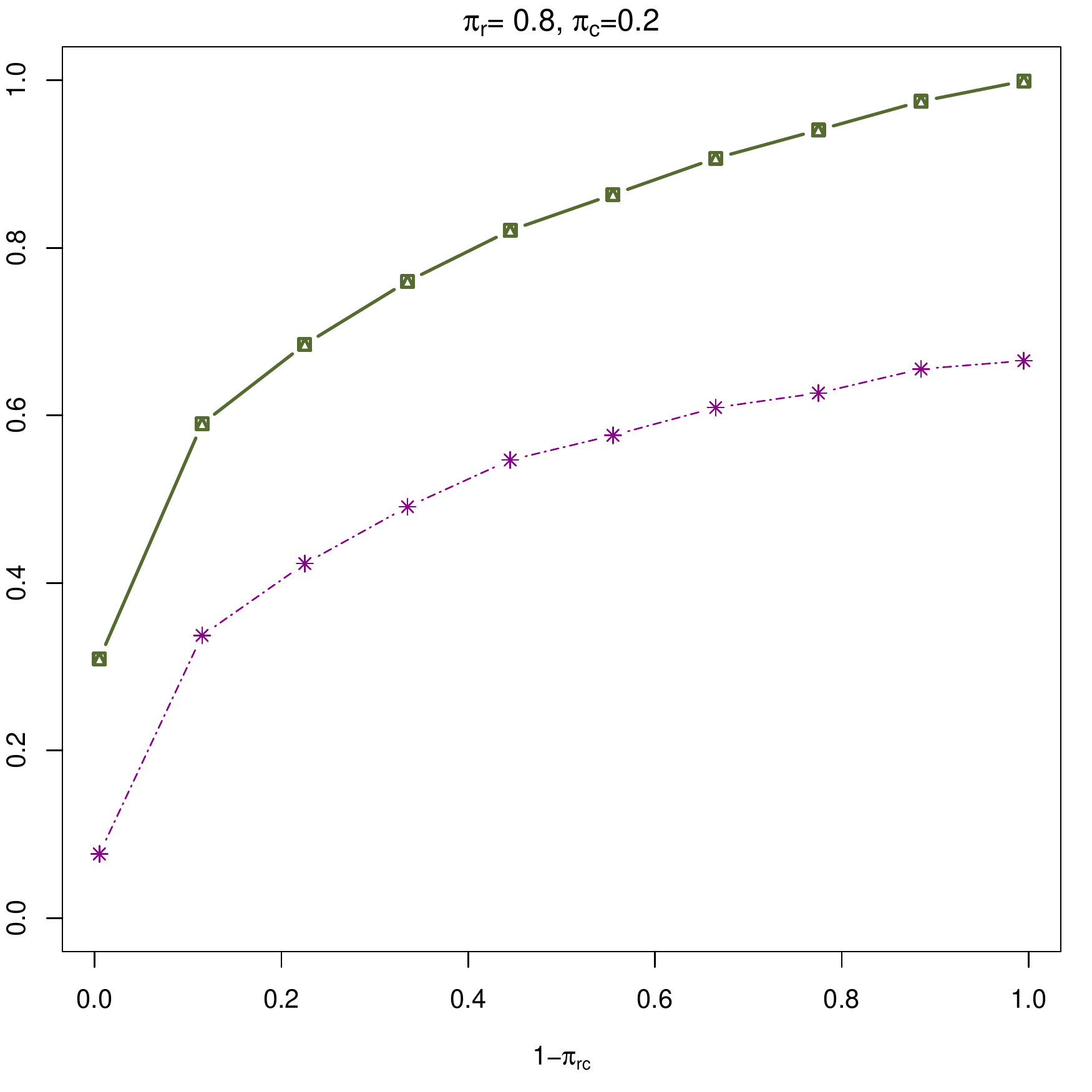}
		\caption{Power Comparisons}
		\label{fig10b}
	\end{subfigure}
	
	\caption{Comparison of the oracle Two-way GBH$_{>1}$ procedure with the BH method, when the hypotheses satisfy PRDS condition. Set of parameters used is $(m = 50, n = 100, p=10, \rho_r =0.3, \rho_c = 0.4, \rho_p =0.2,\pi_r, \pi_c, \pi_{rc})$}\label{fig10}
\end{figure}

\vskip 10 pt
\noindent {\it Comparison of Oracle Procedures:} We wanted to make two types of investigation for Oracle Two-Way GBH$_{>1}$ in Theorem \ref{th5} under both independence and PRDS condition against the usual single group BH - how does it perform in terms of FDR control and power? The findings of these are displayed in Figures \ref{fig9} (for the independent case) and \ref{fig10} (for the PRDS case corresponding to $\rho_r = 0.3$, $\rho_c = 0.4$, and $\rho_p = 0.2$). In either case, the proposed method controls FDR, as expected, and seems to be powerful than the BH.

\begin{figure}
	\centering
	
	\begin{subfigure}[b]{\textwidth}
		\centering
		\includegraphics[width=0.22\textwidth]{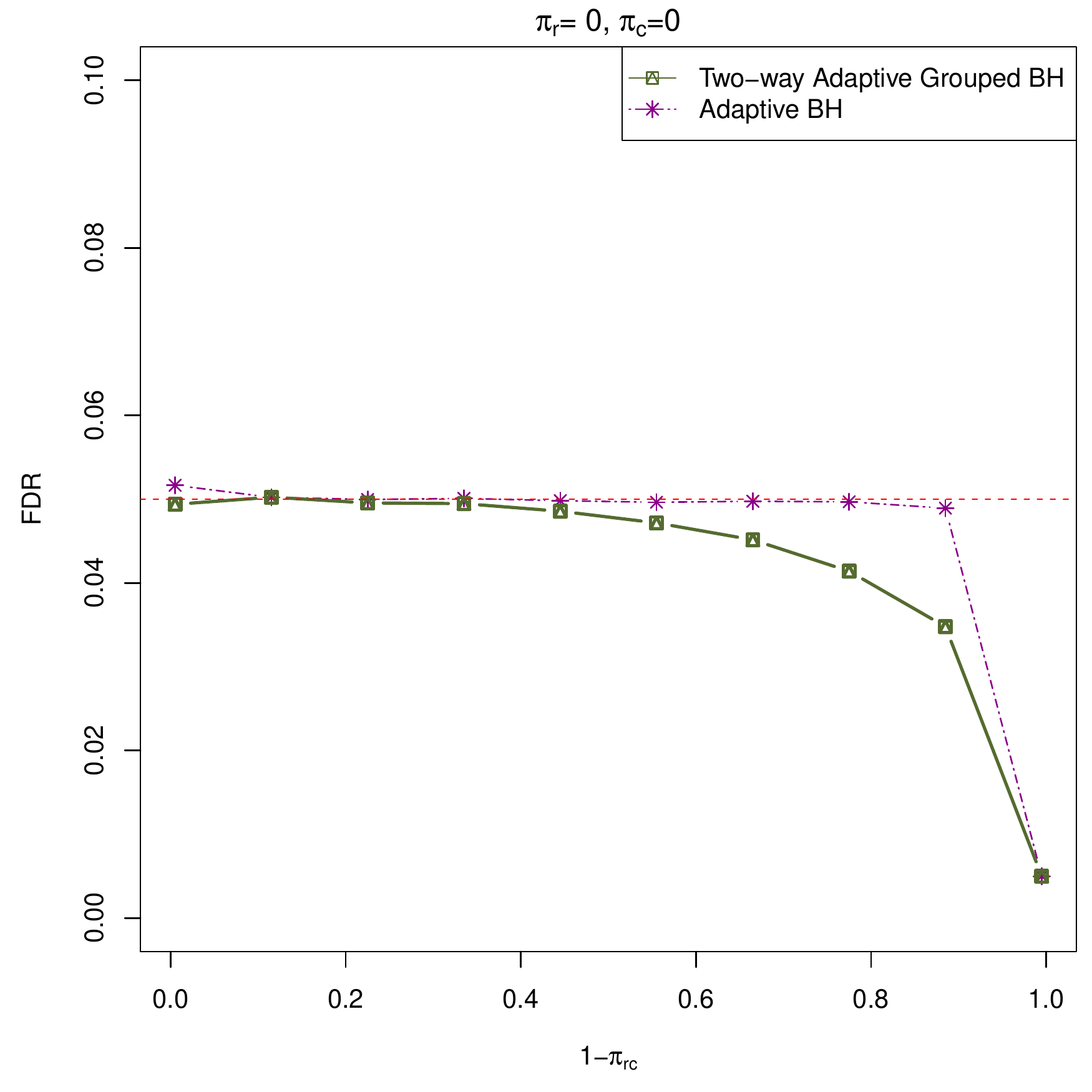}	
		\includegraphics[width=0.22\textwidth]{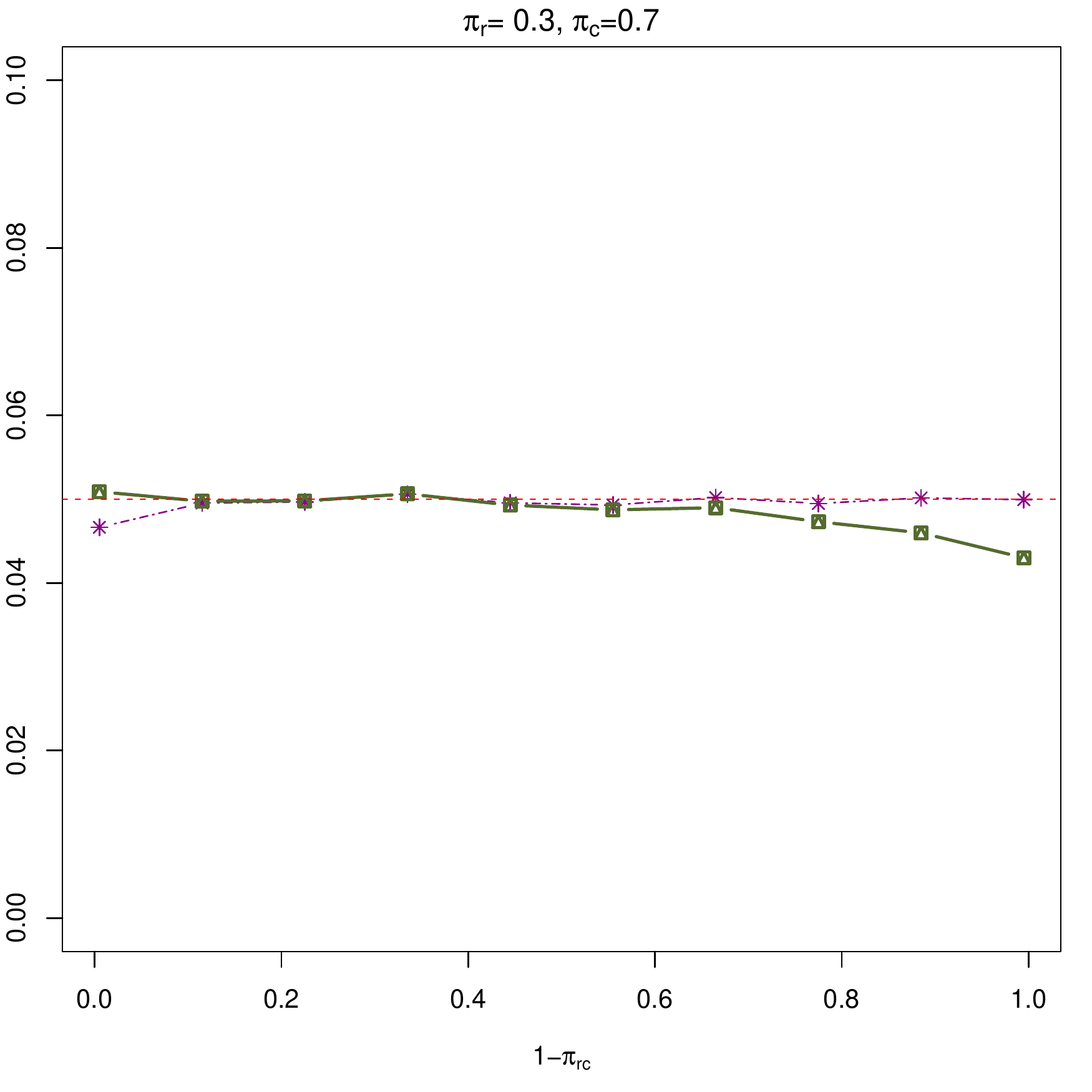}	
		\includegraphics[width=0.22\textwidth]{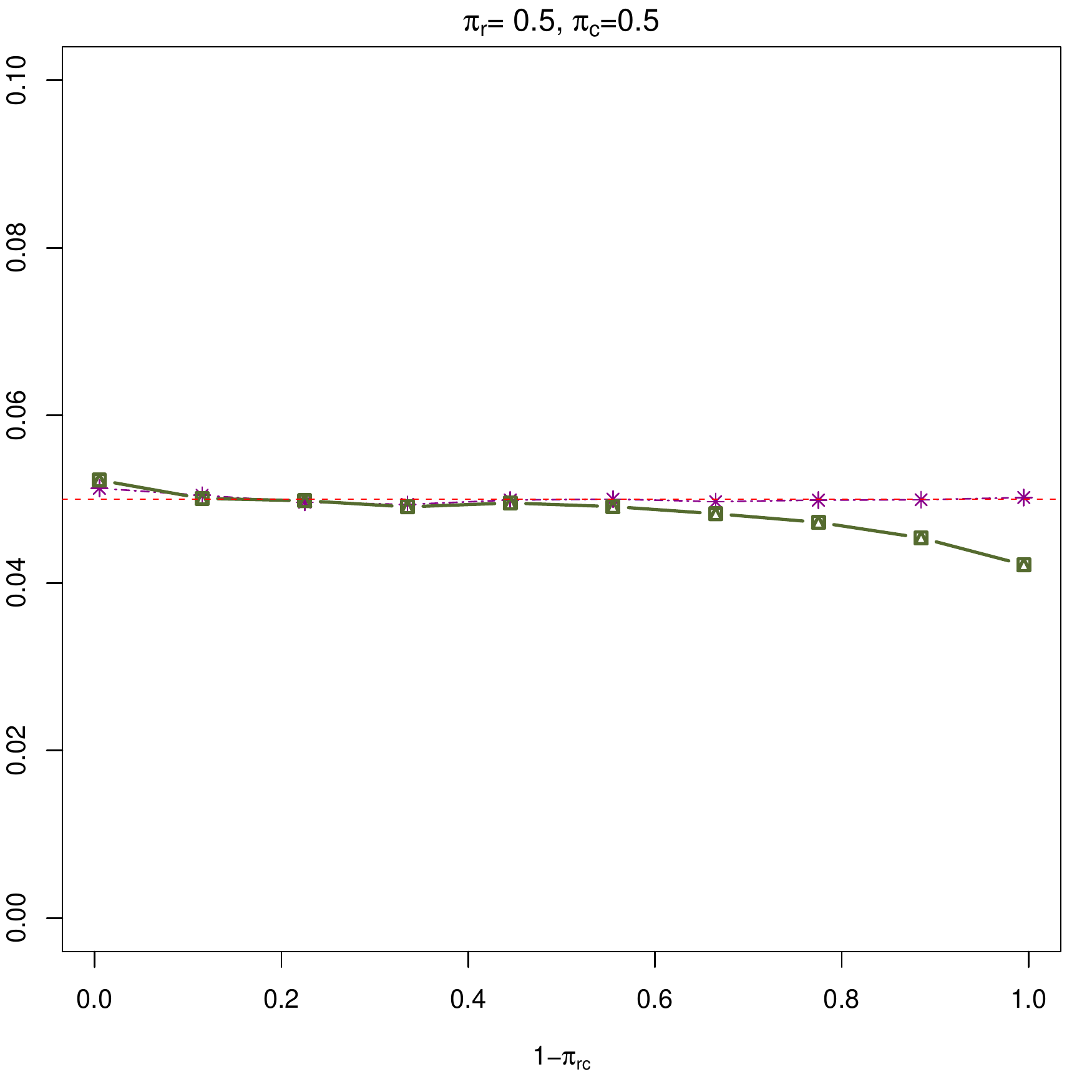}	
		\includegraphics[width=0.22\textwidth]{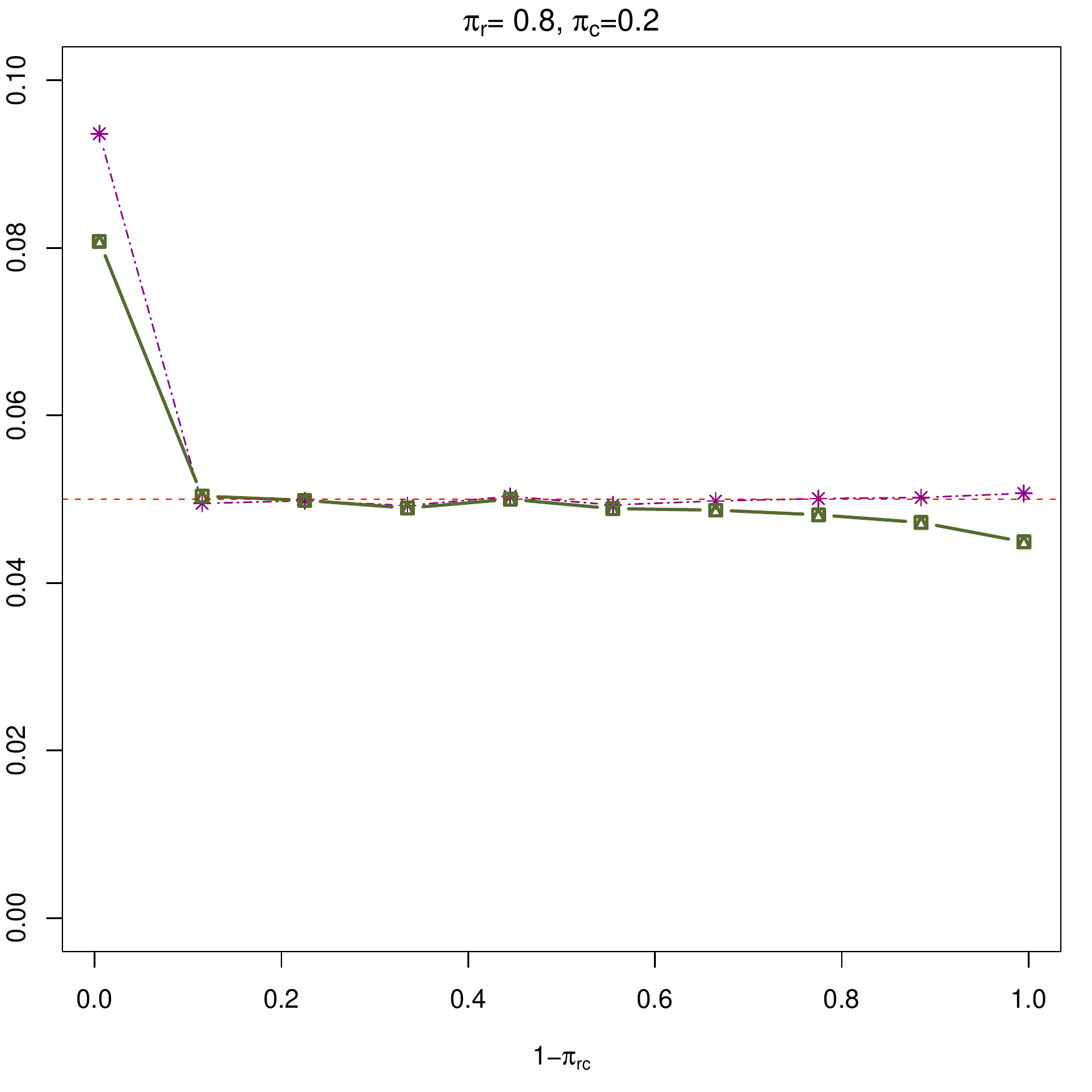}
		\caption{FDR Comparisons}	
		\label{fig11a}
	\end{subfigure}\\
	
	\begin{subfigure}[b]{\textwidth}
		\centering
		\includegraphics[width=0.22\textwidth]{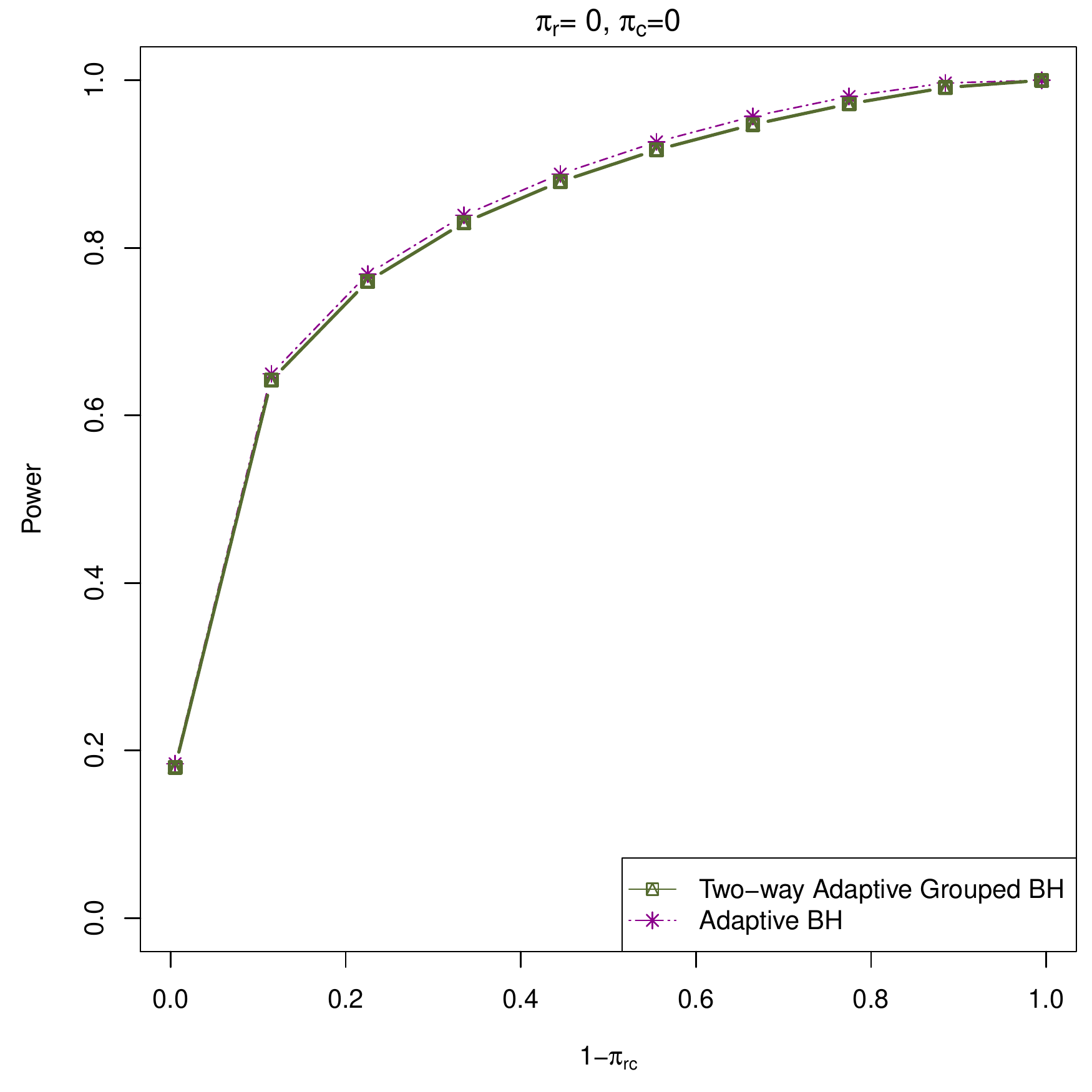}	
		\includegraphics[width=0.22\textwidth]{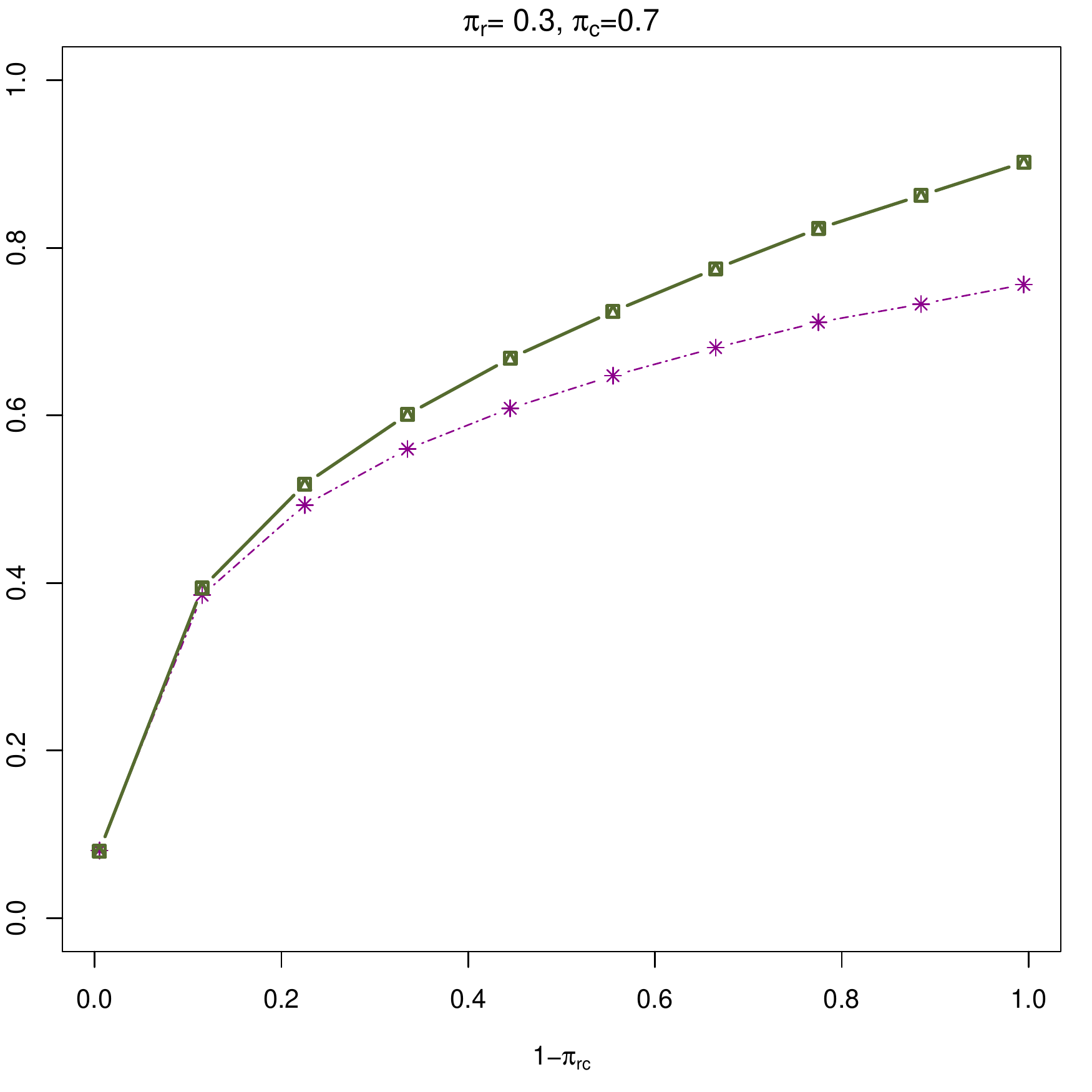}
		\includegraphics[width=0.22\textwidth]{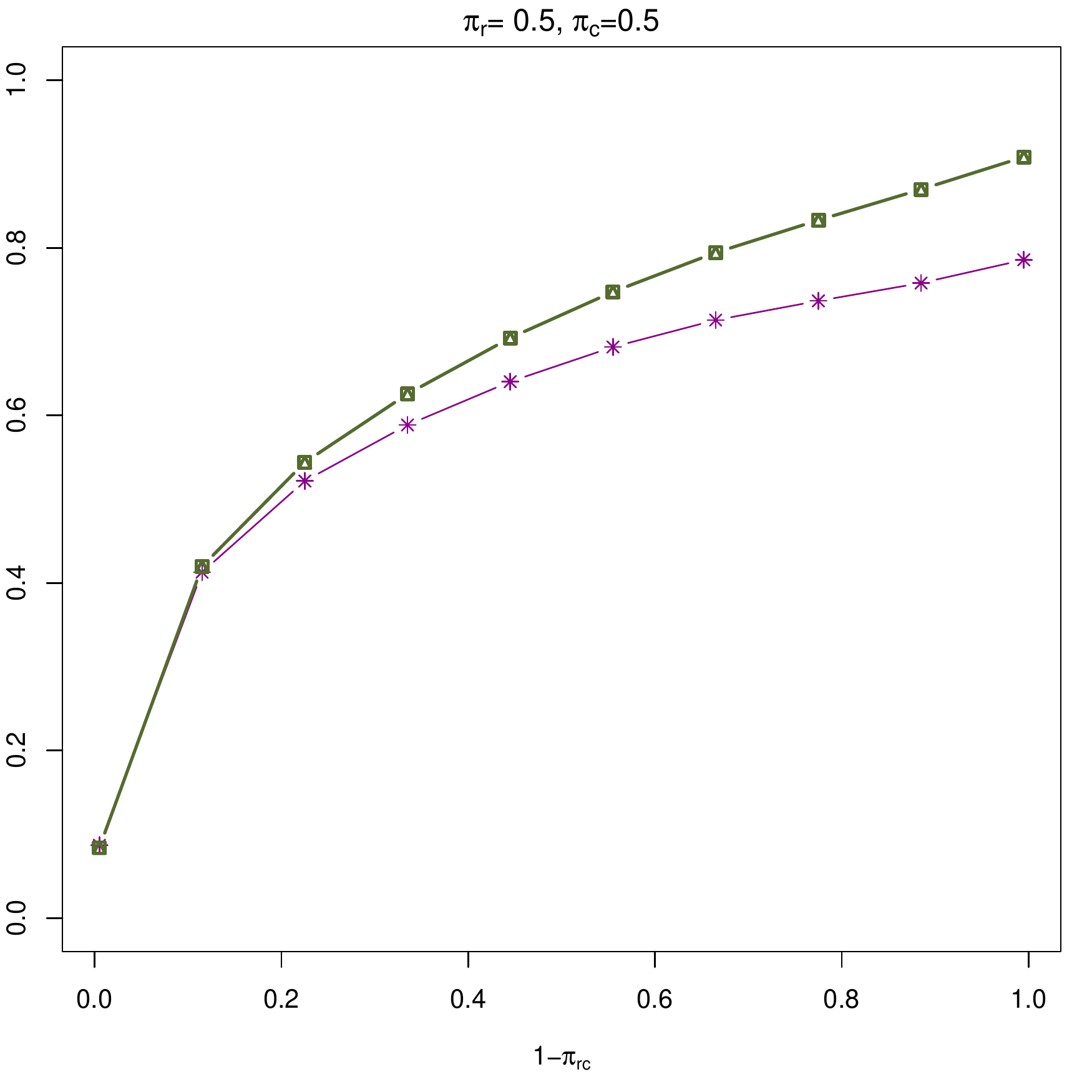}
		\includegraphics[width=0.22\textwidth]{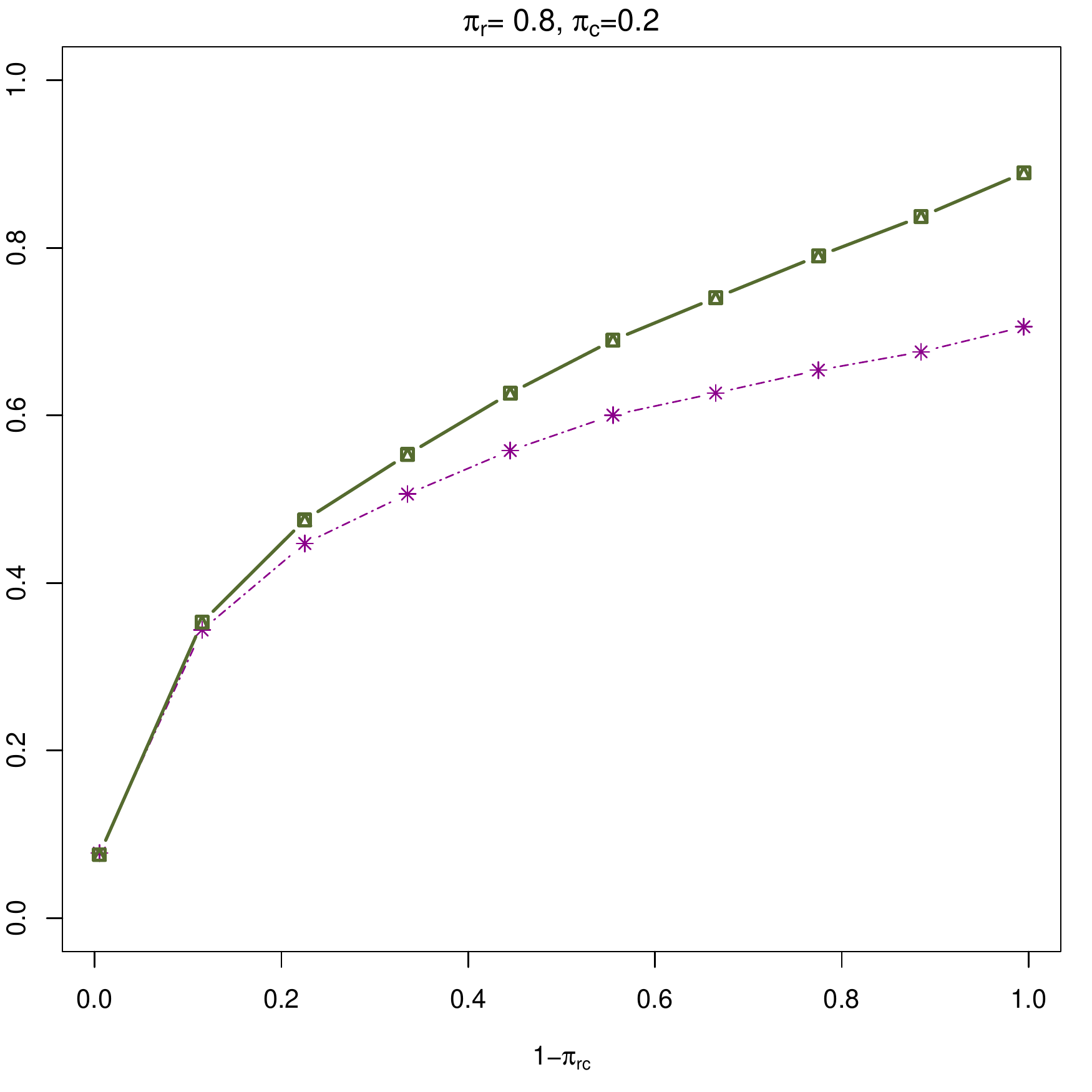}
		\caption{Power Comparisons}
		\label{fig11b}
	\end{subfigure}
	
	\caption{Comparison of the data-adaptive Two-way GBH$_{>1}$ procedure with the naive Adaptive BH method, when the hypotheses are independent. Set of parameters used is $(m = 50, n = 100, p=10,\rho_r =0, \rho_c = 0, \rho_p=0,\pi_r, \pi_c, \pi_{rc})$}\label{fig11}
\end{figure}

\begin{figure}
	\centering
	
	\begin{subfigure}[b]{\textwidth}
		\centering
		\includegraphics[width=0.22\textwidth]{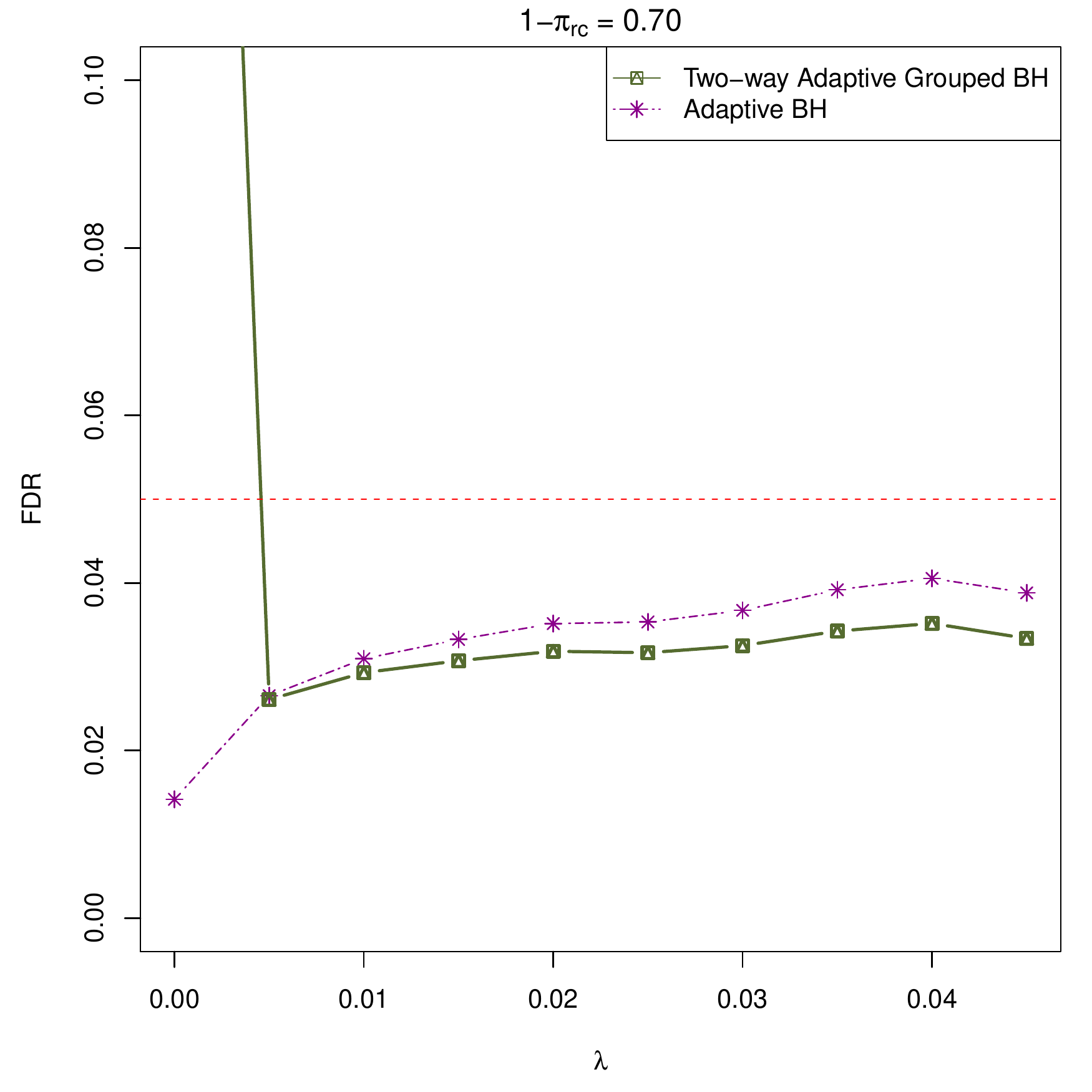}	
		\includegraphics[width=0.22\textwidth]{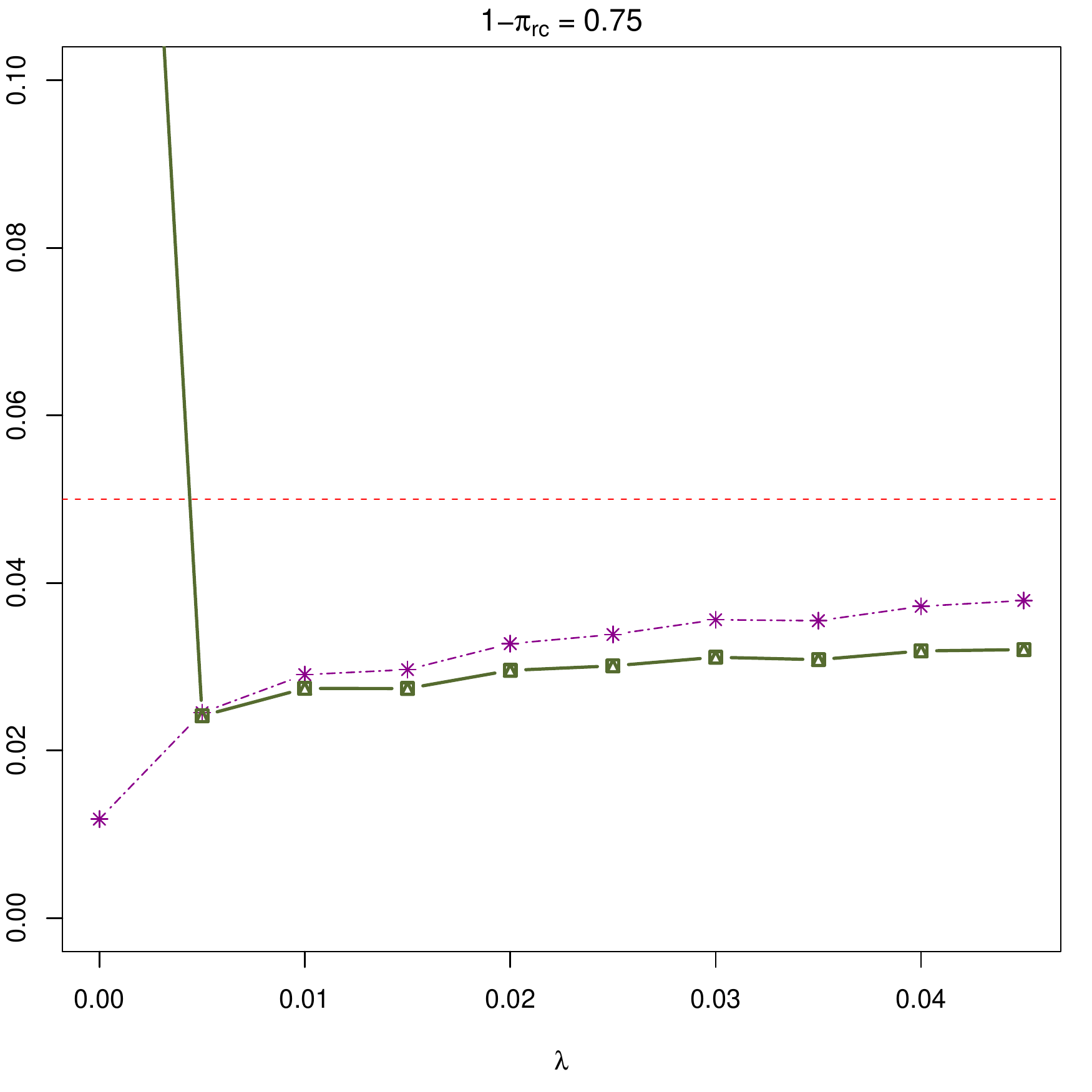}	
		\includegraphics[width=0.22\textwidth]{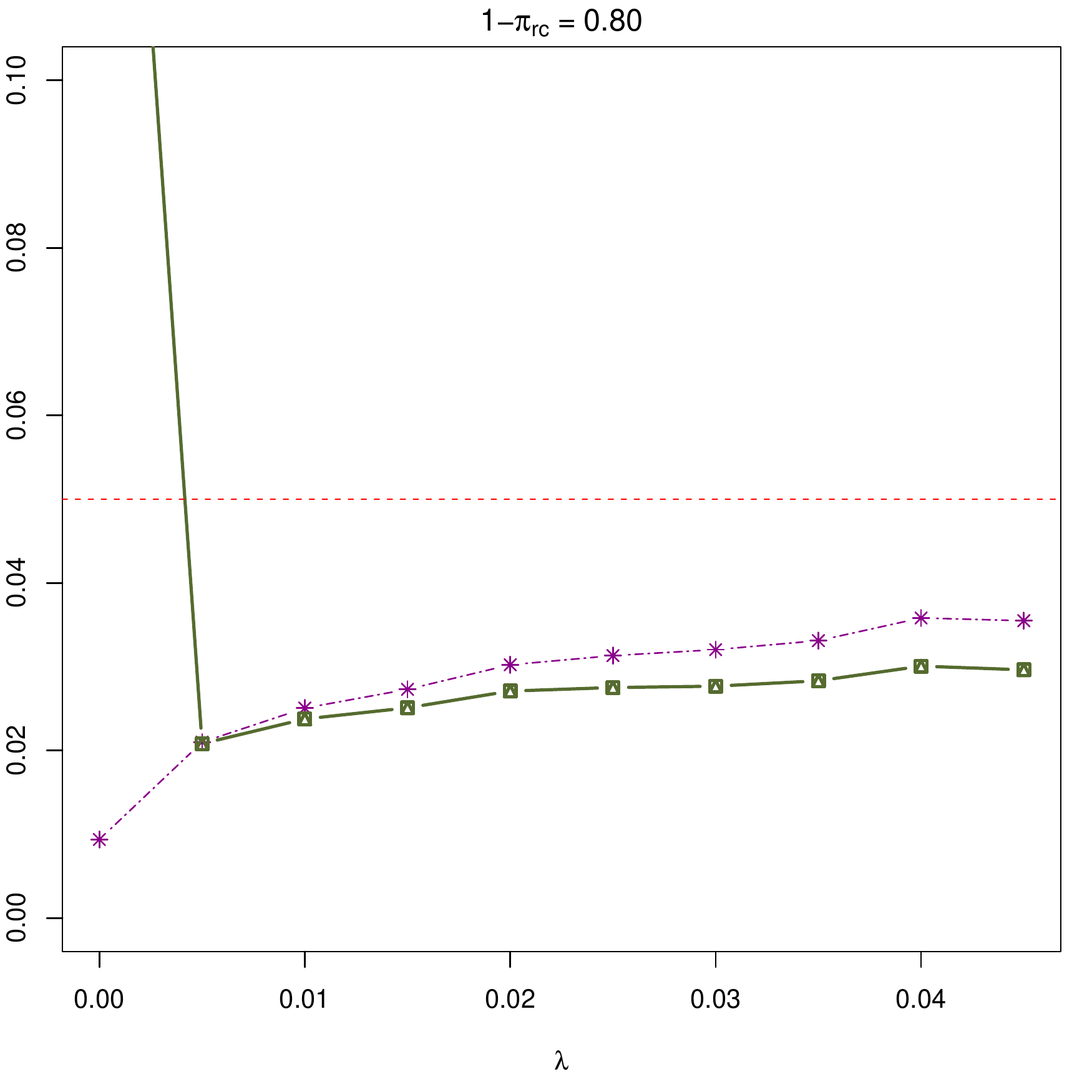}	
		\includegraphics[width=0.22\textwidth]{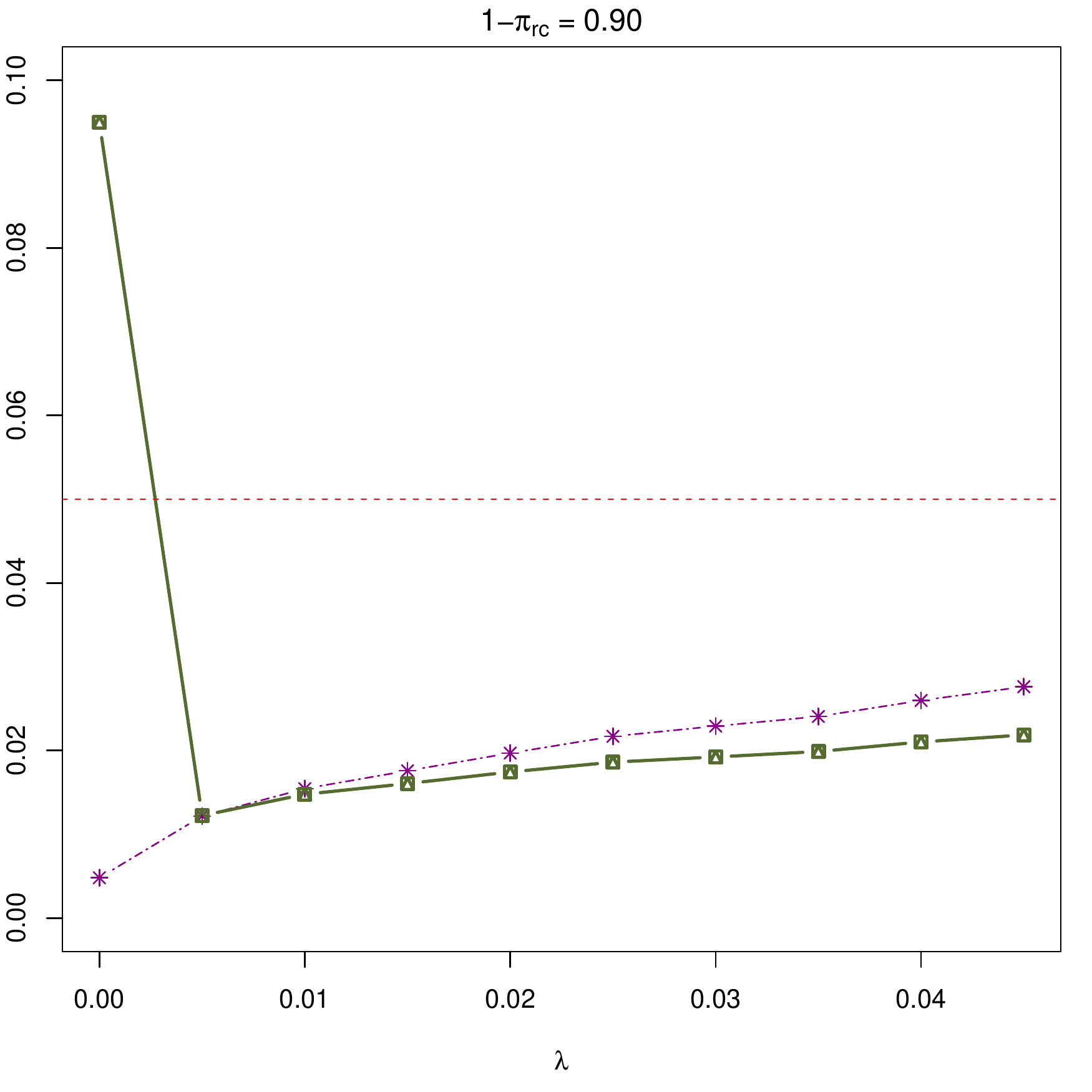}
		\caption{FDR Comparisons}	
		\label{fig12a}
	\end{subfigure}\\
	
	\begin{subfigure}[b]{\textwidth}
		\centering
		\includegraphics[width=0.22\textwidth]{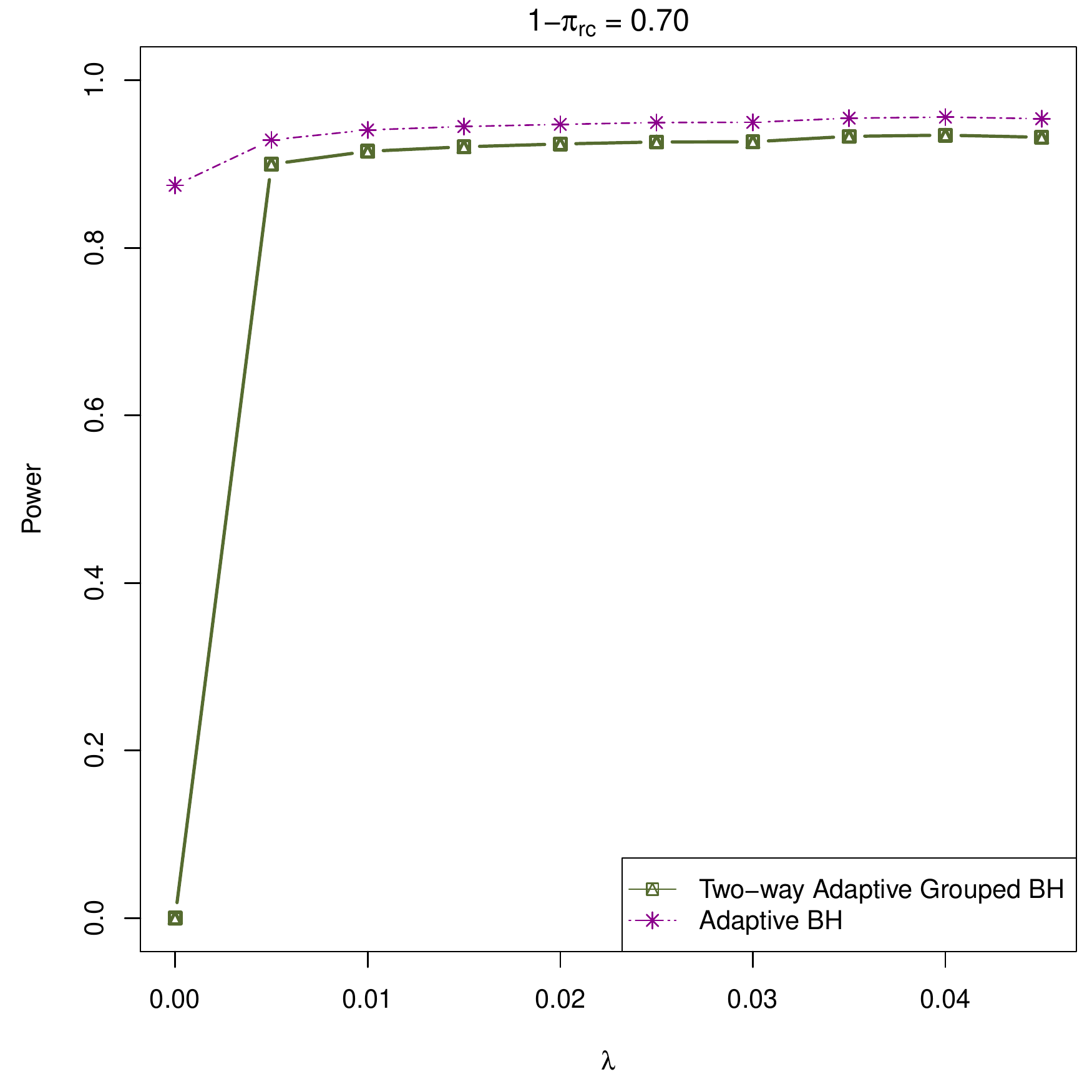}	
		\includegraphics[width=0.22\textwidth]{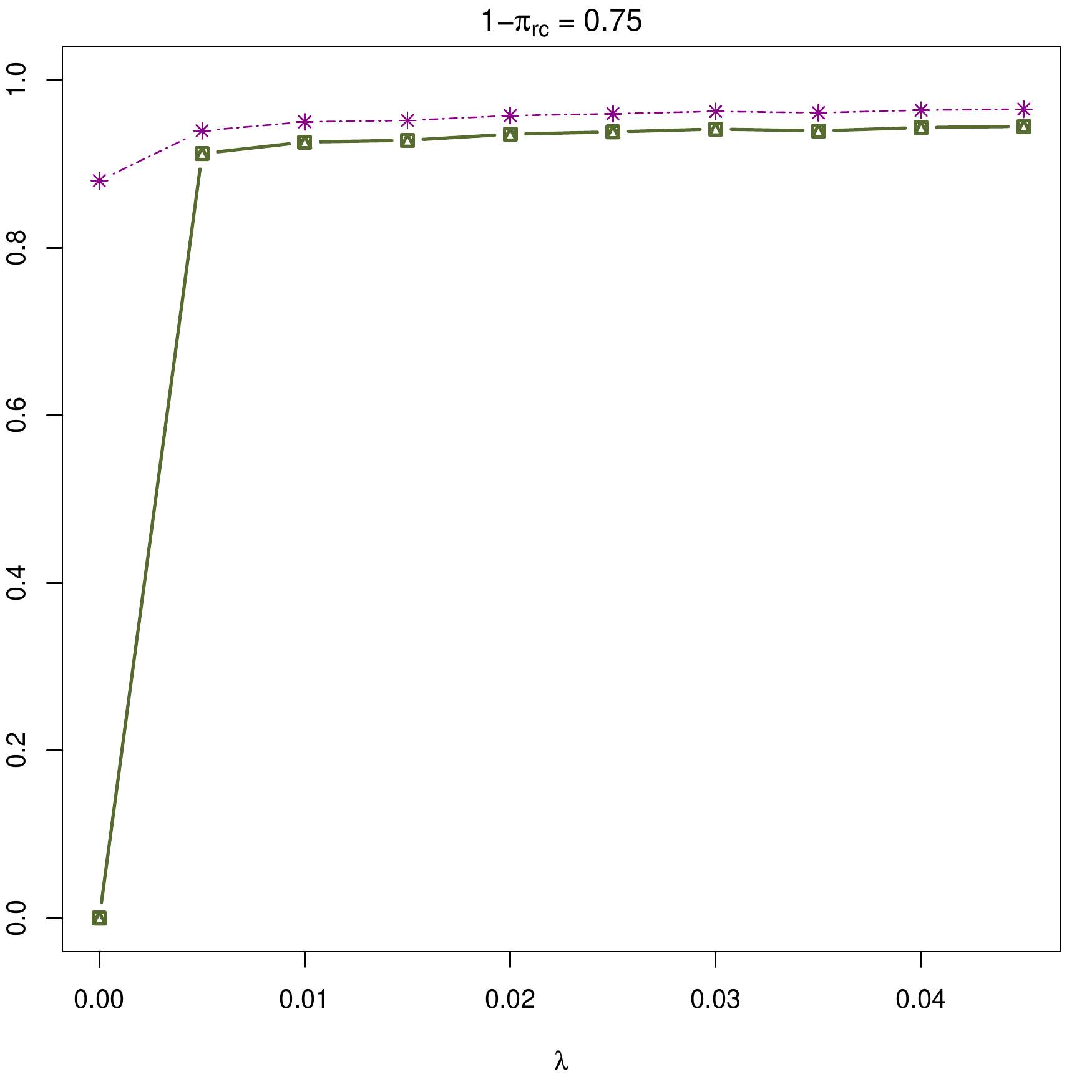}
		\includegraphics[width=0.22\textwidth]{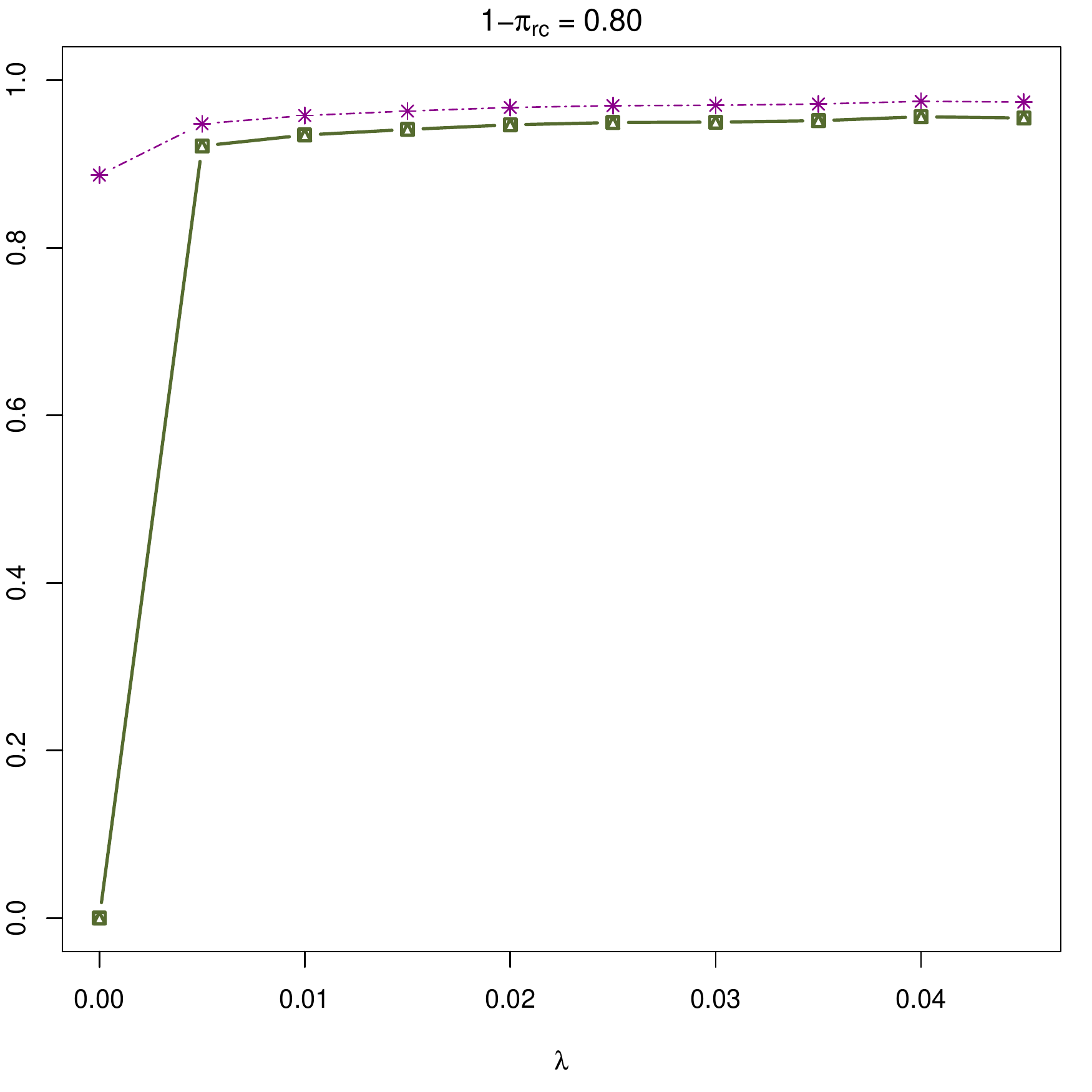}
		\includegraphics[width=0.22\textwidth]{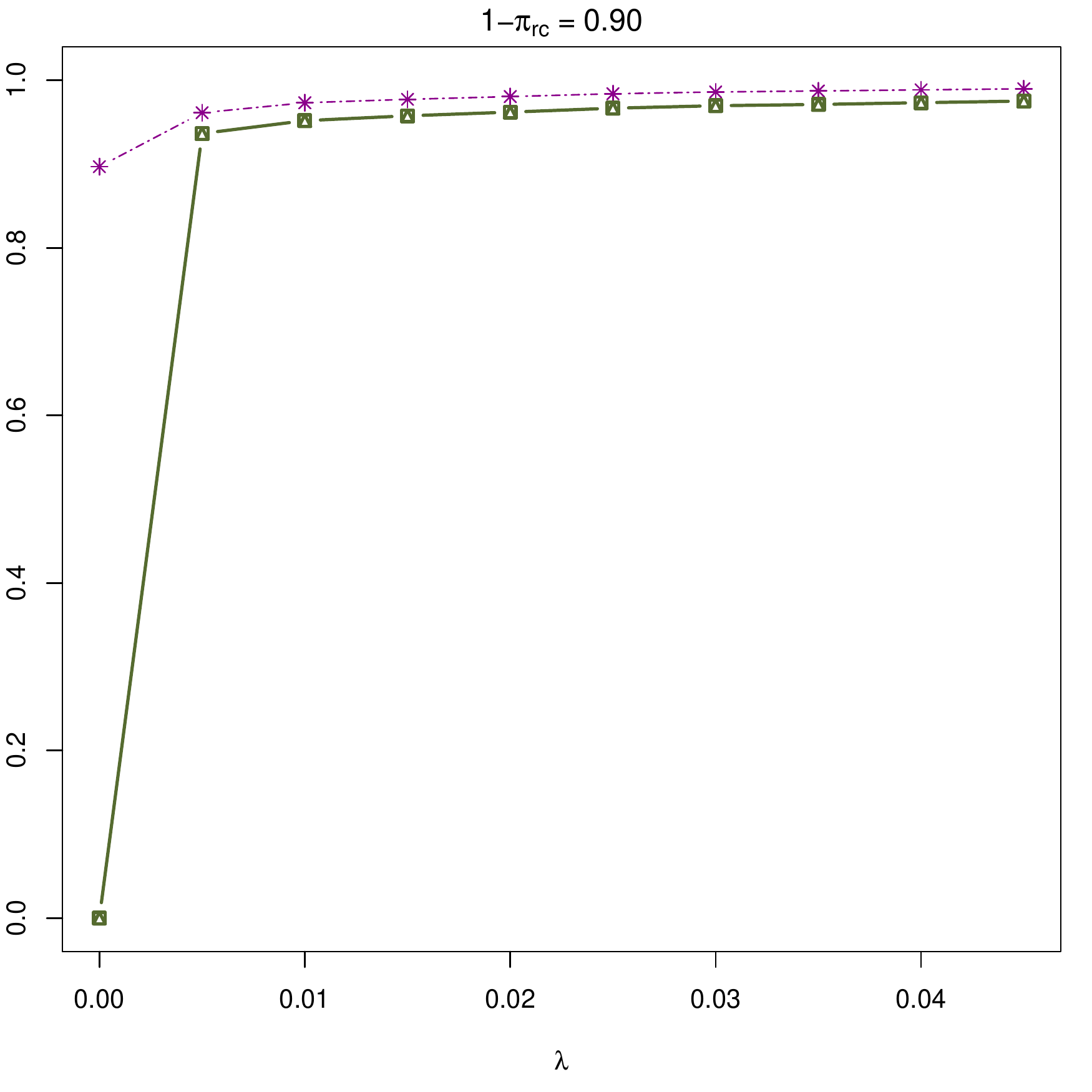}
		\caption{Power Comparisons}
		\label{fig12b}
	\end{subfigure}
	
	\caption{Comparison of the data-adaptive Two-way GBH$_{>1}$ procedure and the naive Adaptive BH procedure, applied to hypotheses with PRDS property, for varying choices of $0<\lambda < 0.05(=\alpha)$. Set of parameters used is $(m = 50, n = 100, p=10, \rho_r =0.3, \rho_c = 0.4, \rho_p=0.2, \pi_r=0, \pi_c=0, \pi_{rc})$}\label{fig12}
\end{figure}

\noindent {\it Comparison of Data-Adaptive Procedures}: As before, our focus in this case was to investigate the following two questions regarding the performance of our proposed Data Adaptive Two-Way GBH$_{>1}$ in Theorem \ref{th6} compared to naive Adaptive BH: (i) How well it performs under independence when both are theoretically known to control FDR? (ii) Can it can possibly control FDR under PRDS in view of the fact that such  control is yet to be theoretically proved for both of these procedures? Figures 11 and 12 display the findings of these investigation. Figure \ref{fig11}, which summarizes the results associated with answering question 1 (with $\lambda =0.5$), indicates that both methods have similar performance when the signals are uniformly distributed over the $m \times n$ grid (which occurs when $\pi_r = \pi_c = 0$). However, our proposed method is more powerful when the
signals are not uniformly distributed, which is displayed for the other combinations of the ($\pi_r, \pi_c$) values. Figure \ref{fig12} says, as in the case of two-way classification with one hypothesis per cell, our proposed Data-Adaptive Two-Way GBH$_{>1}$ can possibly control FDR under PRDS with choices of $\lambda < \alpha$, with a few instances being shown in the figure, when there is a high density of signals across all row and column groups.

\section{Application to Microbiome Data}
We apply  our two-way classified method to a microbial abundance dataset to illustrate its application in real scientific problems. We consider the \texttt{GlobalPatterns} dataset available through the Bioconductor package \texttt{phyloseq}. The data was first studied in \cite{Caporaso2011}
to analyze prevalence of microbial communities in different environments. The data consists of $19216$ microbes identified by their Operational Taxonomic (OTU) Numbers obtained from $26$ samples of $9$ different environments, which includes a mock environment. The environments are characterized by $7$ variables. Classification of the microbes according to their $7$ taxonomic ranks is provided along-with the phylogenetic tree describing the relationships among the microbes. The data records abundance patterns of each microbe across the nine sample environments. Since microbes closely related at the tips of the phylogenetic tree have similar characteristics, it is quite likely that they have similar abundance patterns which renders a positive dependence in the data. A smaller subset of this dataset, consisting of data on only microbes specific to the Chlamydiae bacteria taxon, was studied by \cite{SankaranHolmes14}.
For their analysis, they classified $21$ microbes into four groups formed according to their taxonomic families and invoked the Grouped BH procedure as suggested by \cite{Huetal2010} to find which particular microbes are significantly abundant across the environments.

We perform the analysis on a larger scale on the entire \texttt{GlobalPatterns} dataset. A linear regression is fit from data on each microbe's prevalence to the environment types. Each p-value corresponds to a particular microbe and an environment. The p-value $P_{ij}$, corresponding to the $i$th microbe and $j$th environment answers the question ``Is the $i$th microbe abundantly present in the $j$th environment?" In contrast to the analysis provided in \cite{SankaranHolmes14}, we consider the p-values to be in a two-way classified structure. Considering the microbes as individual groups furnishes $m=19216$ groups and together with $n=9$ environments we obtain a two way structure of dimensions $19216 \times 9$.

Instead of considering the microbes by their individual species, we classify them according to their taxonomic families. While higher taxonomic ranks such as taxonomic class, phyla, etc. can also be utilized for classification of the  microbes, groups formed as such are larger and members have wider variety of characteristics rendering the effect due to grouping vague. After adjusting for missing values and removing hypotheses with missing family labels, we obtain $N = 120942$ hypotheses classified into a grid of $m = 334$ families along rows and $n = 9$ environments along columns. Since there are unequal number of members (min: $1$ and max: $1658$) in each family, we use the data-adaptive method for two-way classified hypotheses with unequal number of hypotheses in each cell with weights as mentioned in expression (\ref{e3.8}).
The method identified $7584$ hypotheses as significant. In comparison, the adaptive BH procedure, applied to the entire set of hypotheses identified $7377$ hypotheses as significant.

\begin{figure}
\centering

\begin{subfigure}[b]{\textwidth}
\centering
\includegraphics[width=0.6\textwidth, keepaspectratio]{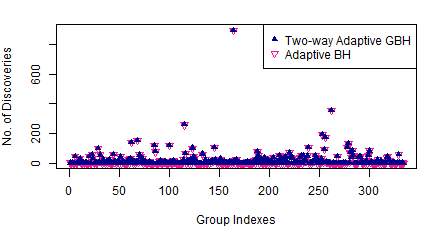}	
\caption{Discoveries made in microbial families}	
\label{fig14a}
\end{subfigure}~	

\begin{subfigure}[b]{\textwidth}
\centering
\includegraphics[width=0.5\textwidth, keepaspectratio]{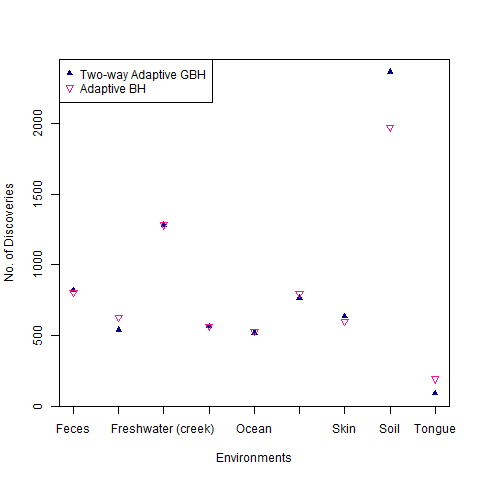}	
\caption{Number of microbes discovered in each environment}
\label{fig14b}
\end{subfigure}

\caption{Comparison of the data-adaptive procedure for two-way classified hypotheses with multiple hypothesis per cell, with the adaptive BH procedure, when applied to the microbiome dataset. }\label{fig14}
\end{figure}
\section{Concluding Remarks}
In this article we have introduced a well-founded framework for one- and two-way classified hypotheses, and an effective yet simple multiple testing method to test such hypotheses. Through simulations and data analysis, we have established that existing multiple testing procedures that do not take into consideration the layout of the hypotheses, are not sufficiently efficient to study such structures. Our proposed method in its oracle form controls FDR for independent hypotheses as well as for positively dependent hypotheses that satisfy the PRDS property. The corresponding data-adaptive procedure maintains control on FDR non-asymptotically for independent hypotheses. Simulation studies show that it is also capable to control FDR for hypotheses with PRDS property, under certain conditions and when the density of signals is high. The method is flexible as it adapts itself to one-way or two-way classification structures depending on the choice of weights. We suggest generic weights suitable for such structures of hypotheses, and these weights can be modified to involve additional information appropriate in any particular situation.

In essence, this article explores an underdeveloped area of multiple testing, which is adapting standard procedures to structures of hypotheses more complex than what these procedures are initially designed under to gain more efficiency. Occurrence of hypotheses exhibiting complex structures, especially in the form of being classified according to multiple criteria, is becoming more and more prevalent in modern statistical investigations with the current boom of Big Data producing massive amounts of data from various sources.  However, research focused on developing methods efficiently accommodating such structural information has been taking place at a pace that is much slower than one would hope for.  Some advances have indeed been made in one-way classification setting (\cite{Huetal2010}, \cite{Liu20161}, and \cite{2017arXiv171205014S}), but there is still scope of making that advancement to a greater extent to provide a fuller coverage of that setting. Moreover, no advancement has been made yet in the direction of adapting methods to two-way classification setting and beyond.  This article makes a significant contribution in this broader domain. There remains a scope of improving the proposed method through specific choices of weights suitable to specific scenarios and produce newer methods that can effectively and efficiently be extended possibly to multi-way classification settings.

We conclude this section with some open issues to be resolved in future research. In extending the Oracle One-Way GBH from one- to two-way classification setting before constructing a data-adaptive version of it, we have proposed using certain specific combinations of the row and column weights (see, (\ref{e8}), (\ref{e10}), (\ref{e13}), (\ref{e14}), (\ref{e15a}) and (\ref{e15b})). However, it would be worthwhile to investigate if these weights can be combined in an optimal manner. The data-adaptive procedures here have been proposed by estimating weights using \cite{Storeyetal2004} type estimates of proportions of true nulls. Developing alternative  data-adaptive procedures using other types of estimates of these proportions would be an important undertaking.
\appendix
\section{Appendix}
\subsection{Proof of Result \ref{result1}}\label{proof1}

The FDR of a stepup procedure based on the weighted p-values and any set of critical constants $c_1 \le \cdots \le c_N$ can be expressed as follows [see, e.g., Sarkar (2002)]:
\begin {eqnarray} \text {FDR} & = & \sum_{r=1}^{N} \sum_{i \in I_0} \frac{1}{r}\text{Pr} \left ( P_{i}^{w} \le c_{r}, R^{w}_{(-i)} = r-1 \right ) \nonumber \\ & = & \sum_{i \in I_0} \text{Pr}(P^w_{i} \le c_1) +\sum_{r=1}^{N-1} \sum_{i \in I_0} \text{E} \left [ \text{Pr} \left (R^{w}_{(-i)} \ge r~|P^{w}_{i} \right ) \left \{ \frac{I(P_{i}^{w} \le c_{r+1})}{r+1} -  \frac{I(P_{i}^{w} \le c_{r})}{r} \right \}  \right ], \nonumber \\ \end{eqnarray} (assuming $c_0=0$ and $0/0=0$), with $R^{w}_{(-i)}$ representing the number of rejections in the stepup procedure based on the weighted p-values $(P^{w}_1, \ldots, P^{w}_N)/\{P^{w}_i\}$ and the critical constants $c_i$, $i=2, \ldots, N$. With  $c_i = i c_1$, $i=1, \ldots, N$, it is bounded above by $\sum_{i \in I_0} \text{Pr} (P^w_{i} \le c_1)$ under PRDS, which can be shown by making use of the following observations for each $i \in I_0$.

\noindent $\bullet$ For each $r=1, \ldots, N-1$,

\begin {eqnarray} \label{e25}& & \text{E} \left [ \text{Pr} \left (R^{w}_{(-i)} \ge r~|P^{w}_{i} \right )
\left \{ \frac{I(P_{i}^{w} \le c_{r+1})}{r+1} -  \frac{I(P_{i}^{w} \le c_{r})}{r} \right \} \right ] \nonumber \\
& \le &  \text{Pr} \left ( R^{w}_{(-i)} \ge r ~|P^{w}_{i} = c_r \right ) \left \{ \frac{\text{Pr}(P_{i}^{w} \le c_{r+1})}{r+1} -  \frac{\text{Pr} (P_{i}^{w} \le c_{r})}{r} \right \} \nonumber \\
& \le & 0. \end{eqnarray} The first inequality in (\ref{e25}) follows from the following two results:

\noindent (i) $\text{Pr} \left (R^{w}_{(-i)} \ge r~|P^{w}_{i} \right ) = \text{E} \left \{ I(R^{w}_{(-i)} \ge r)~|P^{w}_{i} \right \}$ is non-increasing in $P^w_{i}$, since $I(R^{w}_{(-i)} \ge r)$ is a non-increasing function of the weighted p-values, and the PRDS condition on the p-values translates to that on the weighted p-values, and

\noindent (ii) $\frac{I(P_{i}^{w} \le c_{r+1})}{r+1} -  \frac{ I(P_{i}^{w} \le c_{r}))}{r}$ changes sign from - to + at $P^w_i = c_r$ as $P^w_{i}$ increases.

\noindent The second inequality follows from the fact that \begin {eqnarray*} \frac{\text{Pr}(P_{i}^{w} \le c_{r+1})}{r+1} - \frac{ \text{Pr} (P_{i}^{w} \le c_{r})}{r}  = \frac{\min\{(r+1)c_1, 1 \}}{r+1} - \frac{ \min \{rc_1, 1 \}}{r} \le 0. \end {eqnarray*}

\noindent $\bullet$ For $r=0$, the expectation in (\ref{e25}) equals $\text{Pr} (P^w_{i} \le c_1)$, and so with $c_1 = \alpha/N$, $\sum_{i \in I_0} \text{Pr} (P^w_{i} \le c_1) = \sum_{i \in I_0} \min \{\frac{\alpha}{Nw_i}, 1\} \le \frac{\alpha}{N}\sum_{i \in I_0} \frac{1}{w_i}$.

Thus, Result 1 is proved.
\bibliographystyle{Chicago}

\bibliography{two_way_references}

\end{document}